\documentclass{aa}  

\usepackage{graphicx}
\usepackage{txfonts}
\usepackage[table]{xcolor}
\usepackage{lscape}
\usepackage{longtable}
\usepackage{verbatim}
\usepackage{hyperref}
\usepackage{xcolor}
\usepackage{subcaption}
\hypersetup{colorlinks, linkcolor={blue!70!black}, citecolor={blue!70!black}, urlcolor={blue!70!black}
}

\usepackage{color}

\begin{document}

   \title{Substructure in the stellar halo near the Sun. II. Characterisation of independent structures}


   \author{Tom\'{a}s Ruiz-Lara\inst{1}, Tadafumi Matsuno\inst{1}, S. Sofie L\"{o}vdal\inst{1}, Amina Helmi\inst{1}, Emma Dodd\inst{1} \and Helmer H. Koppelman\inst{2}
          }
   \titlerunning{Characterisation of independent halo structures}
   \authorrunning{T. Ruiz-Lara et al.}
   \institute{Kapteyn Astronomical Institute, University of Groningen,
          Landleven 12, 9747 AD Groningen, The Netherlands\\
   \email{tomasruizlara@gmail.com}
    \and
          School of Natural Sciences, Institute for Advanced Study, 1 Einstein Drive, Princeton, NJ 08540, USA\\
             }

   \date{Received --, --; accepted --, --}

 
  \abstract
   {In an accompanying paper, we present a data-driven method for clustering in `integrals of motion' (IoM) space and apply it to a large sample of nearby halo stars with 6D phase-space information. The algorithm identified a large number of clusters, many of which could tentatively be merged into larger groups. }
   {The goal here is to establish the reality of the clusters and groups through a combined study of their stellar populations (average age, metallicity, and chemical and dynamical properties) to gain more insights into  the accretion history of the Milky Way.}
   {To this end, we developed a procedure that quantifies the similarity of clusters based on the Kolmogorov-Smirnov (KS) test using their metallicity distribution functions, and an isochrone fitting method to determine their average age, which is also used to compare the distribution of stars in the colour--absolute magnitude diagram (CaMD). Also taking into consideration how the clusters are distributed in integrals of motion space allows us to group clusters into substructures and to compare substructures with one another.}
   {We find that the 67 clusters identified by our algorithm can be merged into 12 extended substructures and 8 small clusters that remain as such. The large substructures include the previously known {\it Gaia}-Enceladus, Helmi streams, Sequoia, and Thamnos 1 and 2. We  identify a few over-densities that can be associated with the hot thick disc and host a small  metal-poor population. Especially notable is the largest (by number of member stars) substructure in our sample which, although peaking at the  metallicity characteristic of the thick disc, has a very well populated metal-poor component, and dynamics intermediate between the hot thick disc and the halo.  We also identify additional debris in the region occupied by Sequoia with clearly distinct kinematics, likely remnants of three different accretion events with progenitors of similar masses. Although only a small subset of the stars in our sample have chemical abundance information, we are able to identify different trends of [Mg/Fe] versus [Fe/H] for the various substructures, confirming our dissection of the nearby halo.}
   {We find that at least 20\% of the halo near the Sun is associated to substructures. When comparing their global properties, we note that those substructures on retrograde orbits are not only more metal-poor on average but are also older. We provide a table summarising the properties of the substructures, as well as a membership list that can be used for follow-up chemical abundance studies for example.}


   \maketitle
%
\section{Introduction}
\label{intro}

Satellite accretion and galaxy interactions appear to have been 
key actors in the formation and evolution of galaxies. Although mergers must have happened throughout the history of the Universe \citep[][]{1994Natur.370..194I,2015MNRAS.454.1742K, 2018ApJS..237...36P}, at high redshift they were likely much more ubiquitous because of the higher average density of the Universe
\citep[][]{1996ApJS..107....1A, 2007IAUS..235..381C}. This is naturally incorporated into the current cosmological paradigm,  which 
predicts that galaxies grow in mass hierarchically through the accretion of smaller systems \citep[e.g.][]{2017MNRAS.464.1659Q}. For many systems, and particularly for large disc galaxies, their accretion history is expected to have been violent in the early stages, and more quiescent at later times \citep[see e.g.][]{1984Natur.311..517B, 2014ARA&A..52..291C,2017MNRAS.467..179G}. 

Reconstruction of the detailed accretion histories of external galaxies is unfortunately beyond current observational capabilities \citep[although see e.g. ][]{2018NatAs...2..737D,2019Natur.574...69M,
2020MNRAS.496.1579Z,2021MNRAS.507.3089D}. However, our privileged position within our own Galaxy enables the characterisation of its accretion history. This is because individual stars retain information of their formation sites in their dynamics and chemical abundances (see \citealt[][]{2002ARA&A..40..487F} and \citealt[][]{2020ARA&A..58..205H} and references therein). Debris from ancient accretion events is expected to have been deposited mostly in the Galactic halo \citep[][]{1978ApJ...225..357S,2005ApJ...635..931B,2019ApJ...874L..35M}, with the most massive mergers dominating the inner regions \citep[][]{2002PhRvD..66f3502H, 2010MNRAS.406..744C}. 

With the goal of disentangling accretion events in the Galactic halo,  analyses have focused on the space defined by integrals of motion (IoM) such as energy or angular momentum \citep{2000MNRAS.319..657H} or actions \citep[][]{2008MNRAS.390..429M}, based on the fact that stars from accreted systems remain clumped in such a space even after they have phase-mixed \citep[][]{2000MNRAS.319..657H}. The advent of full phase-space information for large samples of stars provided by the {\it Gaia} mission \citep[][]{2016A&A...595A...1G, 2018A&A...616A...1G, 2021A&A...649A...1G}, sometimes in combination with  ground-based spectroscopic surveys (e.g. LAMOST, \citealt[][]{2020ApJS..251...27W, 2019RAA....19...75L}; RAVE, \citealt[][]{2020AJ....160...82S, 2020AJ....160...83S};  GALAH, \citealt[][]{2021MNRAS.506..150B}; or APOGEE \citealt[][]{2020ApJS..249....3A}), has recently made such studies feasible in the halo near the Sun.

The emerging view from these studies \citep[see e.g.][]{2018Natur.563...85H, 2019A&A...630L...4M, 2019A&A...631L...9K, 2019A&A...625A...5K, 2019MNRAS.488.1235M,  2020MNRAS.494.3880B, 2020ApJ...891...39Y} is that the stellar halo indeed contains large amounts of debris associated to different accretion events experienced by the Milky Way. Probably one of the most important events in the early history of the Galaxy is the merger with {\it Gaia}-Enceladus \citep[GE, with a stellar mass of $\sim 6 \times 10^8 M_\odot$ at the time of merging, $\sim$10~Gyr ago,][see also \citealt{Belokurov2018}]{2018Natur.563...85H}. Stars from GE together with an in situ, more metal-rich population that was likely heated up from an ancient disc by this event, dominate the inner halo near the Sun \citep[][]{2018A&A...616A..10G, 2018ApJ...860L..11K,2019NatAs...3..932G, 2020MNRAS.494.3880B}. Apart from GE, several other Galactic building blocks have been uncovered in the inner halo. The Helmi streams discovered by \citet[][]{1999Natur.402...53H} are debris from a system with approximately $10^8 M_\odot$ in stars that was likely accreted to the Milky Way around 5 to 8~Gyr ago \citep[][]{2007AJ....134.1579K, 2019A&A...625A...5K}. More recently, \citet[][]{2019MNRAS.488.1235M} reported the existence of another accreted system, Sequoia. With a stellar mass of $5\times 10^7M_\odot$, Sequoia may have provided the Milky Way with most of the nearby  retrograde halo stars, including several globular clusters \citep{2018MNRAS.478.5449M}. Thamnos, another possible building block of the Milky Way halo and composed of stars with high binding energy, relatively low inclination, and retrograde orbits, was first reported in \citet[][]{2019A&A...631L...9K}. Thamnos was originally divided into two components (Thamnos 1 and Thamnos 2), which could have originated from the same progenitor, a stellar system with $M_\star\leq 5 \times 10^6 M_\odot$. In addition to these events, a growing number of studies have unveiled several small and subtle stellar substructures (Wukong, Arjuna, Rg5, etc.) possibly linked to other minor merger events in the history of our Galaxy \citep[][]{2018MNRAS.478.5449M, 2020ApJ...901...48N, 2020ApJ...891...39Y}. 

Besides studies making use of the distribution of stars in IoM space, the characterisation of the age--metallicity relation of Galactic globular clusters as well as their dynamical and chemical properties led \citealt{2019MNRAS.486.3180K} to postulate the presence of a possible major building block in the inner Galaxy\footnote{This substructure has been named differently by different authors ({\it Kraken} in \citealt{2019MNRAS.486.3180K}, {\it Heracles} in \citealt{2021MNRAS.500.1385H} or simply, Inner Galaxy Structure, IGS). There is currently some debate as to whether or not all these works are referring to the same structure.} \citep[see also][]{2019A&A...630L...4M, 2020MNRAS.498.2472K, 2020MNRAS.493.3363H, 2021MNRAS.500.1385H}. Comparisons with the E-MOSAICS simulations \citep[][]{2018MNRAS.475.4309P, 2019MNRAS.486.3134K} suggest that this `Kraken'-like event might have had a stellar mass of around $2\times10^8 M_\odot$ at the time of the merger (similar to that of GE), around 11~Gyr ago \citep[][]{2020MNRAS.498.2472K}. 

Therefore, through the application of ever more sophisticated techniques to data sets of increasing quality, we have gained a better and more complete picture of the accretion history of the Milky Way. Nevertheless, some important questions remain, such as how many independent merger events our Galaxy has experienced; or what their characteristics were. By answering these seemingly simple questions, not only will we have a better understanding of the role of accretion in galaxy formation \citep{2013MNRAS.434.3348C, 2017MNRAS.464.1659Q, 2018MNRAS.475..648P}, but the characterisation of the different building blocks of our Galaxy will provide key information on galaxy formation and evolution in different environments and at different times. For example, characterisation of the smallest (accreted) systems would allow us to further constrain the subhalo mass fraction, and therefore cosmological models \citep[e.g.][]{2008ApJ...679.1260M, 2009MNRAS.395.1376W, 2013MNRAS.429..633G, 2013ApJ...772..109B, 2019MNRAS.486.4545F}.

To reach these goals, statistically robust identification of the various substructures that may be associated to merger debris is necessary, as well as reliable  determination  of the likelihood of stars to belong to any of the substructures. In this work, we use the outcome of a data-driven hierarchical clustering algorithm \citep[see][paper I from now on]{2022arXiv220102404S} to this end. In paper I, we identified $67$ highly significant clusters in IoM space for nearby stars with halo-like kinematics. Through inspection of the properties of the hierarchical tree of the clusters (via a dendrogram), and comparison to previous identifications of substructures in the nearby halo, we tentatively suggested that it can be divided in six main groups and a number of independent clusters. However, it is only through proper characterisation of the internal properties of all the significant clusters detected by the algorithm, together with a thorough comparison of their properties, that we can establish their independence or relationship in case of a common origin. This is the goal of the present paper. 

The organisation of this paper is as follows. In Sect.~\ref{data}, we present the sample of stars  used for our analyses, and briefly summarise the main characteristics of our clustering approach. Section~\ref{analysis} presents the methodology used to characterise and compare the stellar content of clusters, which is based on isochrone fitting to describe the distribution of stars in the colour--absolute magnitude diagram (CaMD), and on their metallicity distribution functions. These comparisons allow us to define possible independent substructures, which we describe in  Sect.~\ref{results}. We then proceed to  characterise their average age and chemical properties and provide an updated view of the Galactic halo in Sect.~\ref{updated_view}. An interesting result from our analysis is that we find that the building blocks contributing stars with very retrograde orbits  are not only more metal poor but are also on average older than those in prograde orbits. This and other conclusions are provided in Sect.~\ref{conclusions}.  

Please note that in what follows, we mostly use the word {\it cluster} to refer to the different groups of stars identified by the clustering algorithm of paper~I, that is, to clustered data in dynamical space, rather than to gravitationally bound systems, unless stated otherwise. We use the words {\it group} or {\it substructure} to refer to the different associations of clusters found that could potentially correspond to different events in the history of our Galaxy (i.e. different building blocks). 

\begin{figure*}
\centering 
\includegraphics[width = 0.99\textwidth]{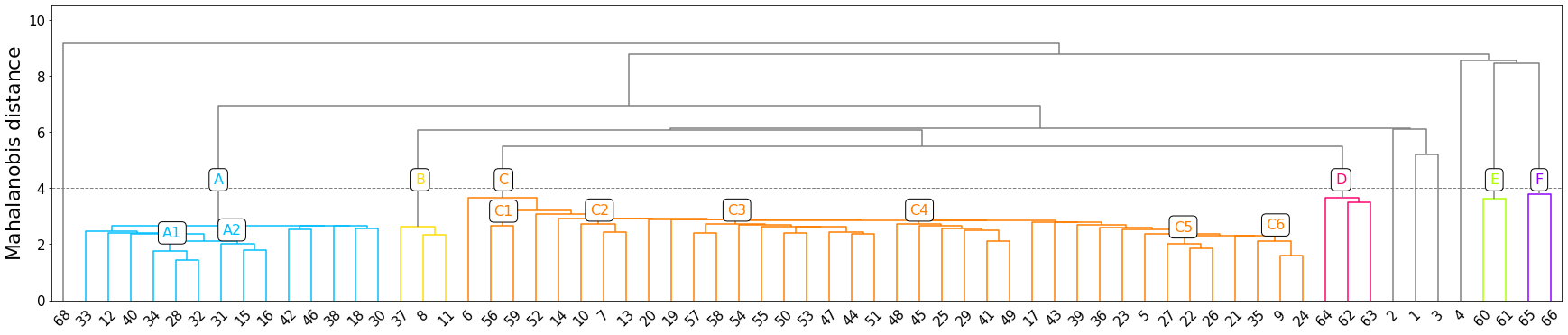}    
\caption{Relationship between the significant clusters according to the single linkage algorithm, depicted according to their Mahalanobis distance in the IoM space (paper I). The different colours represent a tentative classification into different substructures, namely six large groups (named A to F) and a set of individual clusters (in grey; see text for details). The order of the clusters along the x-axis is such that it facilitates the interpretation and readability of the dendrogram, and does not reflect internal properties. In our analyses (e.g. tables, etc), we broadly follow this same order.}
\label{fig:dendrogram} 
\end{figure*}
\section{Data and clustering procedure}
\label{data}

This work is based on the results presented in paper I, where we introduced a data-driven clustering algorithm to systematically identify kinematic structures in the stellar halo near the Sun. For completeness, we provide a short summary below and refer the reader to paper I  for further technical details. 

We use a sample of stars from the local halo (within $2.5$ kpc from the Sun), defining halo stars according to  $|{\bf V}-{\bf V}_{LSR}|$~>~210~km/s, with V$_{VLSR}$ = 232~km/s \citep[][]{2017MNRAS.465...76M}, where ${\bf V}$ is the total velocity vector of a star. The sample is constructed from the {\it Gaia} Early Data Release~3 \citep[EDR3,][]{2021A&A...649A...1G}, and complemented where possible with line-of-sight velocities ({\it vlos}) from GALAH DR3 \citep[][]{2021MNRAS.506..150B}, APOGEE DR16 \citep[][]{2020ApJS..249....3A}, RAVE DR6 \citep[][]{2020AJ....160...82S, 2020AJ....160...83S}, and the LAMOST DR6 medium \citep[][]{2019RAA....19...75L} and low resolution \citep[][]{2020ApJS..251...27W}  surveys. 
We require precise parallaxes (\texttt{parallax\_over\_error} $>5$), low renormalised unit weight error (\texttt{ruwe} $<1.4$), and low {\it vlos} uncertainty (below 20~km/s). This results in a final sample with $51671$ stars.

To this data set we applied (in paper I) the single linkage algorithm in three-dimensional (IoM) space, including total energy ($E$) and the angular momentum in the $z$-direction ($L_z$) and in its perpendicular component ($L_{\perp}$). We refer the reader to Sect.~3.2 of paper I for the exact definitions and computation of these features. The single linkage algorithm defines a series of connected components (or potential clusters) in the data set, as each step of the algorithm forms a new group by connecting the two closest data points not yet in the same group. We then evaluate the significance of each potential cluster produced in this process by comparing the density in an ellipsoidal region around the potential cluster with the density in the same region in an artificially generated halo. This procedure returns a set of (potentially hierarchically overlapping) statistically significant clusters, and final (non-overlapping) clusters are extracted by selecting the subsets with the highest statistical significance. We then refine those clusters by computing Mahalanobis distances ($D_{ij}$) for each star $x_j$ with respect to the centre of every cluster $C_i$ (assuming that this can be fitted with an ellipsoid, i.e. a multivariate Gaussian in 3D characterised by a covariance matrix $\Sigma_i$). The Mahalanobis distance\footnote{The Mahalanobis distance is a dimensionless quantity measuring the distance between a data point (star in our case) and a Gaussian distribution (representative of each cluster) in terms of standard deviations.} between star $x_j$ and cluster $C_i$ is defined as 
\begin{equation}
    D_{ij} = \sqrt{({\bf \mu}_i-{\bf \mu}_j)^T \Sigma_i^{-1} ({\bf \mu}_i-{\bf \mu}_j)},
    \label{eq:Mahalanobis}
    \end{equation}
where ${\bf \mu}_{j}$ denotes the location of star $x_j$ and ${\bf \mu}_i$ that of cluster $C_i$ (mean of the cluster distribution to be precise). In what follows, we consider that a star $x_j$ belongs to cluster $C_i$ (core member) if their Mahalanobis distance is below a certain limit ($D_{ij} <2.13$, which corresponds to the 80$^{th}$ percentile of the distribution of Mahalanobis distances for the stars originally in any cluster; see paper I for more information).

\begin{figure*}
\centering 
\includegraphics[width = 0.98\textwidth]{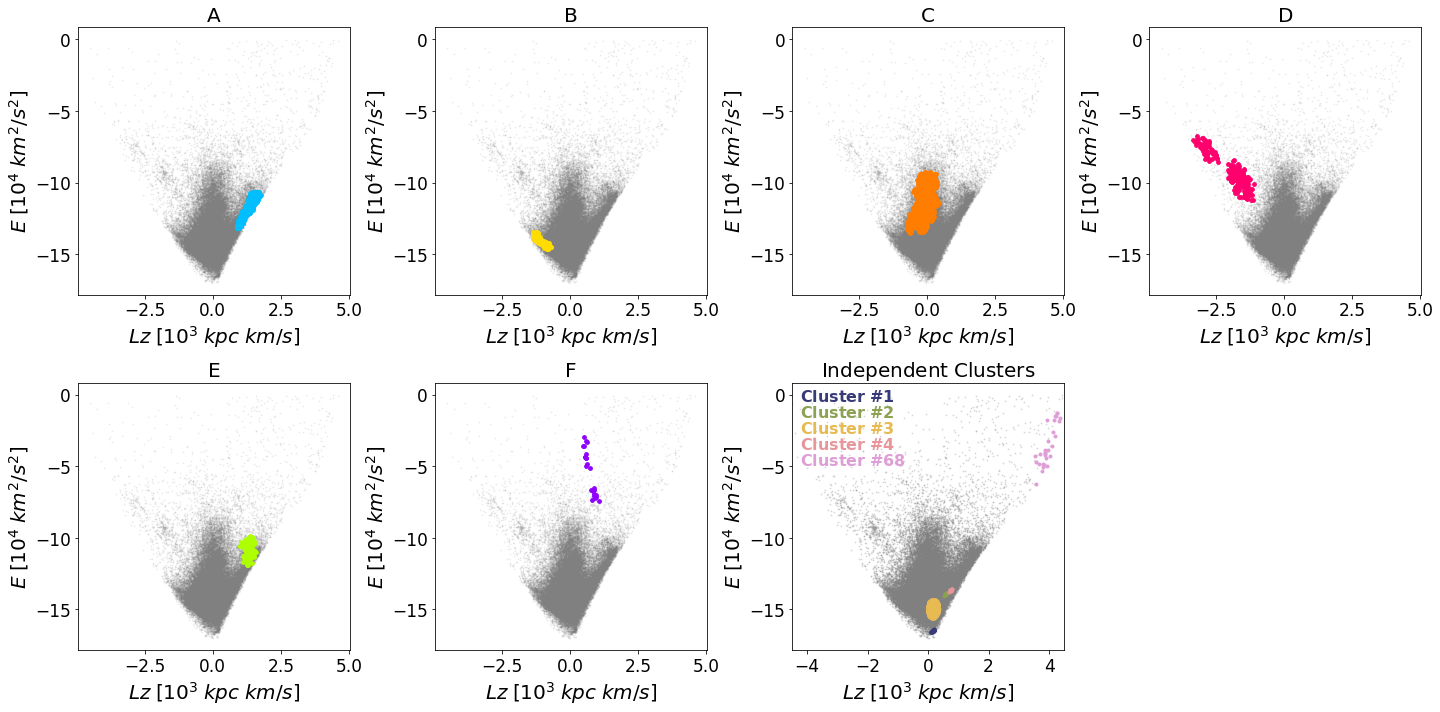} 
 
\includegraphics[width = 0.98\textwidth]{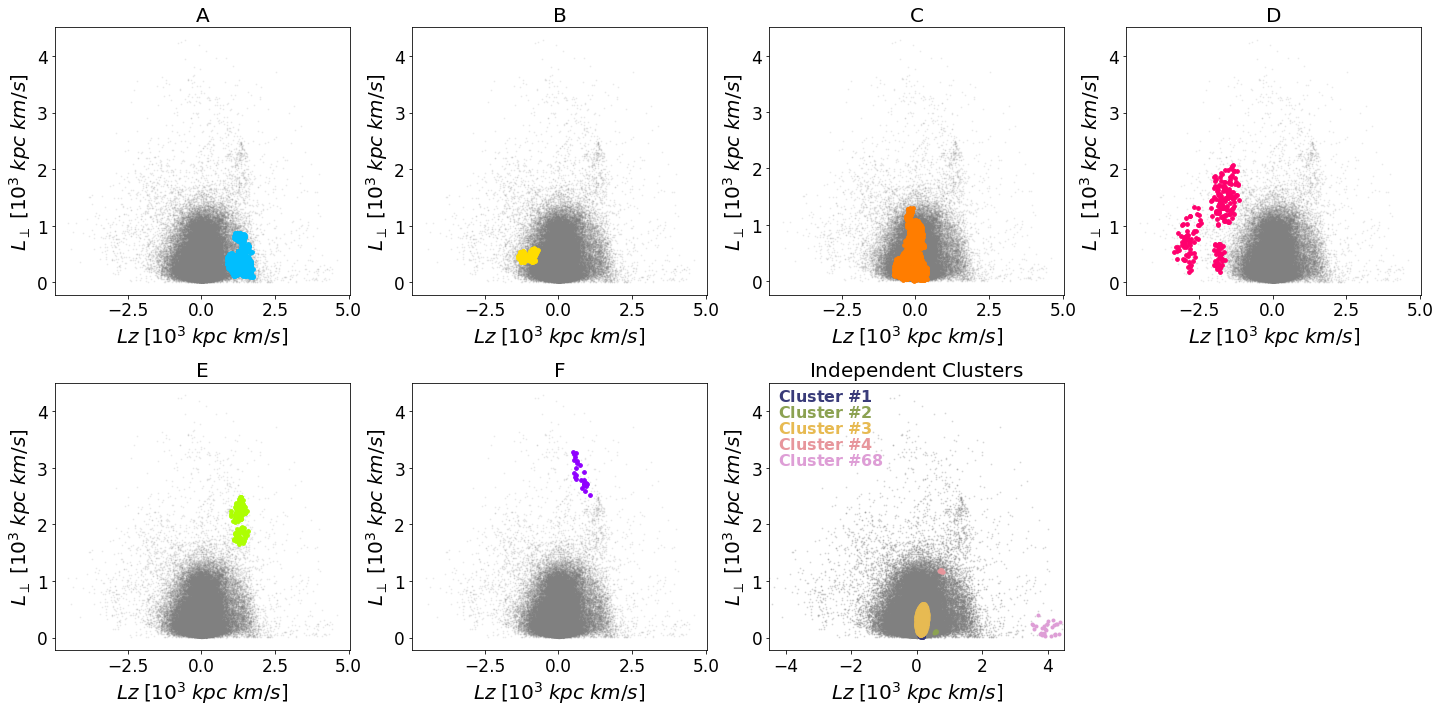}    
\caption{Distribution of stars for the tentative substructures in IoM space: $E$ vs. $L{\rm _z}$  (top two rows) and $L_{\perp}$ vs. $L{\rm _z}$ (bottom two rows). Each panel plots one of the six large groups and the independent clusters from paper I. Colours are as in Fig.~\ref{fig:dendrogram} except for the independent clusters panel. The full original sample upon which this study is built is in grey in all panels.} 
\label{fig:substructures_IoM} 
\end{figure*}

The above procedure identifies $67$ clusters in our data set with a significance of at least 3$\sigma$ above that expected for a randomised halo distribution\footnote{In paper I, 68 clusters were originally detected, but cluster \#67 was discarded because of the large correlation values in the covariance matrix provided by the {\it Gaia} database for the member stars.}. Table~\ref{clusters_prop_tab} in Appendix~\ref{appendix2} provides an overview of the main properties of these refined clusters. Around 13.8\% of the stars in this sample (7127 stars) are within a Mahalanobis distance of 2.13 from any of these clusters\footnote{The threshold of $D_{ij}<2.13$ to select core members was chosen as it clearly improves the purity of the sample, reducing contamination considerably (see paper I for further details)}. To establish possible relations between clusters, we compute the Mahalanobis distance between them using the above definition (Eq.~\ref{eq:Mahalanobis}), but where the covariance matrix is $\Sigma_{ij} = \Sigma_{C_i} + \Sigma_{C_j}$, with $\Sigma_{C_i}$ and $\Sigma_{C_j}$ describing the covariance matrices of the two cluster distributions and ${\bf \mu}_{i}$ and ${\bf \mu}_j$ the mean location of clusters $C_i$ and $C_j$, respectively. 

Figure~\ref{fig:dendrogram} (adapted from paper I) shows the relationship between the $67$ extracted clusters in the form of a dendrogram based on the Mahalanobis distances between the clusters in IoM space. This figure clearly shows that the clusters identified are not fully independent, as they are grouped together in larger structures that even in some cases present a finer hierarchical substructure. Some of these (marked in colour) large structures correspond to previously studied Galactic halo building blocks as reported in paper~I. In that paper,  we tentatively drew the limit for clusters to be linked at a Mahalanobis distance of $\sim$4.0. In Fig.~\ref{fig:substructures_IoM} we show the distribution in IoM of such structures including those clusters that are kept as separate entities (according to our tentative cut). Using this analysis as a starting point, in this paper, we further investigate the independence of the substructures by comparing their stellar population characteristics.

\section{Methodology}
\label{analysis}

In this section, we explain the main tools used to fully assess whether two clusters or groups of clusters ---identified as described above--- can be linked together in a substructure (i.e. they have a common physical origin), or whether they are fully independent. We make use of {\it Gaia} EDR3 astrometric and photometric data as well as information provided by the aforementioned ground-based spectroscopic surveys.

\subsection{Isochrone fitting}
\label{iso_fit}

The precision of the photometric data and distances to individual stars provided by {\it Gaia} allows the construction of accurate CaMDs down to the main sequence turn-off and below. Although issues related to completeness and uncharacterised selection effects prevent us from easily inferring full star formation histories from CaMD fitting techniques \citep[][]{2005ARA&A..43..387G, 2010AdAst2010E...3C, 2019NatAs...3..932G}, the quality of these observed CaMDs enables the estimation of an average age and metallicity via comparison with stellar isochrones. To this end, we use the updated {\tt BaSTI} isochrones \citep[][]{2018ApJ...856..125H} in the {\it Gaia} EDR3 photometric system.  In particular, we employ the $\alpha$-enhanced version of the {\tt BaSTI} models ([$\alpha$/Fe]=0.4), with the canonical He abundance of Y=0.247\footnote{\url{http://basti-iac.oa-abruzzo.inaf.it/index.html}}. 

The observed colour--magnitude diagrams from {\it Gaia} are transposed to the absolute magnitude plane by making use of distances computed by inverting the observed parallax after correcting for an offset of 0.017~mas  \citep[][]{2021A&A...649A...4L}. Although it is possible to improve the characterisation of this parallax offset in {\it Gaia} EDR3 \citep[see e.g.][]{2021A&A...649A...4L,2021ApJ...911L..20R, 2021A&A...654A..20G}, this turns out to have a negligible effect on our analyses. We also correct for the effect of reddening on the magnitudes of individual stars using the dust map from \citet{2018A&A...616A.132L} and the recipes to transform to {\it Gaia} magnitudes provided in \citet{2018A&A...616A..10G}\footnote{The use of different dust maps \citep[e.g.][]{2019ApJ...887...93G} or  photometric recipes \citep[][\url{http://stev.oapd.inaf.it/cgi-bin/cmd_3.4}]{1989ApJ...345..245C, 1994ApJ...422..158O} results in small variations in the magnitude values (below 0.05 mag in most cases), but these do not affect our findings.}. 

\begin{figure*}
\centering 
\includegraphics[width = 0.8\textwidth]{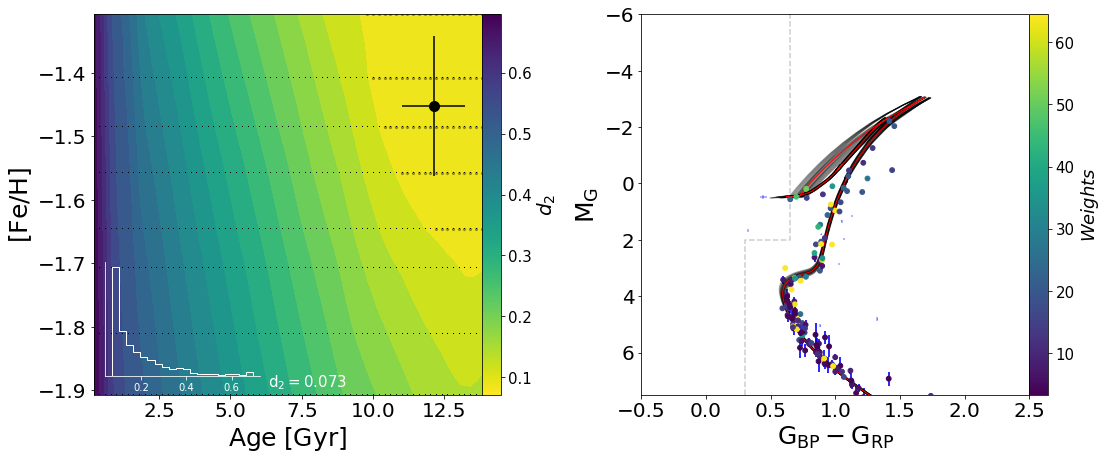}  
\caption{Isochrone fitting procedure applied to clusters \#60 and \#61 together, which are associated with the Helmi streams. {\it Left:} Age--metallicity plane, where the background colours show the distribution of the $d_2$ parameter used for assessing the quality of the fit obtained. The average age and metallicity, together with the corresponding errors, are represented by the black point with errorbars. To compute these average values, we use isochrones with a $d_2$ parameter within its lowest 10$^{th}$ percentile (small black dots). Horizontal dotted lines show the [Fe/H] values adopted for the isochrones. The histogram of $d_2$ values is displayed in the bottom left for all the different isochrones considered. {\it Right:} CaMD for stars in the union of clusters \#60 and \#61. The best-fit isochrone is depicted in red, and all compatible isochrones (black dots in the left-hand panel) are overplotted in black with transparency. Stars bluer than the grey dashed lines (horizontal branch and blue stragglers, shown with transparency) are avoided in the fit. Every star is colour-coded according to its weight in the fit (see text for details).} 
\label{fig:helmi_iso_fit} 
\end{figure*}

Furthermore, we define the observed CaMD by considering only stars fulfilling the extra condition that the BP/RP excess factor (\texttt{phot\_bp\_rp\_excess\_factor}) verifies:
$$
 0.001+0.039 \times \texttt{bp\_rp} < \log(\texttt{phot\_bp\_rp\_excess\_factor})
$$
 and
$$ 
\log(\texttt{phot\_bp\_rp\_excess\_factor}) < 0.12 + 0.039\times \texttt{bp\_rp} 
$$
where \texttt{bp\_rp} is the apparent $(G_{\rm BP}-G_{\rm RP})$ colour. We also avoid stars affected by high extinction, $A_G >0.7$, as their position in the CaMD will be less robust, as well as horizontal branch stars and blue stragglers (that occupy the region to the left of the dashed grey line indicated in the right-hand panel of  Fig.~\ref{fig:helmi_iso_fit}).

To determine the average age and metallicity (hereafter [Fe/H]) of any group of stars, be it individual clusters or several clusters linked together as a group or substructure, we developed an isochrone-fitting procedure which is carried out in a series of steps. The set of isochrones considered in the fitting procedure cover the full range of ages (200~Myr to 14~Gyr, in steps of 0.2~Gyr) but we restrict the range in [Fe/H] to values according to our knowledge from spectroscopic surveys (LAMOST LRS), considering only isochrones corresponding to metallicities from $\langle{\rm [Fe/H]}\rangle-\sigma({\rm [Fe/H]})_{\rm LAMOST,LRS}$ to $\langle{\rm [Fe/H]}\rangle+\sigma({\rm [Fe/H]})_{\rm LAMOST,LRS}$  (with $\langle{\rm [Fe/H]}\rangle$ being the mean and $\sigma({\rm [Fe/H]})_{\rm LAMOST,LRS}$ the standard deviation of the sample). This range encompasses on average 70\% of the stars with [Fe/H] determinations in each of the substructures under analysis\footnote{Similar results are found even without this [Fe/H] range restriction.}; see Sect.~\ref{results}. This constraint is only applied if at least ten stars in the cluster, group, or substructure have [Fe/H] determinations from LAMOST LRS; otherwise the full set of isochrones is used (ranging from [Fe/H]=$-2.5$ to [Fe/H]=0.06\footnote{The exact [Fe/H] values of our set of isochrones are shown in Fig.~\ref{fig:helmi_iso_fit} with horizontal dotted lines.}). 

We then measure how well a given isochrone reproduces the distribution of stars in the CaMD with the following parameter:
\begin{equation}
 d_2 = \sqrt{\frac{\sum_{i=1}^{n}{|X_i-X_{i, iso}|^2}\times W_i}{\sum_{i=1}^{n}{W_i}}} 
 \label{eq:d2}
,\end{equation}
where ${|X_i-X_{i, iso}|^2}$ represents the distance from the CaMD position ($X$) occupied by star {\it i} to the closest point of the isochrone ($X_{i, iso}$) and $W_i$ is the weight given to this particular star, defined as the inverse of the sum in quadrature of the photometric errors (in colour and absolute magnitude). 

To provide an average value and an uncertainty for the age and [Fe/H] of a given structure, we consider all possible isochrones compatible with the observed distribution of stars. To this end, we select those within the lowest 10$^{th}$ percentile of the $d_2$-distribution. The best estimate of the age and [Fe/H] (and its uncertainty) is then computed  using a weighted average (and respectively, dispersion) where the weights are given by the inverse of the $d_2$ values.

We now discuss one specific case as an example of how the outlined procedure works. In paper I, we identified a link between clusters \#60 and \#61 (substructure E in Fig.~\ref{fig:dendrogram}), namely that they are part of the Helmi streams. Figure~\ref{fig:helmi_iso_fit} shows the distribution of the $d_2$ parameter in the age--metallicity plane sampled by the BaSTI isochrones (left), as well as the CaMD of the 143 core member stars from clusters \#60 and \#61 fulfilling our selection criteria. From this isochrone fit, we find that the Helmi stream stars are on average 12.1~$\pm$~1.1 Gyr old and have an average metallicity of [Fe/H]~=~$-1.45~\pm~0.11$ (while the average spectroscopic [Fe/H] of the 100 stars with LAMOST LRS spectra is $-1.6$ with a dispersion of 0.3). These values are in very good agreement with previous studies \citep[e.g.][]{ 2019A&A...625A...5K}.

We note that we expect most of the clusters identified in paper~I to be complex systems, composed of multiple populations. In this sense, the isochrone fitting approach must be understood as an attempt to compute an average age of the system, which will typically not provide a perfect fit to the CaMD.

\subsection{How to assess the independence of halo substructures}
\label{KS_tests}

\begin{figure*}
\centering 
\includegraphics[width = 0.95\textwidth]{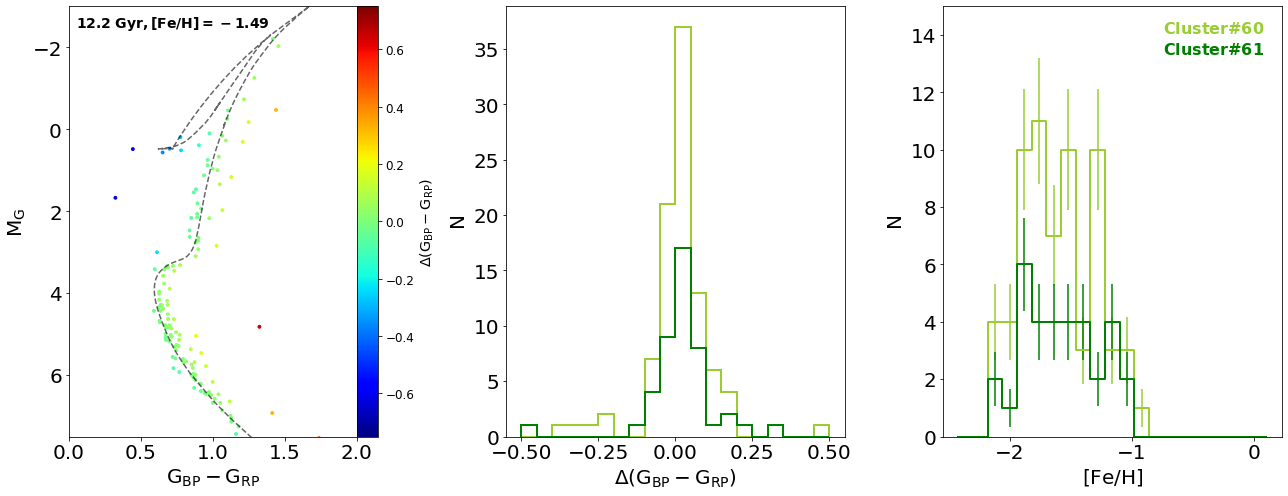} 
\caption{Analysis of the stellar populations of the Helmi streams. {\it Left:} Distribution in the CaMD of the stars classified as part of the Helmi streams (clusters \#60 and \#61) colour-coded according to the colour distance to the best-fitting isochrone (black dashed line, see Sect.~\ref{iso_fit}). {\it Centre:} Histograms of the difference in colour for stars in clusters \#60 and \#61, and showing compatible distributions with a KS p-value of $\sim$~0.6. {\it Right:} Metallicity distribution function from LAMOST LRS for both clusters (p-value of 0.99). Error bars were computed assuming Poisson statistics.} 
\label{fig:colour_distr} 
\end{figure*}

The clustering procedure presented in detail in paper I identified 67 significant clusters in a 3D IoM space. As mentioned above, on the basis of the dendrogram shown in Fig.~\ref{fig:dendrogram} and a comparison to previous work, we established in that
paper that these 67 clusters could tentatively be grouped into six larger substructures (named A to F) and several individual clusters (\#1, \#2, \#3, \#4, and \#68).

Figure~\ref{fig:substructures_IoM} shows the distribution in the $E-L_z$ plane (top two rows) and in $L_\perp-L_z$ (bottom two rows) of the stars belonging to these tentative structures. We can roughly identify substructure A as part of the hot thick disc; associate B with Thamnos1+2 \citep[][]{2019A&A...631L...9K}; C with GE \citep[][see also \citealt{Belokurov2018}]{2018Natur.563...85H}; D with the region reported to be occupied by Sequoia \citep[][]{2019MNRAS.488.1235M}; E as part of the Helmi streams \citep[][]{1999Natur.402...53H}; and find that F is located in the region of the $E-L_z$ plane that is expected to be populated by debris from the Sagittarius dwarf galaxy \citep[Sgr,][]{2020ApJ...901...48N} and that is approximately occupied by globular clusters linked with Sgr in \citet[][]{2019A&A...630L...4M}.

Apart from this rough identification, we note a large amount of complexity in the dendrogram of Fig.~\ref{fig:dendrogram}. For example, substructure A presents two groups of clusters (A1 and A2) linked at small Mahalanobis distances. Substructure C (GE) can be clearly split into several subgroups (C1 to C6). In both cases, several other clusters are linked at larger Mahalanobis distances. Other examples are substructures B (Thamnos1+2) and D (Sequoia region), which show two core members with a third cluster at a larger distance. 

To fully assess if the substructures with similar dynamical properties have a common origin or whether they can be split further, we developed a thorough, exhaustive, and quantitative approach that makes use of their internal stellar population properties. We compare the metallicity distribution function (MDF) of their stars and their colour distribution in the CaMD via Kolmogorov-Smirnov (KS) statistical tests. For the MDF analysis, we concentrate on stars with determinations of [Fe/H] from LAMOST LRS \citep[the spectroscopic survey with the best coverage of our sample,][]{2020ApJS..251...27W}, thereby avoiding heterogeneity from surveys that may use different metallicity scales. For the colour distribution analysis, first we find the isochrone that best represents the group or structure that we are studying (understood as the union of the potential candidate clusters). Then, we compute the colour distance from this isochrone to the individual stars making up each cluster. Specifically, we proceed as follows: 

\renewcommand{\labelitemi}{$\bullet$}
\begin{itemize}
    \item {\bf Step 1:} We identify the smallest units or subgroups within each structure, i.e. composed by clusters linking at the smallest Mahalanobis distances (e.g. A1, A2, C1, ..., C6, B, D, E, F, see Fig.~\ref{fig:dendrogram}). 
    \item {\bf Step 2}: For each (sub)group, we compare the MDFs and colour distributions of each of its constituent clusters with that of the rest of the (sub)group. For example, for the A1 group, composed of clusters \#34, \#28, and \#32, we make three  comparisons: \#34 with (\#28 $\cup$ \#32); \#28 with (\#32 $\cup$ \#34); and \#32 with (\#28 $\cup$ \#34). We decide whether two distributions ($\phi_1$, $\phi_2$) are compatible if the computed KS p-value $P_{\phi_1, \phi_2} > 0.05$ when comparing their MDFs. When comparing the colour distributions, we lower the threshold to 0.03 to claim that two distributions are different (because of the lower discerning power of colours). 
    \subitem {\color{lightgray}$\bullet$}    We note that for clusters with less than 20 stars, Poisson noise might affect the results of the KS tests. In such cases, we first compute $P_{\phi_1, \phi_2}$ as above. To make the comparison meaningful between the two structures ${C}_1$ and ${C}_2$ where $N_2 < 20$, we randomly draw $N_2$ stars from ${C}_1$ ($\phi_1'$). We then run a second KS test comparing $\phi_1'$ with its parent distribution, $\phi_1$, that is, we determine $P_{\phi_1, \phi_1'}$ to assess the effect of small number statistics. This procedure is repeated 100 times. If $P_{\phi_1, \phi_2} < 0.05$, and the recovered $P_{\phi_1,\phi_1'} > 0.05$ for at least 80\% of the cases, we cannot consider random noise to be responsible for the low measured p-value, and therefore we can conclude that $\phi_1$ and $\phi_2$ are indeed drawn from different distributions. 
    \item {\bf Step 3}: Clusters kept as independent because of their larger linking Mahalanobis distances, or those discarded from the groups studied in Step 2, are evaluated to assess their relation with the coherent groups identified, whilst keeping an eye on the distribution of their stars in IoM. No clusters will be added to any group if they are not clearly related in IoM. This step allows us to refine the groups defined in Step 1. 
    \item {\bf Step 4}: All groups and independent clusters (those that remain as such after Step 3) are compared via KS tests again. 
\end{itemize}

As an additional consistency check and whenever possible, we make use of chemical abundance ratios from APOGEE DR16 if there are enough stars in each cluster or substructure with abundance measurements to define a clear trend after removing stars that have \texttt{STAR\_BAD} flag in \texttt{ASPCAPFLAG}, \texttt{FE\_H\_FLAG}$\neq 0$, or \texttt{MG\_FE\_FLAG}$\neq 0$. In this study, we focus on the behaviour of [{Mg}/{Fe}] as a function of [{Fe}/{H}]. 

As in the previous section, we use the Helmi streams again to exemplify the procedure outlined above. The algorithm presented in paper~I associates\footnote{The stars fall within a Mahalanobis distance of 2.13 with respect to the centre of the original clusters, i.e. within 80$^{th}$ percentile of the distribution (core members) .} 95 and 48 stars with clusters \#60 and \#61, respectively.  Of these, 66 and 34 stars respectively have [Fe/H] determinations from LAMOST LRS. The right panel of Fig.~\ref{fig:colour_distr} shows that they present similar metallicity distributions, displaying a p-value  of 0.99   according to the KS test when compared to one another. The left panel of  Fig.~\ref{fig:colour_distr} shows their distribution in the CaMD colour-coded by the distance in colour to the best-fitting isochrone for the whole structure (when the clusters are merged). The distribution of distances from the best-fitting isochrone is shown in the central panel, and is consistent between the clusters, presenting a p-value of 0.6. Thus, as already mentioned in paper I, we can conclude that both clusters are part of the same structure. The origin of the splitting of the Helmi streams into two significant clusters is probably the presence of a resonance induced by the Galactic potential \citep[see][]{2022A&A...659A..61D}. 

\section{Analysis: identification of independent halo substructures}
\label{results}

To obtain a more precise and updated view of the  history of the  Galactic inner halo, in this section we determine which substructures can be considered independent according to all the information at our disposal and using the methodology described above. In Sect.~\ref{updated_view} we use this information to characterise their chemical and stellar population properties and interpret our findings.

\subsection{Substructure A}
\label{thick_disc}

Figure~\ref{fig:dendrogram} shows that substructure A can  first be split into two groups A1 and A2 with clusters (\#34, \#28, \#32) and (\#31, \#15, \#16), respectively,  
linking at small Mahalanobis distances. A further eight clusters are linked at larger Mahalanobis distances. We now proceed to apply the methodology described in Sect.~\ref{KS_tests}.

The results of Steps 1 and 2 (Sect. 3.2) applied to groups A1 and A2 are given in Table~\ref{tab:A1_A2}. According to this table, cluster \#28 from A1 should be considered independent, because the KS test p-value for the  comparison of its MDF to that of the union of clusters \#34 and \#32 (in A1) is $\sim 0.003$. We checked that this value is not affected by Poisson noise given that \#28 has only 19 stars with [Fe/H] determinations from LAMOST LRS. We find such a low value in only 3\% of the random realisations. The comparison of the colour distributions also yields  a relatively low p-value (0.19). Based on this, and the information in  Table~\ref{tab:A1_A2}, we conclude that {\bf A1} = (\#34 $\cup$ \#32) and {\bf A2} =  (\#31 $\cup$ \#15 $\cup$ \#16)\footnote{Once the nature of a tentative substructure is confirmed, we mark this with a letter in boldface.}.

\begin{table}[h] 
\begin{subtable}{0.45\textwidth} 
{\centering 
\small 
\begin{tabular}{lccc} 
\hline 
{\bf } & {\bf 34*} & {\bf  28*} & {\bf  32 } \\ 
{A1} & \cellcolor[HTML]{90d18df}\color{black}0.42 (89\%) & \cellcolor[HTML]{f7fcf5f}\color{red}0.0 (3\%) & \cellcolor[HTML]{f0f9ecf}\color{black}0.05  \\ \hline \hline 
{\bf } & {\bf 31} & {\bf  15} & {\bf  16* } \\ 
{A2} & \cellcolor[HTML]{e5f5e0f}\color{black}0.13 & \cellcolor[HTML]{f0f9ecf}\color{black}0.05 & \cellcolor[HTML]{c6e8bff}\color{black}0.25 (100\%) \\ \hline \hline
\end{tabular} 
\caption{MDF KS test results.}
}
\end{subtable}
\begin{subtable}{0.45\textwidth} 
{\centering 
\small 
\begin{tabular}{lccc} 
\hline 
{\bf } & {\bf 34*} & {\bf  28} & {\bf  32 } \\ 
{A1} & \cellcolor[HTML]{55b567f}\color{black}0.58 (95\%) & \cellcolor[HTML]{d6efd0f}\color{black}0.19 & \cellcolor[HTML]{53b466f}\color{black}0.58  \\ \hline \hline 
{\bf } & {\bf 31} & {\bf  15} & {\bf  16 } \\ 
{A2} & \cellcolor[HTML]{3ca559f}\color{black}0.65 & \cellcolor[HTML]{2a924af}\color{black}0.72 & \cellcolor[HTML]{f1faeef}\color{black}0.04  \\ \hline \hline
\end{tabular} 
\caption{Colour distribution KS test results.} 
} 
\end{subtable}
\caption{Results of the KS tests comparing (a) the MDFs of the clusters in the tentative groups A1 and A2, that are part of possible substructure A, and (b) their colour distributions. The background colour-coding follows the p-values obtained, with dark green meaning more compatible distributions. Values written in red font correspond to p-values below the 0.05 (0.03 in the case of the colour distributions) threshold, meaning that the distributions under comparison are not statistically compatible. For clusters with less than 20 stars with [Fe/H] from LAMOST LRS or low-quality photometric data (highlighted with *), the percentage of random tests with an outcome  identical to the original one is shown in parenthesis. In this and the following tables, the order is that of the dendrogram plotted in Fig.~\ref{fig:dendrogram}.} 
\label{tab:A1_A2}
\end{table}

We now proceed to Step 3 of the method and consider the independent clusters \#33, \#12, \#40, \#42, \#46, \#38, \#18, and \#30, as well as cluster \#28, and whether they should be associated to {\bf A1} or {\bf A2}. The results are given in Table~\ref{tab:A_ind}, where we see that, in addition to cluster \#28, now clusters \#38, and \#40 also present significant differences in their MDF and should be considered independently. Cluster \#12 is a special case, as it displays clear differences in the colour distributions; we return to this cluster below.

\begin{table*}[!h]
{\centering 
\small 
\begin{tabular}{lccccccccccc} 
\hline 
{\bf MDF} & {\bf 28*} & {\bf  33*} & {\bf  12} & {\bf  40*} & {\bf  42} & {\bf  46*} & {\bf  38} & {\bf  18} & {\bf  30* } & {\bf A1} & {\bf A2} \\ 
{\bf A1} & \cellcolor[HTML]{f7fcf5f}\color{red}0.0 (3\%) & \cellcolor[HTML]{006227f}\color{white}0.91 (98\%)& \cellcolor[HTML]{eaf7e6f}\color{black}0.09 & \cellcolor[HTML]{f6fcf4f}\color{red}0.01 (1\%) & \cellcolor[HTML]{157f3bf}\color{white}0.8 & \cellcolor[HTML]{eaf7e6f}\color{black}0.09 (97\%) & \cellcolor[HTML]{f2faf0f}\color{red}0.03 & \cellcolor[HTML]{0a7633f}\color{white}0.84 & \cellcolor[HTML]{a9dca3f}\color{black}0.35 (98\%) & \cellcolor[HTML]{00441bf}\color{white}1.0 & \cellcolor[HTML]{c1e6baf}\color{black}0.27 
\\ 
{\bf A2} & \cellcolor[HTML]{f7fcf5f}\color{red}0.0 (1\%) & \cellcolor[HTML]{005f26f}\color{white}0.92 (93\%) & \cellcolor[HTML]{a4da9ef}\color{black}0.37 & \cellcolor[HTML]{f4fbf1f}\color{red}0.03 (4\%) & \cellcolor[HTML]{a4da9ef}\color{black}0.37 & \cellcolor[HTML]{f0f9edf}\color{black}0.05 (97\%) & \cellcolor[HTML]{f6fcf4f}\color{red}0.01 & \cellcolor[HTML]{7dc87ef}\color{black}0.48 & \cellcolor[HTML]{cfecc9f}\color{black}0.22 (96\%) & \cellcolor[HTML]{c1e6baf}\color{black}0.27 & \cellcolor[HTML]{00441bf}\color{white}1.0 \\ \hline \hline
{\bf CaMD} & {\bf 28} & {\bf  33*} & {\bf  12} & {\bf  40*} & {\bf  42} & {\bf  46} & {\bf  38} & {\bf  18} & {\bf  30* }  & {\bf A1} & {\bf A2} \\ 
{\bf A1} & \cellcolor[HTML]{8ace88f}\color{black}0.44 & \cellcolor[HTML]{4eb264f}\color{black}0.59 (96\%) & \cellcolor[HTML]{f6fcf4f}\color{red}0.0 & \cellcolor[HTML]{8dd08af}\color{black}0.43 (92\%) & \cellcolor[HTML]{006428f}\color{white}0.9 & \cellcolor[HTML]{f4fbf2f}\color{red}0.02 & \cellcolor[HTML]{bce4b5f}\color{black}0.29 & \cellcolor[HTML]{2d954df}\color{black}0.71 & \cellcolor[HTML]{7ac77bf}\color{black}0.48 (99\%) & \cellcolor[HTML]{00441bf}\color{white}1.0 & \cellcolor[HTML]{005924f}\color{white} 0.94
\\ 
{\bf A2} & \cellcolor[HTML]{c8e9c1f}\color{black}0.25 & \cellcolor[HTML]{7dc87ef}\color{black}0.47 (97\%) & \cellcolor[HTML]{f5fbf3f}\color{red}0.01 & \cellcolor[HTML]{006227f}\color{white}0.91 (95\%) & \cellcolor[HTML]{3fa95cf}\color{black}0.63 & \cellcolor[HTML]{e5f5e0f}\color{black}0.13 & \cellcolor[HTML]{005522f}\color{white}0.95 & \cellcolor[HTML]{005c25f}\color{white}0.92 & \cellcolor[HTML]{bee5b8f}\color{black}0.28 (94\%) & \cellcolor[HTML]{005924f}\color{white} 0.94 & \cellcolor[HTML]{00441bf}\color{white} 1.0 \\ \hline \hline 
\end{tabular} 
\caption{Comparison of individual,  more isolated clusters to groups {\bf A1} and {\bf A2} by means of KS tests on their MDFs (top) and their colour distributions (bottom). We add in this table the KS p-values of the comparison between {\bf A1} and {\bf A2}. Information and colour-coding are the same as in  Table~\ref{tab:A1_A2}.\label{tab:A_ind}} 
} 
\end{table*}

\begin{figure*}[!h]
\centering
\includegraphics[width=1.0\textwidth]{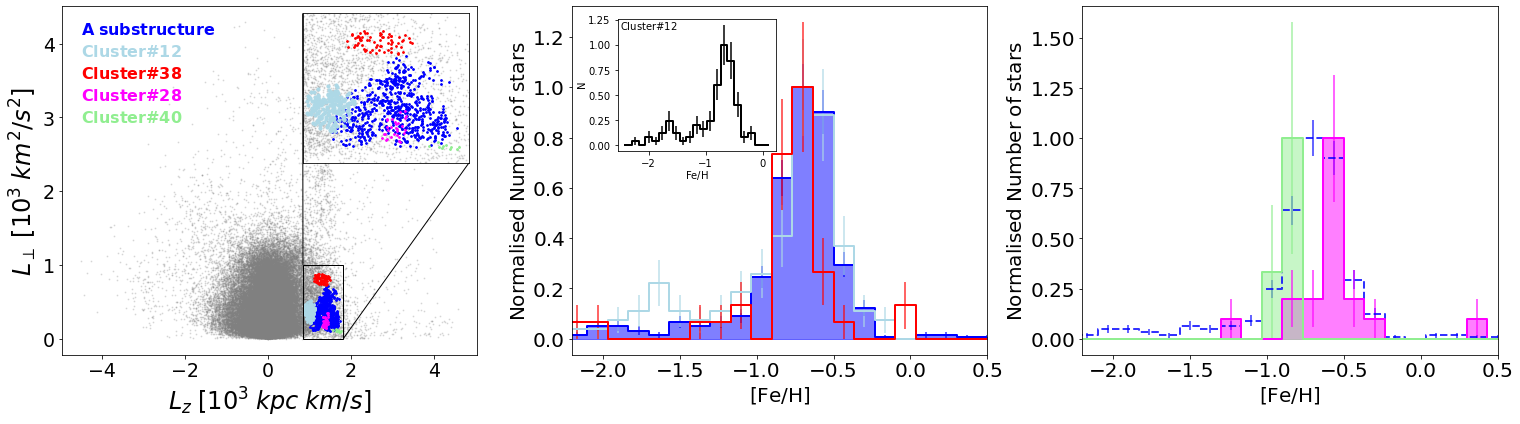}
\caption{Characterisation of substructure {\bf A} and the independent clusters \#12, \#38, \#28, and \#40. {\it Left:} Distribution in the $L_\perp-L_z$ plane. The original sample of halo stars is shown as a grey background. {\it Middle:} Normalised MDFs for the three largest (in number of stars) structures/clusters. As an inset we add the MDF of cluster \#12 alone to highlight its apparently double population. See text for details. {\it Right:} Normalised MDFs for clusters \#28 and \#40, showing a narrow MDF. We add the MDF of substructure A (dashed, blue line) for comparison. Error bars in all plots are computed assuming Poisson statistics.}
\label{fig:A_def}
\end{figure*}

\begin{figure}[!h]
\centering
\includegraphics[width=0.35\textwidth]{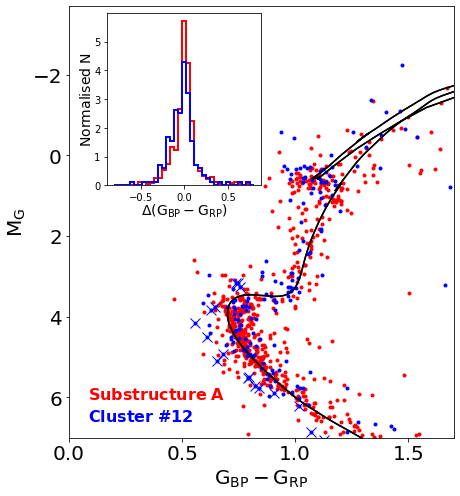}
\caption{Comparison of the distribution of stars in CaMD for cluster \#12 (blue) and substructure {\bf A} (red). The black line represent the isochrone that best reproduces the distribution of stars in the CaMD for substructure {\bf A} (see Sect.~\ref{iso_fit}). The inset panel shows the distributions of colours with respect to the best fit isochrone for both structures. We mark low-metallicity stars ([Fe/H]~$<$~-1.4) with blue crosses.}
\label{fig:12_CaMD}
\end{figure}

The remaining clusters are compatible with both {\bf A1} and {\bf A2}. Indeed, if we compare these two groups (Step 4 of our method), we see that they should form a single substructure, which we refer to as substructure {\bf A}.

Using all this information, Fig.~\ref{fig:A_def} summarises the final characterisation of this substructure. This figure shows the distribution in IoM, as well as the MDFs of substructure {\bf A}  and of clusters \#12, \#38, \#28, and \#40 separately. We can see that these clusters present distributions in IoM and/or MDF that clearly differ from that of {\bf A}. For example, the peak of the MDF of cluster \#38 is shifted to lower metallicities (with this cluster having higher $L_\perp$ than typical of {\bf A}), while the MDFs of clusters \#28 and \#40 are somewhat narrow (peaking at higher and lower metallicities with respect to {\bf A}, respectively). On the other hand, cluster \#12 displays a clear peak at low metallicity. Such stars are responsible for the low KS p-value reported above, and are apparent in Fig.~\ref{fig:12_CaMD} given that they display bluer colours in the CaMD, particularly on the main sequence (represented in the figure as crosses). It seems plausible that cluster \#12 suffers from a large amount of contamination from substructure {\bf A} and that one should consider only the subset of stars associated to the metal-poor peak as truly independent. We return to this possibility later on. 

In summary, substructure {\bf A} is the union of (\#15, \#16, \#46, \#31, \#32, \#30, \#42, \#18, \#33, \#34), with clusters \#12, \#38, \#28, and \#40 to be considered separately. This classification is supported by the final comparison given in Table~\ref{tab:A_def}. We note that substructure {\bf A,} which lies in the hot thick disc, does not overlap substantially with the previously reported Nyx overdensity ---that has prograde thick disc-like kinematics \citep[][]{2020NatAs...4.1078N, 2020ApJ...903...25N}--- because of the dynamical  selection of our sample of halo stars.

The low number of stars populating clusters \#28 and \#40 (29 and 10 stars, respectively) prevents a more thorough characterisation, and therefore we not study them in more detail in the following sections. We deem their origin to be unclear.

\begin{table} 
{\centering 
\small 
\begin{tabular}{lrrrr} 
\hline 
{\bf MDF} & {\bf 12} & {\bf  38} & {\bf  28*} & {\bf  40* } \\ 
{\bf A} & \cellcolor[HTML]{ecf8e8f}\color{black}0.08 & \cellcolor[HTML]{f6fcf4f}\color{red}0.01 & \cellcolor[HTML]{f7fcf5f}\color{red}0.0 (8\%) & \cellcolor[HTML]{f5fbf2f}\color{red}0.02 (8\%) \\ \hline \hline 
{\bf CaMD} & {\bf 12} & {\bf  38} & {\bf  28} & {\bf  40* } \\ 
{\bf A} & \cellcolor[HTML]{f6fcf4f}\color{red}0.01 & \cellcolor[HTML]{78c679f}\color{black}0.49 & \cellcolor[HTML]{87cd86f}\color{black}0.45 & \cellcolor[HTML]{7dc87ef}\color{black}0.48 (97\%) \\ \hline \hline 
\end{tabular} 
\caption{Comparison of clusters \#12, \#38, \#28, and \#40 (considered independent) to substructure {\bf A} based on KS tests of their MDFs (top) and their colour distributions (bottom).} 
\label{tab:A_def}
} 
\end{table}

\subsection{Substructure B: Thamnos 1 and Thamnos 2}
\label{thamnos}

Substructure B (overlapping with Thamnos1 and 2 in IoM) shows a configuration in which two core clusters (\#8, \#11) are linked at a larger Mahalanobis distance to a third one (\#37). 
Table~\ref{tab:thamnos} shows the p-values of KS tests comparing the MDFs of these three clusters, and confirms that there indeed seem to be two distinct subgroups. Clusters 8 and 11 display the more similar MDFs, which are clearly different from that of cluster \#37, although the distributions of stars in the CaMD are compatible. Consequently, we divide substructure B into {\bf B1} (cluster \#37) and {\bf B2} (\#8 $\cup$ \#11).

\begin{figure*}[h]
\centering 
\includegraphics[width = 0.95\textwidth]{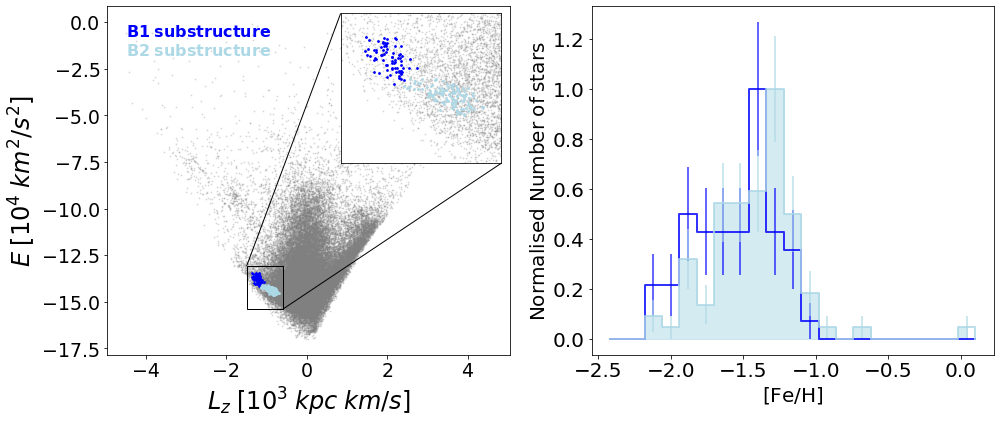} 
\caption{Characterisation of the substructures identified within the tentative structure B. {\it Left:} Distribution in the $E-L_z$ plane. The original sample of halo stars is shown as a grey background. {\it Right:} Normalised MDFs. We use blue for {\bf B1} (Thamnos 1) and cyan for {\bf B2} (Thamnos 2; see text for details). Error bars are computed assuming Poisson statistics.}
\label{fig:Thamnos_IoM} 
\end{figure*}

These substructures are in qualitative agreement with Thamnos 1 and Thamnos 2 respectively, as defined in \citet[][]{2019A&A...631L...9K}, although their mean $E$ and $L_z$ are different from those originally proposed. This likely is due to improvements in the astrometry in {\it Gaia} EDR3 compared to {\it Gaia} DR2 which moves the original selection of Thamnos stars to lower energies and less retrograde orbits. As in \citet[][]{2019A&A...631L...9K}, {\bf B2} (Thamnos 2) shows an overall higher metallicity than {\bf B1} (Thamnos 1). 

Figure~\ref{fig:Thamnos_IoM} shows that there may be some amount of contamination in {\bf B1} from the more metal-rich stars in {\bf B2}. It is likely this contamination that blurs the differences when comparing the colour distributions of these structures.

\begin{table}[h] 
{\centering 
\small 
\begin{tabular}{lrrr} 
\hline 
{\bf MDF} & {\bf 37} & {\bf  8} & {\bf  11 } \\ 
{B} & \cellcolor[HTML]{f5fbf3f}\color{red}0.01 & \cellcolor[HTML]{6dc072f}\color{black}0.52 & \cellcolor[HTML]{edf8e9f}\color{black}0.07  \\ \hline \hline 
{\bf CaMD} & {\bf 37} & {\bf  8} & {\bf  11 } \\ 
{B} & \cellcolor[HTML]{43ac5ef}\color{black}0.62 & \cellcolor[HTML]{c9eac2f}\color{black}0.24 & \cellcolor[HTML]{55b567f}\color{black}0.58  \\ \hline \hline
\end{tabular} 
\caption{Kolmogorov-Smirnov statistical tests comparing metallicity distribution functions of the different clusters identified as possible members of the tentative substructure B. 
\label{tab:thamnos}} 
} 
\end{table}

\subsection{Substructure C: {\it Gaia}-Enceladus}
\label{GE}

{\it Gaia}-Enceladus (GE) is considered the most prominent merger that the Milky Way experienced in its early history \citep[][]{2018Natur.563...85H}. As such, it is no surprise that our clustering algorithm appears to have linked together nearly 40 individual clusters as a single substructure (tentative substructure C) in the region of IoM space thought to be occupied by GE. As in the case of substructure A, our clustering algorithm reveals a rich fine structure within C as shown in Fig.~\ref{fig:dendrogram}, with six groups composed of clusters linked at small Mahalanobis distances\footnote{C3 was further divided into C3a (\#57-\#58), C3b (\#50-\#53), and C3c (\#47-\#44-\#51). After applying a similar process to all groups we concluded that C3a, C3b, and C3c have compatible properties and should be merged to form C3. For clarity, we skip this step in the text.}. Twelve clusters are tentatively
marked as independent\footnote{These individual clusters were not originally joined by the algorithm because of the way the linkage algorithm works and the lack of uniformity in density in the data space, particularly at high-binding energies, as discussed in paper I.} (\#6, \#52, \#14, \#20, \#19, \#17, \#43, \#39, \#36, \#23, \#5, and \#21).

\begin{table*}[h]

\begin{subtable}{0.45\textwidth} 
\centering 
\begin{tabular}{lcc} 
\hline 
{\bf } & {\bf 56} & {\bf  59* } \\ 
{C1} & \cellcolor[HTML]{46ae60f}\color{black}0.61 & \cellcolor[HTML]{46ae60f}\color{black}0.61 (99\%)  \\ \hline \hline 
\end{tabular} 
\end{subtable}
\begin{subtable}{0.45\textwidth} 
\centering 
\begin{tabular}{lccc} 
\hline 
{\bf } & {\bf 10} & {\bf  7} & {\bf  13 } \\ 
{C2} & \cellcolor[HTML]{137d39f}\color{white}0.81 & \cellcolor[HTML]{e3f4def}\color{black}0.13 & \cellcolor[HTML]{73c476f}\color{black}0.5  \\ \hline \hline 
\end{tabular} 
\end{subtable}

\begin{subtable}{0.95\textwidth} 
\centering 
\begin{tabular}{lccccccccc} 
\hline 
{\bf } & {\bf 57*} & {\bf  58*} & {\bf  54*} & {\bf  55} & {\bf  50} & {\bf  53} & {\bf  47} & {\bf  44*} & {\bf  51* } \\ 
{C3} & \cellcolor[HTML]{81ca81f}\color{black}0.46 (94\%)& \cellcolor[HTML]{9ed798f}\color{black}0.38 (97\%)& \cellcolor[HTML]{004e1ff}\color{white}0.97 (99\%)& \cellcolor[HTML]{f4fbf1f}\color{red}0.02 & \cellcolor[HTML]{e1f3dcf}\color{black}0.14 & \cellcolor[HTML]{1f8742f}\color{white}0.76 & \cellcolor[HTML]{9cd797f}\color{black}0.39 & \cellcolor[HTML]{087432f}\color{white}0.85 (96\%)& \cellcolor[HTML]{238b45f}\color{black}0.75 (98\%) \\ \hline \hline
\end{tabular} 
\end{subtable}

\begin{subtable}{0.95\textwidth} 
\centering 
\begin{tabular}{lcccccc} 
\hline 
{\bf } & {\bf 48} & {\bf  45*} & {\bf  25} & {\bf  29} & {\bf  41*} & {\bf  49 } \\ 
{C4} & \cellcolor[HTML]{006b2bf}\color{white}0.88 & \cellcolor[HTML]{004c1ef}\color{white}0.97 (98\%) & \cellcolor[HTML]{70c274f}\color{black}0.51 & \cellcolor[HTML]{5bb86af}\color{black}0.56 & \cellcolor[HTML]{d0edcaf}\color{black}0.21 (98\%)& \cellcolor[HTML]{6dc072f}\color{black}0.52  \\ \hline \hline
\end{tabular} 
\end{subtable}

\begin{subtable}{0.45\textwidth} 
\centering 
\begin{tabular}{lccc} 
\hline 
{\bf } & {\bf 27} & {\bf  22} & {\bf  26 } \\ 
{C5} & \cellcolor[HTML]{00451cf}\color{white}0.99 & \cellcolor[HTML]{b6e2aff}\color{black}0.31 & \cellcolor[HTML]{1d8640f}\color{white}0.77  \\ \hline \hline
\end{tabular} 
\end{subtable}
\begin{subtable}{0.45\textwidth} 
\centering 
\begin{tabular}{lccc} 
\hline 
{\bf } & {\bf 35} & {\bf  9*} & {\bf  24 } \\ 
{C6} & \cellcolor[HTML]{f5fbf2f}\color{red}0.02 & \cellcolor[HTML]{228a44f}\color{white}0.75 (96\%)& \cellcolor[HTML]{f5fbf2f}\color{red}0.02  \\ \hline \hline
\end{tabular} 
\end{subtable}

\caption{Results of the KS tests applied to the MDFs of the clusters and groups tentatively defining substructure C. }
\label{tab:MDF_Cs}
\end{table*}

\begin{table*}[h]

\begin{subtable}{0.45\textwidth} 
\centering 
\begin{tabular}{lcc} 
\hline 
{\bf } & {\bf 56} & {\bf  59 } \\ 
{C1} & \cellcolor[HTML]{81ca81f}\color{black}0.46 & \cellcolor[HTML]{81ca81f}\color{black}0.46  \\ \hline \hline 
\end{tabular} 
\end{subtable}
\begin{subtable}{0.45\textwidth} 
\centering 
\begin{tabular}{lccc} 
\hline 
{\bf } & {\bf 10} & {\bf  7} & {\bf  13 } \\ 
{C2} & \cellcolor[HTML]{2f974ef}\color{black}0.7 & \cellcolor[HTML]{b7e2b1f}\color{black}0.3 & \cellcolor[HTML]{1e8741f}\color{white}0.77  \\ \hline \hline 
\end{tabular} 
\end{subtable}

\begin{subtable}{0.95\textwidth} 
\centering 
\begin{tabular}{lccccccccc} 
\hline 
{\bf } & {\bf 57} & {\bf  58} & {\bf  54} & {\bf  55} & {\bf  50} & {\bf  53} & {\bf  47} & {\bf  44} & {\bf  51 } \\ 
{C3} & \cellcolor[HTML]{2e964df}\color{black}0.71 & \cellcolor[HTML]{99d595f}\color{black}0.4 & \cellcolor[HTML]{005f26f}\color{white}0.92 & \cellcolor[HTML]{18823df}\color{white}0.79 & \cellcolor[HTML]{309950f}\color{black}0.7 & \cellcolor[HTML]{5ab769f}\color{black}0.56 & \cellcolor[HTML]{005a24f}\color{white}0.93 & \cellcolor[HTML]{bbe4b4f}\color{black}0.29 & \cellcolor[HTML]{cfecc9f}\color{black}0.22  \\ \hline \hline
\end{tabular} 
\end{subtable}

\begin{subtable}{0.95\textwidth} 
\centering 
\begin{tabular}{lcccccc} 
\hline 
{\bf } & {\bf 48} & {\bf  45} & {\bf  25} & {\bf  29} & {\bf  41} & {\bf  49 } \\ 
{C4} & \cellcolor[HTML]{127c39f}\color{white}0.81 & \cellcolor[HTML]{9ed798f}\color{black}0.38 & \cellcolor[HTML]{7cc87cf}\color{black}0.48 & \cellcolor[HTML]{42ab5df}\color{black}0.62 & \cellcolor[HTML]{005522f}\color{white}0.95 & \cellcolor[HTML]{2e964df}\color{black}0.7  \\ \hline \hline 
\end{tabular} 
\end{subtable}

\begin{subtable}{0.45\textwidth} 
\centering 
\begin{tabular}{lccc} 
\hline 
{\bf } & {\bf 27} & {\bf  22} & {\bf  26 } \\ 
{C5} & \cellcolor[HTML]{f1faeef}\color{black}0.04 & \cellcolor[HTML]{e1f3dcf}\color{black}0.14 & \cellcolor[HTML]{258d47f}\color{black}0.74  \\ \hline \hline 
\end{tabular} 
\end{subtable}
\begin{subtable}{0.45\textwidth} 
\centering 
\begin{tabular}{lccc} 
\hline 
{\bf } & {\bf 35} & {\bf  9*} & {\bf  24 } \\ 
{C6} & \cellcolor[HTML]{6bc072f}\color{black}0.52 & \cellcolor[HTML]{147e3af}\color{white}0.8 (98\%) & \cellcolor[HTML]{3ba458f}\color{black}0.65  \\ \hline \hline 
\end{tabular} 
\end{subtable}
\caption{Same as Table~\ref{tab:MDF_Cs}, now for  the colour distributions.} 
\label{tab:CaMD_Cs}
\end{table*}

Tables~\ref{tab:MDF_Cs} and~\ref{tab:CaMD_Cs} study the similarities in MDF and CaMD, respectively, between the six groups in substructure C. Based on these tables, we see that the six subgroups are indeed composed of the clusters with a probable common origin. The only notable exceptions being cluster \#55 in group C3 and group C6 on the basis of their MDFs. Therefore, in Step 3, where we evaluate the independent or more isolated clusters, we consider the clusters originally in C6 (\#9, \#24, \#35) 
  and \#55 separately.  

\begin{table*} 
{\centering 
\scriptsize
\begin{tabular}{lcccccccccccccccc} 
\hline 
{\bf MDF} & {\bf 6} & {\bf  52} & {\bf  55} & {\bf  14} & {\bf  20*} & {\bf  19} & {\bf  17} & {\bf  43} & {\bf  39} & {\bf  36} & {\bf  23} & {\bf  5} & {\bf  21} & {\bf  35} & {\bf  9*} & {\bf  24 } \\ 
{\bf C1} & \cellcolor[HTML]{f1faeef}\color{red}0.04 & \cellcolor[HTML]{0c7735f}\color{white}0.83 & \cellcolor[HTML]{f4fbf2f}\color{red}0.02 & \cellcolor[HTML]{63bc6ef}\color{black}0.54 & \cellcolor[HTML]{137d39f}\color{white}0.81 (98\%) & \cellcolor[HTML]{e8f6e4f}\color{black}0.1 & \cellcolor[HTML]{3ea75af}\color{black}0.64 & \cellcolor[HTML]{005321f}\color{white}0.95 & \cellcolor[HTML]{83cb82f}\color{black}0.46 & \cellcolor[HTML]{026f2ef}\color{white}0.87 & \cellcolor[HTML]{b8e3b2f}\color{black}0.3 & \cellcolor[HTML]{b1e0abf}\color{black}0.32 & \cellcolor[HTML]{94d390f}\color{black}0.41 & \cellcolor[HTML]{cfecc9f}\color{black}0.22 & \cellcolor[HTML]{0a7633f}\color{white}0.84 (98\%) & \cellcolor[HTML]{a2d99cf}\color{black}0.37  \\ {\bf C2} & \cellcolor[HTML]{b8e3b2f}\color{black}0.3 & \cellcolor[HTML]{147e3af}\color{white}0.8 & \cellcolor[HTML]{cbeac4f}\color{black}0.24 & \cellcolor[HTML]{f1faeef}\color{red}0.04 & \cellcolor[HTML]{16803cf}\color{white}0.79  (93\%) & \cellcolor[HTML]{c2e7bbf}\color{black}0.27 & \cellcolor[HTML]{f4fbf1f}\color{red}0.02 & \cellcolor[HTML]{a4da9ef}\color{black}0.37 & \cellcolor[HTML]{39a257f}\color{black}0.66 & \cellcolor[HTML]{e7f6e3f}\color{black}0.11 & \cellcolor[HTML]{87cd86f}\color{black}0.45 & \cellcolor[HTML]{bbe4b4f}\color{black}0.29 & \cellcolor[HTML]{f3faf0f}\color{red}0.03 & \cellcolor[HTML]{f5fbf2f}\color{red}0.02 & \cellcolor[HTML]{afdfa8f}\color{black}0.33 (98\%) & \cellcolor[HTML]{b1e0abf}\color{black}0.32  \\ {\bf C3} & \cellcolor[HTML]{f7fcf5f}\color{red}0.0 & \cellcolor[HTML]{c6e8bff}\color{black}0.26 & \cellcolor[HTML]{f4fbf1f}\color{red}0.02 & \cellcolor[HTML]{5ab769f}\color{black}0.57 & \cellcolor[HTML]{4bb062f}\color{black}0.6 (95\%) & \cellcolor[HTML]{f4fbf1f}\color{red}0.02 & \cellcolor[HTML]{005622f}\color{white}0.94 & \cellcolor[HTML]{2e964df}\color{black}0.71 & \cellcolor[HTML]{9bd696f}\color{black}0.39 & \cellcolor[HTML]{d4eecef}\color{black}0.2 & \cellcolor[HTML]{def2d9f}\color{black}0.16 & \cellcolor[HTML]{e3f4def}\color{black}0.14 & \cellcolor[HTML]{6bc072f}\color{black}0.52 & \cellcolor[HTML]{ddf2d8f}\color{black}0.16 & \cellcolor[HTML]{004e1ff}\color{white}0.97 (98\%) & \cellcolor[HTML]{f6fcf4f}\color{red}0.0  \\ {\bf C4} & \cellcolor[HTML]{f7fcf5f}\color{red}0.0 & \cellcolor[HTML]{55b567f}\color{black}0.58 & \cellcolor[HTML]{e6f5e1f}\color{black}0.12 & \cellcolor[HTML]{dcf2d7f}\color{black}0.16 & \cellcolor[HTML]{0c7735f}\color{white}0.83 (97\%) & \cellcolor[HTML]{e1f3dcf}\color{black}0.14 & \cellcolor[HTML]{65bd6ff}\color{black}0.54 & \cellcolor[HTML]{005b25f}\color{white}0.93 & \cellcolor[HTML]{2f974ef}\color{black}0.7 & \cellcolor[HTML]{006328f}\color{white}0.9 & \cellcolor[HTML]{86cc85f}\color{black}0.45 & \cellcolor[HTML]{b2e0acf}\color{black}0.32 & \cellcolor[HTML]{def2d9f}\color{black}0.15 & \cellcolor[HTML]{f2faf0f}\color{red}0.03 & \cellcolor[HTML]{6dc072f}\color{black}0.52 (99\%) & \cellcolor[HTML]{d8f0d2f}\color{black}0.18  \\ {\bf C5} & \cellcolor[HTML]{f1faeef}\color{black}0.06 & \cellcolor[HTML]{006729f}\color{white}0.89 & \cellcolor[HTML]{ebf7e7f}\color{black}0.09 & \cellcolor[HTML]{ddf2d8f}\color{black}0.16 & \cellcolor[HTML]{004d1ff}\color{white}0.97 (97\%) & \cellcolor[HTML]{cbebc5f}\color{black}0.23 & \cellcolor[HTML]{ceecc8f}\color{black}0.22 & \cellcolor[HTML]{006c2cf}\color{white}0.88 & \cellcolor[HTML]{006729f}\color{white}0.89 & \cellcolor[HTML]{63bc6ef}\color{black}0.54 & \cellcolor[HTML]{5ab769f}\color{black}0.56 & \cellcolor[HTML]{76c578f}\color{black}0.49 & \cellcolor[HTML]{e9f7e5f}\color{black}0.09 & \cellcolor[HTML]{ecf8e8f}\color{black}0.08 & \cellcolor[HTML]{4eb264f}\color{black}0.59 (97\%) & \cellcolor[HTML]{107a37f}\color{white}0.82  \\ \hline \hline 
{\bf CaMD} & {\bf 6} & {\bf  52} & {\bf  55} & {\bf  14} & {\bf  20*} & {\bf  19} & {\bf  17} & {\bf  43} & {\bf  39} & {\bf  36} & {\bf  23} & {\bf  5} & {\bf  21} & {\bf  35} & {\bf  9*} & {\bf  24 } \\ 
{\bf C1} & \cellcolor[HTML]{3aa357f}\color{black}0.65 & \cellcolor[HTML]{c3e7bcf}\color{black}0.26 & \cellcolor[HTML]{a0d99bf}\color{black}0.38 & \cellcolor[HTML]{1d8640f}\color{white}0.77 & \cellcolor[HTML]{cbebc5f}\color{black}0.23 (96\%) & \cellcolor[HTML]{19833ef}\color{white}0.78 & \cellcolor[HTML]{9cd797f}\color{black}0.39 & \cellcolor[HTML]{2b934bf}\color{black}0.72 & \cellcolor[HTML]{9ed798f}\color{black}0.38 & \cellcolor[HTML]{00471cf}\color{white}0.99 & \cellcolor[HTML]{d3eecdf}\color{black}0.2 & \cellcolor[HTML]{268e47f}\color{black}0.74 & \cellcolor[HTML]{16803cf}\color{white}0.8 & \cellcolor[HTML]{004a1ef}\color{white}0.98 & \cellcolor[HTML]{e4f5dff}\color{black}0.13 (97\%) & \cellcolor[HTML]{7dc87ef}\color{black}0.47  \\ {\bf C2} & \cellcolor[HTML]{d2edccf}\color{black}0.2 & \cellcolor[HTML]{f5fbf3f}\color{red}0.02 & \cellcolor[HTML]{edf8e9f}\color{black}0.07 & \cellcolor[HTML]{bee5b8f}\color{black}0.28 & \cellcolor[HTML]{c6e8bff}\color{black}0.26 (97\%) & \cellcolor[HTML]{005020f}\color{white}0.96 & \cellcolor[HTML]{edf8e9f}\color{black}0.07 & \cellcolor[HTML]{eaf7e6f}\color{black}0.09 & \cellcolor[HTML]{daf0d4f}\color{black}0.17 & \cellcolor[HTML]{a3da9df}\color{black}0.37 (94\%) & \cellcolor[HTML]{005f26f}\color{white}0.92 & \cellcolor[HTML]{39a257f}\color{black}0.66 & \cellcolor[HTML]{309950f}\color{black}0.69 & \cellcolor[HTML]{cbeac4f}\color{black}0.24 & \cellcolor[HTML]{a2d99cf}\color{black}0.37 & \cellcolor[HTML]{1f8742f}\color{white}0.77  \\ {\bf C3} & \cellcolor[HTML]{98d594f}\color{black}0.4 & \cellcolor[HTML]{7cc87cf}\color{black}0.48 & \cellcolor[HTML]{107a37f}\color{white}0.82 & \cellcolor[HTML]{228a44f}\color{white}0.75 & \cellcolor[HTML]{8ace88f}\color{black}0.44 (94\%) & \cellcolor[HTML]{e0f3dbf}\color{black}0.15 & \cellcolor[HTML]{005c25f}\color{white}0.92 & \cellcolor[HTML]{8dd08af}\color{black}0.43 & \cellcolor[HTML]{caeac3f}\color{black}0.24 & \cellcolor[HTML]{9bd696f}\color{black}0.39 & \cellcolor[HTML]{d4eecef}\color{black}0.2 & \cellcolor[HTML]{d5efcff}\color{black}0.2 & \cellcolor[HTML]{c6e8bff}\color{black}0.25 & \cellcolor[HTML]{4bb062f}\color{black}0.6 & \cellcolor[HTML]{e3f4def}\color{black}0.14 (95\%) & \cellcolor[HTML]{f3faf0f}\color{black}0.03  \\ {\bf C4} & \cellcolor[HTML]{278f48f}\color{black}0.73 & \cellcolor[HTML]{a9dca3f}\color{black}0.35 & \cellcolor[HTML]{63bc6ef}\color{black}0.54 & \cellcolor[HTML]{107a37f}\color{white}0.82 & \cellcolor[HTML]{349d53f}\color{black}0.68 (96\%) & \cellcolor[HTML]{abdda5f}\color{black}0.34 & \cellcolor[HTML]{53b466f}\color{black}0.58 & \cellcolor[HTML]{b5e1aef}\color{black}0.31 & \cellcolor[HTML]{48ae60f}\color{black}0.61 & \cellcolor[HTML]{3fa85bf}\color{black}0.63 & \cellcolor[HTML]{78c679f}\color{black}0.49 & \cellcolor[HTML]{9bd696f}\color{black}0.39 & \cellcolor[HTML]{72c375f}\color{black}0.5 & \cellcolor[HTML]{00692af}\color{white}0.89 & \cellcolor[HTML]{bee5b8f}\color{black}0.28 (93\%) & \cellcolor[HTML]{bde5b6f}\color{black}0.28  \\ {\bf C5} & \cellcolor[HTML]{e4f5dff}\color{black}0.13 & \cellcolor[HTML]{c3e7bcf}\color{black}0.27 & \cellcolor[HTML]{cbebc5f}\color{black}0.23 & \cellcolor[HTML]{75c477f}\color{black}0.5 & \cellcolor[HTML]{29914af}\color{black}0.72 (95\%) & \cellcolor[HTML]{75c477f}\color{black}0.5 & \cellcolor[HTML]{e5f5e0f}\color{black}0.13 & \cellcolor[HTML]{e5f5e1f}\color{black}0.12 & \cellcolor[HTML]{208843f}\color{white}0.76 & \cellcolor[HTML]{90d18df}\color{black}0.42 & \cellcolor[HTML]{42ab5df}\color{black}0.62 & \cellcolor[HTML]{6ec173f}\color{black}0.52 & \cellcolor[HTML]{005c25f}\color{white}0.93 & \cellcolor[HTML]{6ec173f}\color{black}0.51 & \cellcolor[HTML]{aadda4f}\color{black}0.35 (94\%) & \cellcolor[HTML]{7cc87cf}\color{black}0.48  \\ \hline \hline 
\end{tabular} 
\caption{Comparison of individual, more isolated clusters to groups C1 to C5 based on a KS test of their MDFs (top) and their colour distributions (bottom).
\label{tab:C_ind}} 
} 
\end{table*}

\begin{figure*}
\centering
\includegraphics[width=0.95\textwidth]{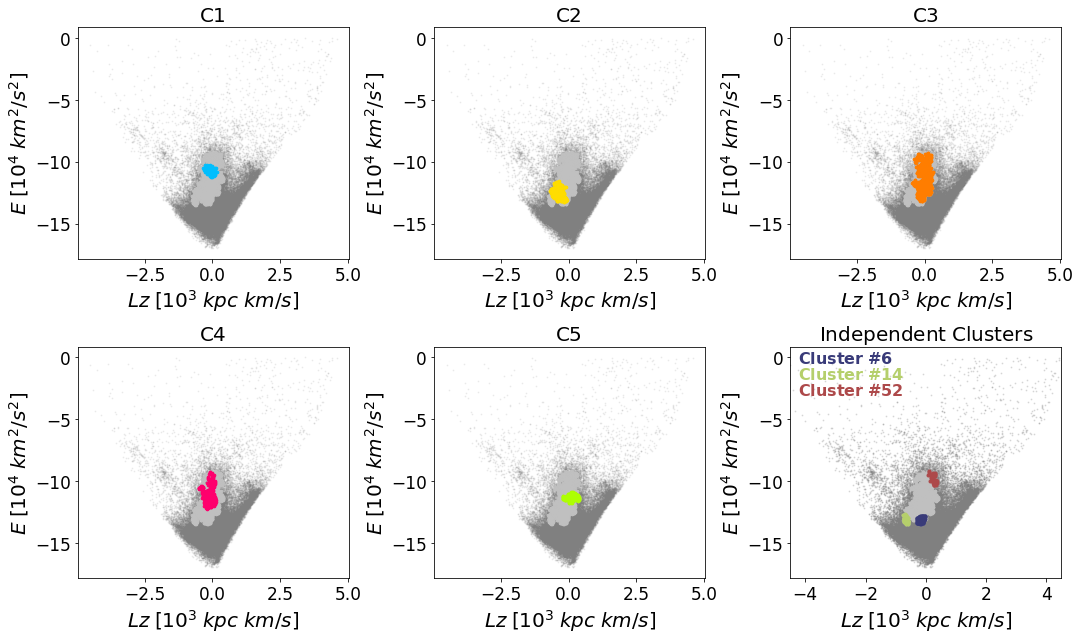} 

\includegraphics[width=0.95\textwidth]{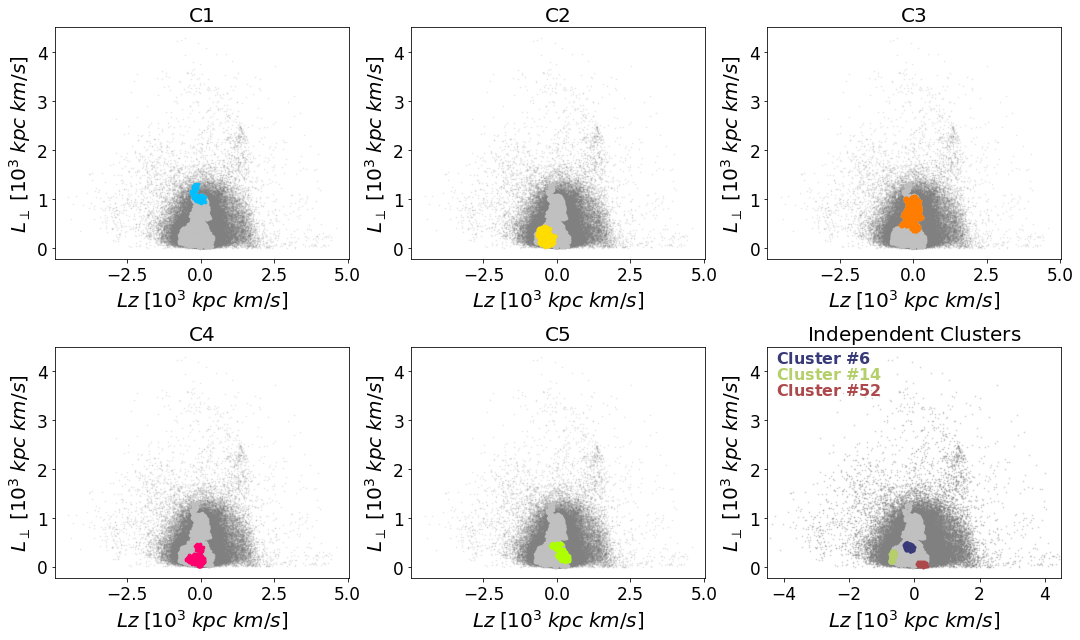}
\caption{$E-L_z$ (top) and $L_\perp-L_z$ (bottom) distribution of the subgroups making up GE (substructure {\bf C}). As in Fig.~\ref{fig:substructures_IoM}, the panels present each of the groups identified and the three independent clusters. The full original sample upon which this study is built is represented as a grey background in all panels. The full extension of substructure {\bf C} is highlighted in silver.}
\label{fig:GE_IoM}
\end{figure*}

This is exactly what is shown in Table~\ref{tab:C_ind}, where we compare all these clusters  with groups C1 to C5. Cluster \#6 is particularly striking as it has an MDF that is clearly different from that of groups C1 to C5 (with the only exception being C2). In the remaining cases, visual inspection of their position in IoM and similarities between their MDFs and CaMDs allow us to assign them to groups C1 to C5, establishing the following subgroup configuration for substructure~C:
\begin{itemize}
    \item {\bf C1}: Clusters \#56 and \#59, 
    \item {\bf C2}: Clusters \#10, \#7, \#13, \#20, \#19, and \#43, 
    \item {\bf C3}: \#57, \#58, \#54, \#50, \#53, \#47, \#44, \#51, \#39, \#35, \#9, \#24, \#5, \#23, \#36, \#21, and \#17, 
    \item {\bf C4}: \#48, \#45, \#25, \#29, \#41, \#49, and \#55,
    \item {\bf C5}: \#27, \#22, and \#26.
\end{itemize}
Apart from cluster \#6 (as commented above), the more isolated cluster \#52 is too far in the IoM space from any of these five subgroups in C (despite the similarity in MDFs and CaMDs). This is 
shown in Figure~\ref{fig:GE_IoM}, which plots their distribution in the $E-L_z$ and $L_\perp-L_z$ planes. Cluster \#14 on the other hand is close to subgroup {\bf C2} with both depicting the more retrograde motions of substructure C, but their MDFs are not consistent. Their position in $E-L_z$ space roughly coincides with that of the Rg5 structure identified in \citet[][;see also paper I, and \citealt{2020ApJ...891...39Y}; notably \citealt{2020ApJ...901...48N} associate this structure to Thamnos]{2019MNRAS.488.1235M}. However, the $L_\perp$ found by these latter authors appears quite different ($<$ 500 kpc km/s compared to $>$ 1000~kpc~km/s of Rg5 stars in \citealt{2020ApJ...891...39Y}). Therefore, either {\bf C2} or \#14 is not related at all to Rg5, or they all stem from a fairly large progenitor (in which case, we would have expected it to also extend further in $E-L_z$). 
As we see below, all evidence suggests that both {\bf C2} and cluster \#14 are part of GE.  

Now we compare (pairwise) the MDFs of all the subgroups and independent clusters, again using KS statistical tests (our Step 4 of Sect.~\ref{KS_tests}). As can be seen from Table~\ref{tab:GE_def}, all groups and independent clusters can be considered compatible with each other. The only cluster that remains independent is \#6 (with 122 stars, 88 with [Fe/H] determinations from LAMOST LRS), whose MDF is clearly different from those of the others\footnote{We note here that cluster \#6 occupies a region in IoM that is consistent with 
an overdensity of globular clusters that was reported by \citet{2022ApJ...926..107M}  after our work was submitted, and which the authors suggest is associated to an independent accretion event, {\it Pontus}.}. This analysis suggests that  substructure {\bf C}, which unites all of the groups, can be firmly associated with GE.

\begin{table} 
{\centering 
\small 
\begin{tabular}{lrrrrrrrr} 
\hline 
{\bf } & {\bf C1} & {\bf C2} & {\bf C3} & {\bf C4} & {\bf C5} & {\bf 14} & {\bf 52} & {\bf 6} \\ 
 {\bf C1} & \cellcolor[HTML]{008000}\color{black}1.0 & \cellcolor[HTML]{bce8bc}\color{black}0.2 & \cellcolor[HTML]{3aa03a}\color{black}0.74 & \cellcolor[HTML]{9dd79d}\color{black}0.32 & \cellcolor[HTML]{4dab4d}\color{black}0.68 & \cellcolor[HTML]{6cbc6c}\color{black}0.54 & \cellcolor[HTML]{339c33}\color{black}0.83 & \cellcolor[HTML]{dcfadc}\color{red}0.04 \\ 
{\bf C2} & \cellcolor[HTML]{c1ebc1}\color{black}0.2 & \cellcolor[HTML]{008000}\color{black}1.0 & \cellcolor[HTML]{e1fde1}\color{red}0.02 & \cellcolor[HTML]{c3ecc3}\color{black}0.15 & \cellcolor[HTML]{9dd79d}\color{black}0.36 & \cellcolor[HTML]{cff3cf}\color{black}0.12 & \cellcolor[HTML]{108910}\color{black}0.94 & \cellcolor[HTML]{cef2ce}\color{black}0.1 \\ 
{\bf C3} & \cellcolor[HTML]{3da23d}\color{black}0.74 & \cellcolor[HTML]{e5ffe5}\color{red}0.02 & \cellcolor[HTML]{008000}\color{black}1.0 & \cellcolor[HTML]{d5f6d5}\color{black}0.08 & \cellcolor[HTML]{8bcd8b}\color{black}0.43 & \cellcolor[HTML]{76c176}\color{black}0.5 & \cellcolor[HTML]{7ac47a}\color{black}0.6 & \cellcolor[HTML]{e5ffe5}\color{red}0.0 \\ 
{\bf C4} & \cellcolor[HTML]{a4dba4}\color{black}0.32 & \cellcolor[HTML]{c7eec7}\color{black}0.15 & \cellcolor[HTML]{d5f6d5}\color{black}0.08 & \cellcolor[HTML]{008000}\color{black}1.0 & \cellcolor[HTML]{4dab4d}\color{black}0.68 & \cellcolor[HTML]{daf9da}\color{black}0.08 & \cellcolor[HTML]{6fbd6f}\color{black}0.63 & \cellcolor[HTML]{e5ffe5}\color{red}0.0 \\ 
{\bf C5} & \cellcolor[HTML]{4daa4d}\color{black}0.68 & \cellcolor[HTML]{96d396}\color{black}0.36 & \cellcolor[HTML]{82c882}\color{black}0.43 & \cellcolor[HTML]{49a849}\color{black}0.68 & \cellcolor[HTML]{008000}\color{black}1.0 & \cellcolor[HTML]{c6eec6}\color{black}0.16 & \cellcolor[HTML]{209220}\color{black}0.89 & \cellcolor[HTML]{d8f8d8}\color{black}0.06 \\ 
{\bf 14} & \cellcolor[HTML]{6fbd6f}\color{black}0.54 & \cellcolor[HTML]{cff3cf}\color{black}0.12 & \cellcolor[HTML]{73c073}\color{black}0.5 & \cellcolor[HTML]{d5f6d5}\color{black}0.08 & \cellcolor[HTML]{cef2ce}\color{black}0.16 & \cellcolor[HTML]{008000}\color{black}1.0 & \cellcolor[HTML]{c6eec6}\color{black}0.35 & \cellcolor[HTML]{e1fde1}\color{red}0.02 \\ 
{\bf 52} & \cellcolor[HTML]{299729}\color{black}0.83 & \cellcolor[HTML]{0d870d}\color{black}0.94 & \cellcolor[HTML]{5db35d}\color{black}0.6 & \cellcolor[HTML]{55af55}\color{black}0.63 & \cellcolor[HTML]{1a8e1a}\color{black}0.89 & \cellcolor[HTML]{9ad59a}\color{black}0.35 & \cellcolor[HTML]{008000}\color{black}1.0 & \cellcolor[HTML]{afe1af}\color{black}0.24 \\ 
{\bf 6} & \cellcolor[HTML]{e5ffe5}\color{red}0.04 & \cellcolor[HTML]{d3f5d3}\color{black}0.1 & \cellcolor[HTML]{e5ffe5}\color{red}0.0 & \cellcolor[HTML]{e5ffe5}\color{red}0.0 & \cellcolor[HTML]{e5ffe5}\color{black}0.06 & \cellcolor[HTML]{e5ffe5}\color{red}0.02 & \cellcolor[HTML]{e5ffe5}\color{black}0.24 & \cellcolor[HTML]{008000}\color{black}1.0 \\ 
\hline \hline 
\end{tabular} 
\caption{Kolmogorov-Smirnov statistical tests comparing metallicity distribution functions of the different clusters and subgroups identified as Gaia-Enceladus.} 
\label{tab:GE_def} }
\end{table}

\begin{figure}
\centering
\includegraphics[width=0.5\textwidth]{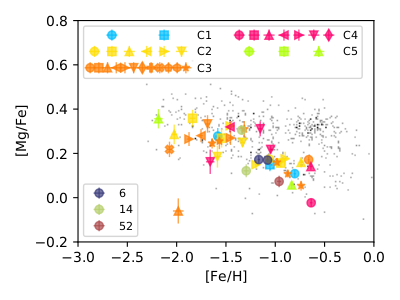}
\caption{[{Mg}/{Fe}] abundance ratios as a function of [Fe/H] for substructure {\bf C} (GE) from APOGEE DR16. We use different symbols for the different clusters. Colours for the groups (C1 to C5) follow Fig.~\ref{fig:GE_IoM}. 
The small dots show abundances of all the halo stars within $2.5\,\mathrm{kpc}$ in APOGEE DR16.}
\label{GEMgFe}
\end{figure}

Abundance determinations from APOGEE are available for a large number of the stars belonging to {\bf C}/GE according to our algorithm. This allows us to study their [Mg/Fe] abundance ratios as a function of [Fe/H] as extra proof of their common origin.  As shown in Fig.~\ref{GEMgFe}, all of the subgroups seem to follow the same chemical abundance track, which is compatible with that known for GE \citep{2018Natur.563...85H,2019MNRAS.482.3426M}. Therefore, the chemical abundance ratios
also support our analysis and interpretation. Interestingly, Fig.~\ref{GEMgFe} shows that the $\langle$[Mg/Fe]$\rangle$ at low [Fe/H] is lower than that of other stars in our sample at similar metallicity (see also Sect.~\ref{chemistry}). 

\begin{figure}
\centering 
\includegraphics[width = 0.45\textwidth]{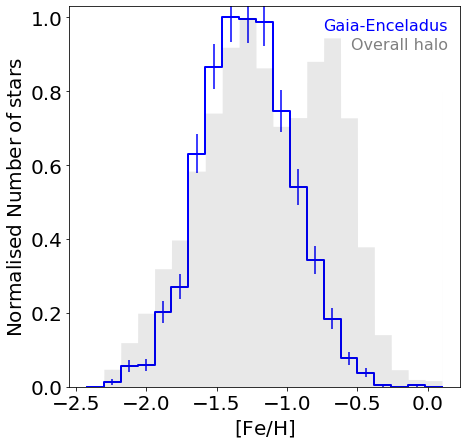} 
\caption{Combined metallicity distribution function of all clusters in substructure {\bf C}/{\it Gaia}-Enceladus. We overlay the [Fe/H] distribution for all 34532 stars with [Fe/H] from LAMOST LRS in our initial halo sample (containing 51671 stars). We note that our selection of GE stars based on our clustering algorithm naturally does not suffer from contamination from the hot thick disc (red sequence). Error bars were computed assuming Poisson statistics.} 
\label{fig:GE_MDF} 
\end{figure}

Now that our clustering algorithm has clearly identified GE,  in Fig.~\ref{fig:GE_MDF} we compare the MDF of all the stars that we have just established belong to GE with the overall MDF of our halo sample. This halo sample is composed of both accreted and in situ stars as discussed in previous works \citep[e.g. \citealt{2007Natur.450.1020C}, and especially the blue and red inner halo sequences reported in][]{ 2018A&A...616A..10G, 2018ApJ...863..113H, 2019NatAs...3..932G, 2019A&A...632A...4D}. The presence of two main components is clearly reflected in the bimodal shape of the halo MDF (shaded area). It is reassuring that the MDF of the stars identified as GE corresponds to only the more metal-poor component, which provides further support to the good performance of our approach in identifying accreted structures. 

\subsection{Substructure D: not only Sequoia}
\label{sequoia}

Substructure D contains some of the most retrograde stars in our halo sample and, based on the location in IoM, may be associated (at least in part) with Sequoia \citep[][]{2019MNRAS.488.1235M}. A quick visual inspection of Fig.~\ref{fig:substructures_IoM} reveals three clusters in this region: two large ones with different $E$, $L_z$, and $L_\perp$ (\#63 and \#64, the latter being the most retrograde) and a third one (\#62) with low $L_\perp$ but similar $L_z$ and energy to those of the larger cluster \#63 but easily distinguishable in the $L_\perp-L_z$ plane in Fig.~\ref{fig:substructures_IoM}. 

\begin{figure*}
\centering 
\includegraphics[width = 0.95\textwidth]{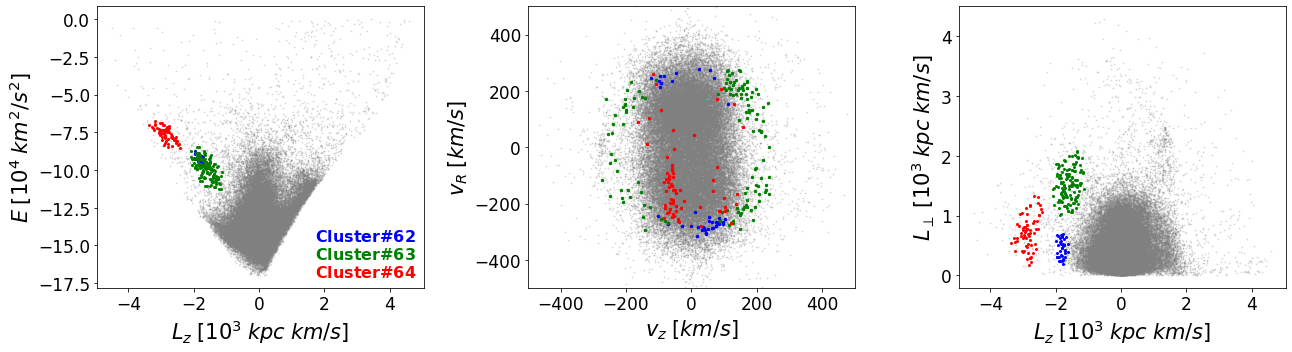} 
\caption{Detailed distribution in the $E-L_z$ (left), the $v_R-v_z$ (middle), and the $L_\perp-L_z$ (right) planes of the tentative substructure D, which is colour-coded according to the different clusters associated to it. The original sample of halo stars is shown in grey in all panels.} 
\label{fig:seq_IoM} 
\end{figure*}

Even though Table~\ref{tab:D_table} suggests that the three clusters can be linked (p-values above 0.05, with cluster \#62 being the most different according to their MDFs), their position in IoM and velocity space suggests otherwise. The right-hand and middle panels of Fig.~\ref{fig:seq_IoM} show that cluster \#64 (in red) is detached from clusters \#62 (in blue) and \#63 (in green). In particular, the kinematics of its stars in the $v_R-v_\phi$ plane (middle panel) are clearly different from the rest. Furthermore, the three clusters identified in the Sequoia region are disconnected in IoM as shown in the rightmost panel of the figure. Therefore, we decided to divide substructure D into the three clusters identified: (a) Cluster \#64, the most retrograde; (b) cluster \#63, which we associate with what has been reported until now as Sequoia \citep[][]{2019MNRAS.488.1235M}; and (c) cluster \#62, which displays a higher average [Fe/H] when compared to clusters \#63 and \#64 (see Fig.~\ref{fig:all_MDFs}). Given that cluster \#62 seems to be complementing cluster \#63 in the $v_z - v_R$ plane, the option that cluster \#62 is really a substream of Sequoia cannot be discarded. 

\begin{table}[h] 
{\centering 
\small 
\begin{tabular}{lccc} 
\hline 
{\bf MDF} & {\bf 64} & {\bf  63} & {\bf  62 } \\ 
{D} & \cellcolor[HTML]{81ca81f}\color{black}0.46 & \cellcolor[HTML]{99d595f}\color{black}0.4 & \cellcolor[HTML]{e7f6e2f}\color{black}0.11  \\ \hline \hline 
{\bf CaMD} & {\bf 64} & {\bf  63} & {\bf  62 } \\ 
{D} & \cellcolor[HTML]{e5f5e0f}\color{black}0.15 & \cellcolor[HTML]{087432f}\color{white}0.85 & \cellcolor[HTML]{d8f0d2f}\color{black}0.18  \\ \hline \hline 
\end{tabular} 
\caption{Results of the KS tests applied to the MDFs (top) and colour distribution (bottom) of the clusters forming the tentative substructure D. \label{tab:D_table}} 
} 
\end{table}

\subsection{Substructure E: Helmi streams}

Substructure {\bf E}, which is related to the Helmi streams, is discussed in Sect.~\ref{analysis} and is used to show our methodology. We refer the reader to that section (as well as to Figs.~\ref{fig:helmi_iso_fit} and~\ref{fig:colour_distr}) for more information.

\subsection{Substructure F}
\label{sgr}

Clusters \#65 and \#66 could be grouped together according to their distribution in IoM at a Mahalanobis distance of $\sim$3.9. Unfortunately, both clusters have very few stars, with cluster \#65 comprising eight stars with [Fe/H] measurements, and cluster \#66 nine such core members. Comparisons of both MDFs and CaMDs suggest that they are drawn from different distributions; see Table~\ref{tab:F_table}.  Although this conclusion is tentative because of the small number of stars, it is supported by their different average metallicities of $\langle$[Fe/H]$\rangle = -1.7$ and $-1.3,$ respectively. Given these results, we do not consider substructure F to be a physically independent structure.

\begin{table}[h] 
{\centering 
\small 
\begin{tabular}{lcc} 
\hline 
{\bf MDF} & {\bf 65*} & {\bf  66* } \\ 
{F} & \cellcolor[HTML]{f6fcf4f}\color{red}0.01 (0\%) & \cellcolor[HTML]{f6fcf4f}\color{red}0.01 (0\%) \\ \hline \hline 
{\bf CaMD} & {\bf 65*} & {\bf  66* } \\ 
{F} & \cellcolor[HTML]{f6fcf4f}\color{red}0.0 (0\%) & \cellcolor[HTML]{f6fcf4f}\color{red}0.0 (0\%)  \\ \hline \hline 
\end{tabular} 
\caption{Results of the KS tests applied to the MDFs (top) and colour distribution (bottom) of the clusters tentatively forming substructure F. \label{tab:F_table}} 
} 
\end{table}

 It is interesting to note that the location in the $E-L_z$ plane of  clusters \#65 and \#66 overlaps with the region that would be associated to debris from the Sagittarius dwarf galaxy \citep[][]{2020ApJ...901...48N} and indeed with some of the globular clusters linked with Sgr in \citet[][]{2019A&A...630L...4M}. However, that is not the case in other IoM projections, with the clusters identified in this work displaying a higher $L_\perp$ than the globular clusters associated with Sgr in \citet[][]{2019A&A...630L...4M}. With the information at hand, it is not possible to properly further assess  the nature of the clusters or their tentative association to Sgr, except to note that the average metallicity of cluster \#66 is only slightly more metal-poor than that reported for the leading arm of Sgr in \citet{2020ApJ...889...63H}.

\subsection{Cluster \#3}
\label{low_energy}

Cluster \#3  is not only detached from any other structure according to the dendrogram of Fig.~\ref{fig:dendrogram} and our tentative Mahalanobis distance threshold, but is also composed of a large number of stars, which drives us to consider it as a potential substructure on its own. Cluster \#3 is located in the low-energy part of the densely packed region where halo and thick disc cohabit. It is the largest cluster detected by the algorithm, with 3032 stars in total, and presents a statistical significance of 6.3$\sigma$ (see paper I). Of its 3032 stars, 2075  have metallicity determinations from LAMOST LRS.  

Cluster \#3 displays a complex MDF (see Fig.~\ref{fig:all_MDFs}). While most of the stars are compatible with being thick-disc stars ([Fe/H]~$>-1.0$, peaking at $\sim-0.7$), there exists a non-negligible population of metal-poor stars (with [Fe/H] from $-1.0$ to $-2.2$). Thus, cluster \#3 contains a mix of populations, with the more metal-poor one possibly of accreted origin, although we cannot rule out the possibility that this population is associated to an old in situ halo or disc. Comparing the energy and angular momentum distributions of the globular clusters associated to Kraken/Heracles with cluster \#3, we find that cluster \#3 is a more weakly bound (displaying higher energies) system than Kraken/Heracles, and thus, they are not related to one another \citep[see][] {2019MNRAS.486.3180K, 2019A&A...630L...4M, 2021MNRAS.500.1385H}.

\subsection{Individual small clusters}
\label{ind_clusters}

From the analysis presented so far in the previous sections, we identify 12 large substructures, and isolate 4 small clusters that are not easily associated to any of those large structures. These are clusters \#28 and \#40 (originally associated with substructure A), and clusters \#65 and \#66 (originally postulated to form substructure F). Apart from these, there are 4 additional clusters that are not linked in IoM by our approach unless we allow a large Mahalanobis distance for the union (see Fig.~\ref{fig:dendrogram}). These clusters are \#1, \#2, \#4, and \#68, and are shown in the bottom right panels of  Fig.~\ref{fig:substructures_IoM}. We briefly comment on each of them below:
\begin{itemize}
    \item Cluster \#1 is formed by 33 stars that tightly clump together, not only in IoM (very low energy), but especially in velocity space and on the sky. As discussed in paper I, these stars are likely members of the globular cluster M4. 
 
    \item Cluster \#2 is composed of 12 stars (8 with [Fe/H] values from LAMOST LRS) located in between the hot thick disc and the low-energy half of the halo population, with $L_z \sim 500$~kpc~km/s. Its stars show an overall metallicity of [Fe/H]~$\sim -0.86$, with a wide dispersion of 0.4 dex. Approximately 63\% of its stars have a narrower metallicity range, namely [Fe/H]=[$-0.75$, $-0.4$]. The velocity distribution is particularly tight in the  $v_\phi$ velocity component.
 
    \item Cluster \#4 is also located in the region dominated by thick-disc stars (close to cluster \#2), and contains only 14 stars, of which 9 have [Fe/H] determinations from LAMOST LRS. Their average [Fe/H]~$\sim -1.2$ and dispersion is 0.4~dex.

    \item Cluster \#68 contains 29 stars (represented in pink in Fig.~\ref{fig:substructures_IoM}, panels for independent clusters) and seems to be related to the outer disc (given its $z$-angular momentum and low $L_\perp$). \end{itemize}

We do not further consider the details of these small individual clusters (\#1, \#2, \#4, \#28, \#40, \#65, \#66, and \#68) because of their limited sample size and given that we are interested in unveiling the main constituents of the Milky Way halo. We will revise these small individual clusters together with other tight clumps in IoM and velocity space in Balbinot et al. (in prep.). 

\section{An updated view of the Galactic halo}
\label{updated_view}

The thorough analysis carried out in the previous section allows us to add support to some of the tentative conclusions of paper I, where we identified 67 clusters with high significance in the nearby halo with a single-linkage-based clustering algorithm. By comparing the properties of these clusters, we are able to isolate 12 extended (in IoM) separate substructures, some of which are constituted by a single cluster. The substructures include the Helmi streams \citep[substructure {\bf E},][]{1999Natur.402...53H}; three dynamically distinct clusters in the Sequoia region \cite[clusters \#62, \#63, and \#64, see also][]{2019MNRAS.488.1235M}; Thamnos~1 and Thamnos~2 \citep[substructures {\bf B1} and {\bf B2} respectively, see also ][]{2019A&A...631L...9K}; GE \citep[substructure {\bf C}, ][]{2018Natur.563...85H}; one independent structure with similar angular momentum to GE but higher binding energy (cluster \#6); a high-binding-energy substructure in a region of overlap between the hot thick disc and halo (cluster \#3); and three subgroups (substructure {\bf A}, and clusters \#12 and \#38) in the region occupied by the hot thick disc. In Appendix~\ref{appendix1}, we present summary plots of their distributions in IoM and in velocity space. In those figures, the colour coding represents the constituent clusters from paper~I.

Apart from these 12 substructures, we  discern eight individual small clusters that, based on their IoM distribution or internal properties, we are not able to assign to any of the identified structures.  Their small sizes and in some cases their tightness in IoM or velocity space suggests a possible link to disrupted star clusters. These will be investigated in detail in Balbinot et al. (in prep.). Therefore, we do not consider these small clusters in the remainder of this paper. 

We now proceed to the full characterisation of the 12 substructures. To this end, in Sect.~\ref{comparing_substructures} we first compare their global MDFs as well as their CaMDs using our isochrone-fitting approach, which includes determination of their average age and [Fe/H]. In Sect.~\ref{sub_IoM} we characterise their dynamical properties via a principal component analysis (PCA), which permits the identification of additional member stars. Lastly, in  Sect.~\ref{chemistry}, we make a first attempt at characterising their chemical properties using APOGEE data. This combined approach provides an updated view of independent substructures in the nearby Galactic halo, as discussed in Sect.~\ref{new_view}. 

\subsection{MDF and CaMDs of the substructures}
\label{comparing_substructures}

\begin{table*} 
{\centering 
\small 
\begin{tabular}{lrrrrrrrrrrrrr} 
\hline 
{\bf } & {\bf Helmi} & {\bf 64} & {\bf Sequoia} & {\bf 62} & {\bf Thamnos1} & {\bf Thamnos2} & {\bf GE} & {\bf 6} & {\bf 3} & {\bf A} & {\bf 12} & {\bf 38} \\ 
{\bf Helmi} & \cellcolor[HTML]{00441bf}\color{white}1.0 & \cellcolor[HTML]{70c274f}\color{black}0.51 & \cellcolor[HTML]{5eb96bf}\color{black}0.55 & \cellcolor[HTML]{f2faeff}\color{red}0.04 & \cellcolor[HTML]{72c375f}\color{black}0.51 & \cellcolor[HTML]{f7fcf5f}\color{red}0.0 & \cellcolor[HTML]{f7fcf5f}\color{red}0.0 & \cellcolor[HTML]{f7fcf5f}\color{red}0.0 & \cellcolor[HTML]{f7fcf5f}\color{red}0.0 & \cellcolor[HTML]{f7fcf5f}\color{red}0.0 & \cellcolor[HTML]{f7fcf5f}\color{red}0.0 & \cellcolor[HTML]{f7fcf5f}\color{red}0.0  \\ {\bf 64} & \cellcolor[HTML]{70c274f}\color{black}0.51 & \cellcolor[HTML]{00441bf}\color{white}1.0 & \cellcolor[HTML]{1a843ff}\color{white}0.78 & \cellcolor[HTML]{e7f6e3f}\color{black}0.11 & \cellcolor[HTML]{e4f5dff}\color{black}0.13 & \cellcolor[HTML]{f7fcf5f}\color{red}0.0 & \cellcolor[HTML]{f7fcf5f}\color{red}0.0 & \cellcolor[HTML]{f7fcf5f}\color{red}0.0 & \cellcolor[HTML]{f7fcf5f}\color{red}0.0 & \cellcolor[HTML]{f7fcf5f}\color{red}0.0 & \cellcolor[HTML]{f7fcf5f}\color{red}0.0 & \cellcolor[HTML]{f7fcf5f}\color{red}0.0  \\ {\bf Sequoia} & \cellcolor[HTML]{5eb96bf}\color{black}0.55 & \cellcolor[HTML]{1a843ff}\color{white}0.78 & \cellcolor[HTML]{00441bf}\color{white}1.0 & \cellcolor[HTML]{e9f7e5f}\color{black}0.1 & \cellcolor[HTML]{88ce87f}\color{black}0.44 & \cellcolor[HTML]{f7fcf5f}\color{red}0.0 & \cellcolor[HTML]{f7fcf5f}\color{red}0.0 & \cellcolor[HTML]{f7fcf5f}\color{red}0.0 & \cellcolor[HTML]{f7fcf5f}\color{red}0.0 & \cellcolor[HTML]{f7fcf5f}\color{red}0.0 & \cellcolor[HTML]{f7fcf5f}\color{red}0.0 & \cellcolor[HTML]{f7fcf5f}\color{red}0.0  \\ {\bf 62} & \cellcolor[HTML]{f2faeff}\color{red}0.04 & \cellcolor[HTML]{e7f6e3f}\color{black}0.11 & \cellcolor[HTML]{e9f7e5f}\color{black}0.1 & \cellcolor[HTML]{00441bf}\color{white}1.0 & \cellcolor[HTML]{a7dba0f}\color{black}0.36 & \cellcolor[HTML]{b5e1aef}\color{black}0.31 & \cellcolor[HTML]{f6fcf4f}\color{red}0.01 & \cellcolor[HTML]{f7fcf5f}\color{red}0.0 & \cellcolor[HTML]{f7fcf5f}\color{red}0.0 & \cellcolor[HTML]{f7fcf5f}\color{red}0.0 & \cellcolor[HTML]{f7fcf5f}\color{red}0.0 & \cellcolor[HTML]{f7fcf5f}\color{red}0.0  \\ {\bf Thamnos1} & \cellcolor[HTML]{72c375f}\color{black}0.51 & \cellcolor[HTML]{e4f5dff}\color{black}0.13 & \cellcolor[HTML]{88ce87f}\color{black}0.44 & \cellcolor[HTML]{a7dba0f}\color{black}0.36 & \cellcolor[HTML]{00441bf}\color{white}1.0 & \cellcolor[HTML]{f5fbf3f}\color{red}0.01 & \cellcolor[HTML]{f7fcf5f}\color{red}0.0 & \cellcolor[HTML]{f7fcf5f}\color{red}0.0 & \cellcolor[HTML]{f7fcf5f}\color{red}0.0 & \cellcolor[HTML]{f7fcf5f}\color{red}0.0 & \cellcolor[HTML]{f7fcf5f}\color{red}0.0 & \cellcolor[HTML]{f7fcf5f}\color{red}0.0  \\ {\bf Thamnos2} & \cellcolor[HTML]{f7fcf5f}\color{red}0.0 & \cellcolor[HTML]{f7fcf5f}\color{red}0.0 & \cellcolor[HTML]{f7fcf5f}\color{red}0.0 & \cellcolor[HTML]{b5e1aef}\color{black}0.31 & \cellcolor[HTML]{f5fbf3f}\color{red}0.01 & \cellcolor[HTML]{00441bf}\color{white}1.0 & \cellcolor[HTML]{f7fcf5f}\color{red}0.0 & \cellcolor[HTML]{f7fcf5f}\color{red}0.0 & \cellcolor[HTML]{f7fcf5f}\color{red}0.0 & \cellcolor[HTML]{f7fcf5f}\color{red}0.0 & \cellcolor[HTML]{f7fcf5f}\color{red}0.0 & \cellcolor[HTML]{f7fcf5f}\color{red}0.0  \\ {\bf GE} & \cellcolor[HTML]{f7fcf5f}\color{red}0.0 & \cellcolor[HTML]{f7fcf5f}\color{red}0.0 & \cellcolor[HTML]{f7fcf5f}\color{red}0.0 & \cellcolor[HTML]{f6fcf4f}\color{red}0.01 & \cellcolor[HTML]{f7fcf5f}\color{red}0.0 & \cellcolor[HTML]{f7fcf5f}\color{red}0.0 & \cellcolor[HTML]{00441bf}\color{white}1.0 & \cellcolor[HTML]{f6fcf4f}\color{red}0.0 & \cellcolor[HTML]{f7fcf5f}\color{red}0.0 & \cellcolor[HTML]{f7fcf5f}\color{red}0.0 & \cellcolor[HTML]{f7fcf5f}\color{red}0.0 & \cellcolor[HTML]{f7fcf5f}\color{red}0.0  \\ {\bf 6} & \cellcolor[HTML]{f7fcf5f}\color{red}0.0 & \cellcolor[HTML]{f7fcf5f}\color{red}0.0 & \cellcolor[HTML]{f7fcf5f}\color{red}0.0 & \cellcolor[HTML]{f7fcf5f}\color{red}0.0 & \cellcolor[HTML]{f7fcf5f}\color{red}0.0 & \cellcolor[HTML]{f7fcf5f}\color{red}0.0 & \cellcolor[HTML]{f6fcf4f}\color{red}0.0 & \cellcolor[HTML]{00441bf}\color{white}1.0 & \cellcolor[HTML]{f7fcf5f}\color{red}0.0 & \cellcolor[HTML]{f7fcf5f}\color{red}0.0 & \cellcolor[HTML]{f7fcf5f}\color{red}0.0 & \cellcolor[HTML]{f7fcf5f}\color{red}0.0  \\ {\bf 3} & \cellcolor[HTML]{f7fcf5f}\color{red}0.0 & \cellcolor[HTML]{f7fcf5f}\color{red}0.0 & \cellcolor[HTML]{f7fcf5f}\color{red}0.0 & \cellcolor[HTML]{f7fcf5f}\color{red}0.0 & \cellcolor[HTML]{f7fcf5f}\color{red}0.0 & \cellcolor[HTML]{f7fcf5f}\color{red}0.0 & \cellcolor[HTML]{f7fcf5f}\color{red}0.0 & \cellcolor[HTML]{f7fcf5f}\color{red}0.0 & \cellcolor[HTML]{00441bf}\color{white}1.0 & \cellcolor[HTML]{f7fcf5f}\color{red}0.0 & \cellcolor[HTML]{f7fcf5f}\color{red}0.0 & \cellcolor[HTML]{f7fcf5f}\color{red}0.0  \\ {\bf A} & \cellcolor[HTML]{f7fcf5f}\color{red}0.0 & \cellcolor[HTML]{f7fcf5f}\color{red}0.0 & \cellcolor[HTML]{f7fcf5f}\color{red}0.0 & \cellcolor[HTML]{f7fcf5f}\color{red}0.0 & \cellcolor[HTML]{f7fcf5f}\color{red}0.0 & \cellcolor[HTML]{f7fcf5f}\color{red}0.0 & \cellcolor[HTML]{f7fcf5f}\color{red}0.0 & \cellcolor[HTML]{f7fcf5f}\color{red}0.0 & \cellcolor[HTML]{f7fcf5f}\color{red}0.0 & \cellcolor[HTML]{00441bf}\color{white}1.0 & \cellcolor[HTML]{ecf8e8f}\color{black}0.08 & \cellcolor[HTML]{f6fcf4f}\color{red}0.01  \\ {\bf 12} & \cellcolor[HTML]{f7fcf5f}\color{red}0.0 & \cellcolor[HTML]{f7fcf5f}\color{red}0.0 & \cellcolor[HTML]{f7fcf5f}\color{red}0.0 & \cellcolor[HTML]{f7fcf5f}\color{red}0.0 & \cellcolor[HTML]{f7fcf5f}\color{red}0.0 & \cellcolor[HTML]{f7fcf5f}\color{red}0.0 & \cellcolor[HTML]{f7fcf5f}\color{red}0.0 & \cellcolor[HTML]{f7fcf5f}\color{red}0.0 & \cellcolor[HTML]{f7fcf5f}\color{red}0.0 & \cellcolor[HTML]{ecf8e8f}\color{black}0.08 & \cellcolor[HTML]{00441bf}\color{white}1.0 & \cellcolor[HTML]{f1faeef}\color{black}0.06  \\ {\bf 38} & \cellcolor[HTML]{f7fcf5f}\color{red}0.0 & \cellcolor[HTML]{f7fcf5f}\color{red}0.0 & \cellcolor[HTML]{f7fcf5f}\color{red}0.0 & \cellcolor[HTML]{f7fcf5f}\color{red}0.0 & \cellcolor[HTML]{f7fcf5f}\color{red}0.0 & \cellcolor[HTML]{f7fcf5f}\color{red}0.0 & \cellcolor[HTML]{f7fcf5f}\color{red}0.0 & \cellcolor[HTML]{f7fcf5f}\color{red}0.0 & \cellcolor[HTML]{f7fcf5f}\color{red}0.0 & \cellcolor[HTML]{f6fcf4f}\color{red}0.01 & \cellcolor[HTML]{f1faeef}\color{black}0.06 & \cellcolor[HTML]{00441bf}\color{white}1.0  \\ \hline \hline 
\end{tabular} 
\caption{Kolmogorov-Smirnov statistical tests comparing the metallicity distribution functions of all of the 12 extended substructures (including the large independent clusters) identified in this work. Colours and information are the same as in Table~\ref{tab:A1_A2}} \label{KS_met_tab_ALL}
} 
\end{table*} 

Table~\ref{KS_met_tab_ALL} displays the p-values of the KS tests comparing the MDFs of every pair of possible independent substructures. This table reveals clear quantitative differences in the metallicity distributions of the various substructures, particularly  of substructures that are located in the region of the hot thick disc (substructure {\bf A} and clusters \#3, \#12, and \#38) compared to the rest. 

We also see from Table~\ref{KS_met_tab_ALL} that some substructures that occupy very different regions of IoM present similar MDFs, such as Sequoia, the Helmi streams, and Thamnos 1. Some of this is expected, because the progenitors of Sequoia and the Helmi streams for example have likely been systems of comparable size \citep[see e.g.][]{2019A&A...631L...9K}, possibly obeying a mas-s-metallicity relation \citep[e.g.][]{2013ApJ...779..102K}. The MDFs of substructures that are closer in IoM, as in the case of Sequoia and cluster \#64, could be similar because they suffer from contamination from each other (see Sect.~\ref{sub_IoM} for further discussion). Also, interesting similarities are present between cluster \#62 (which is tentatively independent of Sequoia) and Thamnos 1 and 2. Although we have not fully ruled out an association between cluster \#62 and Sequoia, the MDF of the former appears to be somewhat more similar to that of smaller structures like Thamnos 1 and 2. 

\begin{figure*}[!h]
\centering 
\includegraphics[width = 0.99\textwidth]{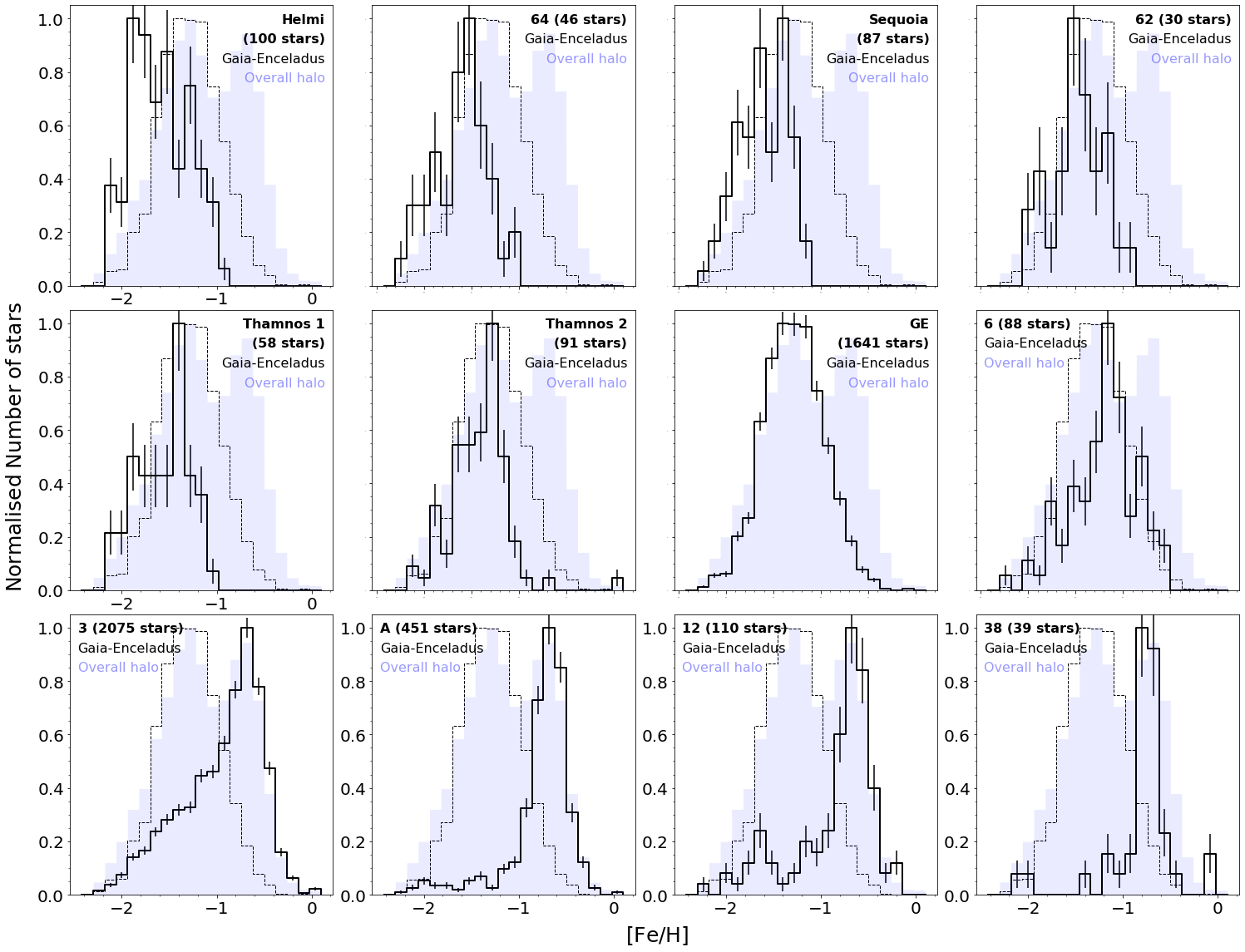}
\caption{Metallicity distribution functions of the 12 extended substructures identified in this work. These MDFs are compared to that of the whole halo population under analysis (blue shaded area) and to that of {\it Gaia}-Enceladus (grey dashed histogram). Substructure ID and number of associated stars with a [Fe/H] determination from LAMOST LRS are listed. Error bars were computed assuming Poisson statistics.} 
\label{fig:all_MDFs} 
\end{figure*}

Figure~\ref{fig:all_MDFs} shows the MDFs of the various substructures. There clearly exists a correlation between the location of structures in the $E-L_z$ plane and the properties  of their MDF. Structures overlapping with the hot thick disc region (substructure {\bf A} and clusters \#3, \#12, and \#38) show an important contribution of higher metallicity stars (with peak [Fe/H]~$\sim -0.6$) associated to the halo red sequence \citep[see e.g.][]{2018ApJ...863..113H}.  Cluster \#3 occupies a region in IoM that is intermediate between accreted halo stars from GE and in situ hot thick disc stars. Despite showing average values of age and metallicity that could be compatible with a thick-disc population ($\sim$ 10.5~Gyr and [Fe/H]$\sim-0.75$), Fig.~\ref{fig:all_MDFs} clearly reveals a very important contribution  from metal-poor
stars to its MDF. In comparison, substructure {\bf A} is much more dominated by a thick disc population, with their MDF peaking at [Fe/H]$\sim-0.6$, and displaying a more tenuous tail of metal-poor stars. This is also apparent for cluster \#12, but a possible interpretation in this case is that it suffers from contamination from substructure {\bf A} and that the peak at lower metallicity should be considered as an independent subgroup (this is further supported by the CaMD analysis; see Table~\ref{age_met_tab}). 

We interpret these findings as hints that an accreted, metal-poor population might be buried in the region that overlaps with the hot thick disc, although we cannot rule out the possibility of this being a relic of the old, in situ halo or disc. The chemical analysis presented in Sect.~\ref{chemistry} supports this interpretation to some extent, but more information is necessary to establish this firmly.  

\begin{figure*}[!h]
\centering 
\includegraphics[width = 0.93\textwidth]{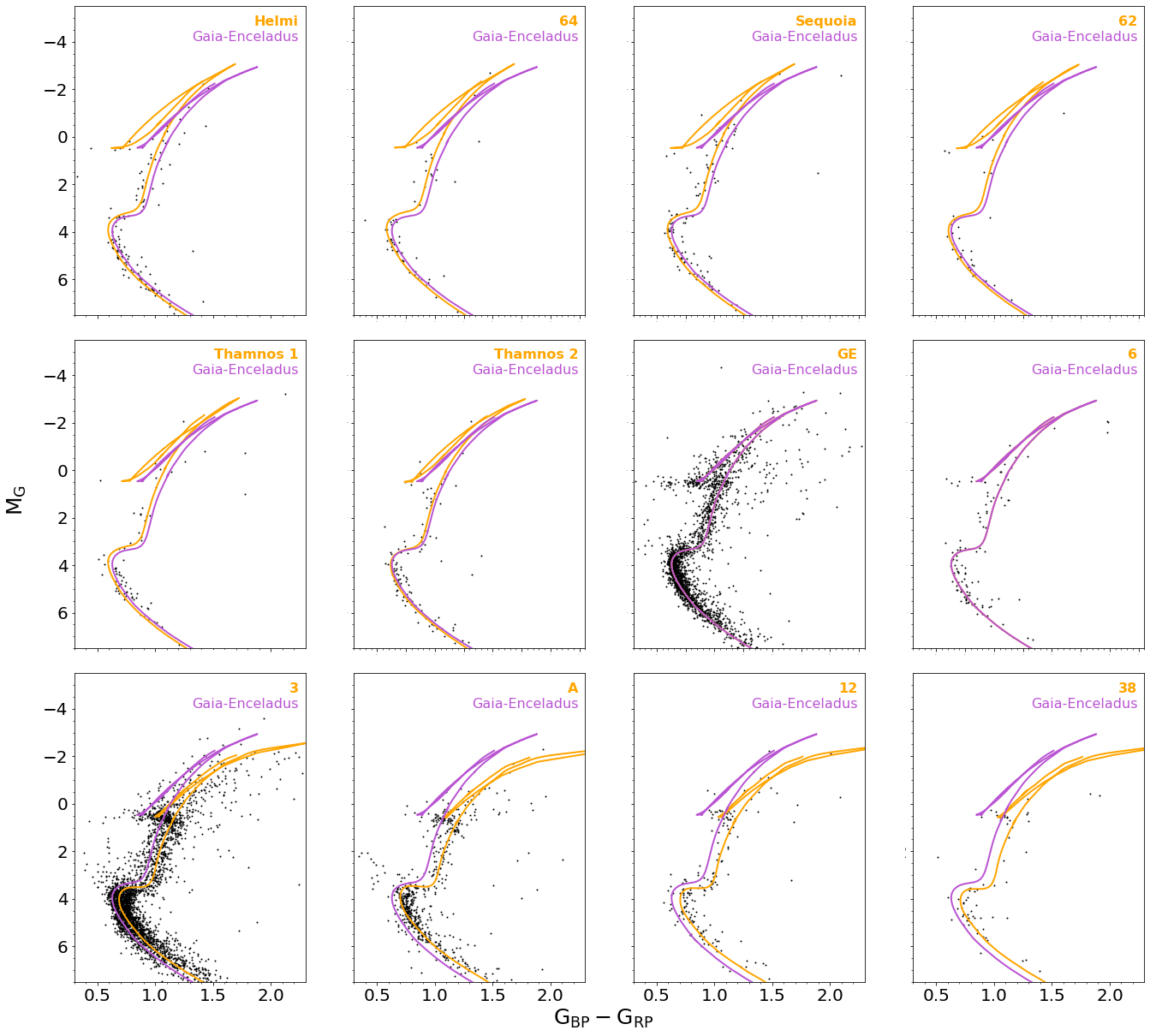} 
\caption{CaMDs for the 12 extended substructures identified in this work. We include the best isochrone from our isochrone-fitting approach (see Sect.~\ref{iso_fit}) to each structure (orange solid line) as well as that for the combined {\it Gaia}-Enceladus substructure (purple solid line).} 
\label{fig:all_CMDs} 
\end{figure*}

Figure~\ref{fig:all_MDFs} shows that, with our definition of Sequoia, its MDF might hint at the existence of three metallicity peaks. These have similar properties to those reported in \citet[][which led these authors to divide Sequoia into three, namely L'itoi, Sequoia proper, and Arjuna]{2020ApJ...901...48N}, but the peaks are not statistically significant in our sample (see also the discussion in paper~I) and there is no correspondence between Arjuna, I’itoi, and Sequoia in \citet[][]{2020ApJ...901...48N} and our clusters \#62, \#63, and \#64. The remaining substructures shown in Fig.~\ref{fig:all_MDFs} display MDFs with peaks at lower metallicities, compatible with an origin in accreted systems of lower mass.

We now focus on the CaMDs of the various substructures. The results of the isochrone-fitting procedure for each of the substructures are given in Table~\ref{age_met_tab}. This lists for each substructure its average age and [Fe/H], where the latter is in good agreement with the estimates based on the LAMOST LRS data. As stated above, here we also observe a clear dichotomy between accreted structures (old and metal-poor) and those with more disky dynamics (younger and more metal-rich). Figure~\ref{fig:all_CMDs} shows the CaMDs of all 12 independent substructures with their best-fit isochrone overplotted with an orange solid line. To compare with one of the main constituents making up the inner accreted halo, we also overplot the GE best-fit isochrone. We see that there are only subtle differences in the distribution of stars in the CaMD, which are nevertheless picked up with a proper quantitative analysis (via the fitting of isochrones) as can be seen in Table~\ref{age_met_tab}.

\begin{table} 
{\centering
\begin{tabular}{lrr}
\hline
 Substructure &   Age  &  [Fe/H]  \\
           &   (Gyr)  &  (dex) \\
\hline
    Helmi ({\bf E}) &  12.1 $\pm$ 1.1 & -1.45 $\pm$ 0.11 \\
       64 &  11.7 $\pm$ 1.4 & -1.47 $\pm$ 0.06 \\
  Sequoia (\# 63) &  12.2 $\pm$ 1.2 & -1.50 $\pm$ 0.09 \\
       62 &  12.3 $\pm$ 1.1 & -1.36 $\pm$ 0.13 \\
 Thamnos1 ({\bf B1}) &  11.6 $\pm$ 1.4 & -1.42 $\pm$ 0.09 \\
 Thamnos2 ({\bf B2}) &  12.3 $\pm$ 1.0 & -1.30 $\pm$ 0.14 \\
       GE ({\bf C}) &  11.4 $\pm$ 1.4 & -1.12 $\pm$ 0.10 \\
        6 &  11.3 $\pm$ 1.4 & -1.10 $\pm$ 0.07 \\
        3 &  10.5 $\pm$ 1.9 & -0.75 $\pm$ 0.11 \\
        {\bf A} &   8.9 $\pm$ 2.1 & -0.55 $\pm$ 0.10 \\
       12 &  10.4 $\pm$ 1.9 & -0.64 $\pm$ 0.10 \\
       38 &  10.7 $\pm$ 1.6 & -0.64 $\pm$ 0.11 \\
\hline \hline
\end{tabular}
\caption{Isochrone fitting age and metallicity ([Fe/H]) for the different independent substructures identified in this work. If cluster \#12 is split into two populations, i.e. the low- and the high-metallicity peaks seen in Fig.~\ref{fig:all_MDFs}, our isochrone fits estimate a mean age of $12.2\pm 1$ Gyr for the low-metallicity stars, and $8.8 \pm 0.52$ Gyr for those with high metallicity. The latter is consistent with that of substructure {\bf A}.} 
\label{age_met_tab}
}
\end{table}

\subsection{Substructure characterisation in IoM}
\label{sub_IoM}

In this subsection we re-examine the extent of each of the individual substructures in IoM. To do this, we follow a similar approach to that presented in paper I for individual clusters, and compute the covariance matrix of the members of each of the 12 independent  substructures. Here the members we refer to consist of the union of the original members from clusters that are associated to each given substructure. The results are given in Table~\ref{table:substructure_description}, along with the significance of each substructure, which is calculated by comparing the density of member stars to that from randomised data sets within the PCA ellipsoidal contour encompassing 95.4\% of the members (see Sect.~3.3 of paper I for a full description). To transform the values listed in this table to the corresponding physical quantities, for the means of the ellipsoids we can use that
\begin{equation}
    \langle \mu_i \rangle = \frac{2}{\Delta_i} (\langle I_{i}\rangle-I_{i,min}) -1
    \label{eq:transf_m}
,\end{equation}
where $\Delta_i = I_{i,max}-I_{i,min}$, with $i=0..2$ and $I_{i}$ corresponding to $E$, $L_\perp$,  and $L_z$, respectively, with the minimum and maximum values given by $E=[-170000, 0]$ km$^2$/s$^2$, $L_\perp=[0, 4300]$ kpc km/s, and  $L_z=[-4500,4600]$ kpc km/s, which roughly correspond to the minimum and maximum values in the halo set. For the covariance matrix, these definitions lead to
\begin{equation}
    \sigma_{ij} = \frac{4}{\Delta_i \Delta_j}\Sigma_{I_{i,j}}
    \label{eq:transf_s}
,\end{equation}
where $\Sigma_{I_{i,j}}$ is the covariance matrix in IoM space. 

\begin{figure}
\centering 
\includegraphics[width = 0.45\textwidth]{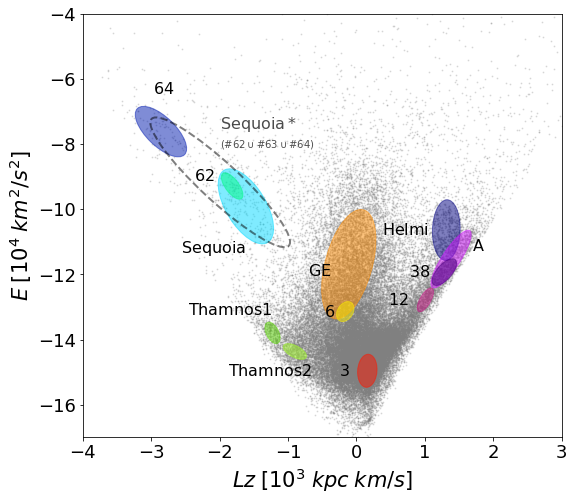}
\caption{Ellipsoidal boundaries of the 12 independent extended substructures identified in this work, using a PCA. The black dashed ellipse represents the ellipsoid that would be representative of Sequoia if defined as the union of clusters \#62, \#63, and \#64. The original sample of halo stars is shown as a grey background.} 
\label{fig:all_ellipses} 
\end{figure}

Figure~\ref{fig:all_ellipses} shows the ellipsoidal boundaries of such a characterisation of the various substructures but now in the $E-L_z$ plane. These were determined by a PCA analysis in this plane, where the contours indicate the 1.77$\sigma$ extent along the principal axes of the structure (roughly encompassing 80\% of a perfect Gaussian distribution). 

We observe in Fig.~\ref{fig:all_ellipses} that the orientation of the ellipsoid of GE is somewhat odd, as it does not conform to the expectation that the ellipsoids' tilt approximately follows the circularity parameter \citep[see e.g.][]{2000MNRAS.319..657H}; that is, typically stars from a given accreted object ought to have higher $|L_z|$ as they have more energetic or less-bound orbits. The tilt is likely in part due to how the random sets to assess the significance of a cluster have been constructed. Whereas in the original dataset, GE stars and the hot thick disc are clearly separated (see e.g. the distribution of grey points in Fig.~\ref{fig:all_ellipses}), in our random sets they are not (see also Fig.~3 of paper~I). This is because stars from both these dominant components are being moved by the randomisation process, which consists in reshuffling their velocities to the region in between. A better way to assess the significance would therefore be to model the presence of the two components.  However, this can be remedied  to a certain extent by considering clusters with lower significance as well (down to 2$\sigma$ in comparison to the randomised datasets; see paper I). Furthermore, given the distribution of stars in IoM, especially the increase in density towards higher binding energies and slightly prograde orbits (i.e. positive $L_z$), the identification of  highly significant clusters is more challenging in that area, and could therefore cause part of the tilt.

The analysis presented in Sect.~\ref{sequoia} prompted us to deem the clusters located in the Sequoia region  as independent, mainly based on their different dynamical properties (position in IoM planes as well as velocity distributions). Another argument against their union is that this would require a very massive progenitor  \citep[see e.g.][]{2019A&A...631L...9K}. The black dashed ellipse in Fig.~\ref{fig:all_ellipses} represents the extent in the $E-L_z$ plane of an alternative definition of Sequoia including clusters \#62, \#63, and \#64. Such a large system would occupy an area similar to that of GE in the $E-L_z$ plane (translated into a very massive system). This would be inconsistent with the average low metallicity of the three clusters ([Fe/H]~$\sim -1.6$, with a dispersion of 0.3 dex for clusters \#63 and \#64; and [Fe/H]~$\sim-1.47$, with a dispersion of 0.25 dex for cluster \#62, and which can be roughly compared to the $\langle$[Fe/H]$\rangle \sim-1.3$ of GE, with a dispersion of 0.3~dex), if we assume they follow the known mass--metallicity relation \citep[from the local to the high-redshift Universe, e.g.][]{2006ApJ...644..813E, 2013ApJ...779..102K, 2016MNRAS.456.2140M}. 

Interestingly, cluster \#64 (the most retrograde structure in the Sequoia region) presents a distribution of stars in the CaMD that is almost `parallel' to that of GE towards the more metal-poor end (as shown by its lower metallicity compared to GE; see Table~\ref{age_met_tab}). This could be consistent with the possibility that this substructure originates in part from the metal-poor tail of GE (indeed they show compatible ages, with cluster \#64 being marginally older than GE), an interpretation that would be favoured by the simulations of \citet{2019A&A...631L...9K}. However, as recently shown in \citet{2021arXiv211115423M}, some stars that would be associated to cluster \#64 have chemical properties that appear to be different from those of GE, and more similar to those of Sequoia. As shown in that paper, there is tentative evidence of contamination by GE in that region of IoM, at a level of 10\%-20\%. To draw firmer conclusions, a more profound characterisation of the chemistry of these clusters is much needed.

Figure~\ref{fig:all_ellipses} is similar to figures presented in \citet{2019A&A...631L...9K} and \citet{ 2020ApJ...901...48N} based on {\it Gaia} DR2 data, but reveals many more substructures, with an important number in the region populated by the hot thick disc.  However, as pointed out in paper I we also miss some known substructures, such as Wukong/LMS-1 (\citealt{2020ApJ...901...48N} and \citealt{2020ApJ...898L..37Y}), likely because our sample is concentrated on the local halo, and Aleph \citep[][]{2020ApJ...901...48N} because of our kinematic selection. 

The characterisation of the 3D ellipsoids in IoM listed in Table~\ref{table:substructure_description} enables us to identify additional tentative members of the structures. To do this, as in paper~I and following the refinement of individual clusters in paper I and Sect.~\ref{data}, we consider stars that are located within a Mahalanobis distance $D_{\rm cut} \le 2.13$ of the centre of the different independent structures as additional tentative members. There are a total of 10477 stars within the $D_{\rm cut}$ Mahalanobis distance  to one of the 12 substructures, implying that approximately 20.3\% of the total sample under analysis can likely be associated to these large substructures. The final number of stars associated to each independent substructure (obtained as described immediately above) is listed in Table \ref{table:substructure_description} in parentheses.

\subsection{Substructure chemical characterisation}
\label{chemistry}

\begin{figure*}
\centering
\includegraphics[width=\textwidth]{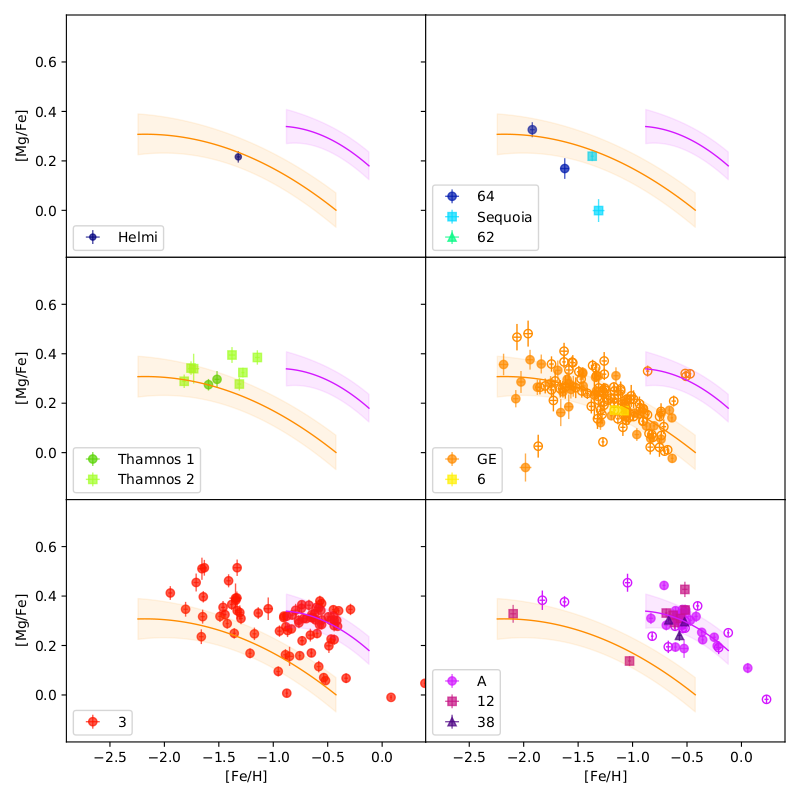}
\caption{[{Mg}/{Fe}] and [{Fe}/{H}] abundances of the different substructures identified in this work, using data from APOGEEDR16. Filled symbols represent core members of the clusters associated with the substructure considered, while smaller open symbols are for stars that are "added" based on their Mahalanobis distance  to each substructure (see Sect.~\ref{sub_IoM} and Table~\ref{table:substructure_description}).
Chemical abundance trends of A (thick disc; purple) and GE (yellow) are shown in all panels to guide the eye. Colours are as in Fig.~\ref{fig:all_ellipses}.
\label{fig:Mg_ALL}}
\end{figure*}

We now study the behaviour of [{Mg}/{Fe}] with [Fe/H] using data from APOGEE (see Sect.~\ref{KS_tests}) for the different substructures identified. Following the procedure described in the previous section, we consider members of the different substructures to be those stars with Mahalanobis distances $D_{\rm cut} \le  2.13$ to the centre of the different substructures. To guide the eye, we fit the chemical evolution trends of substructure {\bf A} and {\it Gaia}-Enceladus (representative of the hot thick disc and of the accreted population, respectively) with quadratic polynomial functions. For the fitting itself, we use the subset of core members of the clusters  that were merged to form substructure {\bf A} or {\it Gaia}-Enceladus, respectively (see Sects.~\ref{thick_disc} and \ref{GE}). After a first fit is obtained, we remove $3\sigma$ chemical outliers, where $\sigma$ includes both the uncertainty in the fit (estimated through bootstrapping) and the residual scatter. The results are shown in Fig.~\ref{fig:Mg_ALL}, where the shaded region denotes the 1-$\sigma$ width described above. In this figure, we also plot the abundances of the individual stars associated to the various identified substructures.

The bottom right panel of Fig.~\ref{fig:Mg_ALL} presents chemical abundances of substructure {\bf A}, and clusters \#12 and \#38. Substructure {\bf A} has a high [{Mg}/{Fe}] ratio as expected from the suggested hot thick disc origin. While both clusters \#12 and \#38 overlap with the track followed by {\bf A}, cluster \#12 includes a few stars on the Gaia-Enceladus track. The additional chemical dimension of Mg abundance enables us to see chemical differences between the two stellar populations  more
clearly, highlighting the importance of studying abundance ratios. 

Compared to substructure {\bf A}, the substructure associated to GE has a lower [{Mg}/{Fe}] (middle right panel of Fig.~\ref{fig:Mg_ALL}), which is consistent with the well-known chemical abundance differences \citep[first reported as high-Mg and low-Mg sequences respectively,  in][]{Nissen2010}. In comparison to Figure~\ref{GEMgFe}, which shows only the core members of the clusters associated to GE, we note a broader chemical sequence which is the result of adding extra tentative members likely associated with {\it Gaia}-Enceladus based on their Mahalanobis distance to the substructure. For instance, above $[\mathrm{Fe/H}]>-0.9$, there are no contaminant stars (i.e. with high [{Mg}/{Fe}] $>0.2$) among cluster core members (9 stars in this metallicity range), while this fraction is 19\% (5 out of 27 stars) when adding members based on the Mahalanobis distance to the whole structure. 
This might suggest that the actual distribution of stars from {\it Gaia}-Enceladus is better captured by those stars belonging to the clusters associated with it, and that adding members to substructures assuming these follow a Gaussian distribution might lead to a lower purity.  We also include substructure \#6 in the same panel as GE. The two stars from this cluster seem to follow the same chemical track as GE (but small number statistics prevent us from discerning the differences seen in their MDFs).

\begin{table*}
\centering
\begin{tabular}{l l c l l l l}
\hline 
                        name & signif. & N clusters & N members & $\mu_0$ [$10^{-3}$] & $\mu_1$ [$10^{-3}$] & $\mu_2$ [$10^{-3}$]   \\
                        \hline
                        Helmi & 9.352 & 2 & 143 (159) & -248.543 & -14.405 & 272.765 \\
                        64    & 4.180 & 1 & 67 (67) & 101.313 & -655.568 & -639.765 \\
                        Sequoia & 9.285 & 1 & 135 (135) & -169.435 & -283.902 & -361.180 \\
                        62    & 5.371 & 1 & 42 (42) & -93.511 & -782.321 & -407.653 \\
                        Thamnos 1 & 3.631 & 1 & 78 (79) & -622.767 & -788.847 & -281.345 \\
                        Thamnos 2 & 3.624 & 2 & 120 (142) & -688.164 & -782.274 & -211.327 \\
                        GE    & 22.415 & 37 & 2336 (5641) & -366.742 & -775.033 & -34.410 \\
                        6     & 3.011 & 1 & 122 (81) & -546.248 & -808.798 & -46.820 \\
                        3     & 6.342 & 1 & 3032 (3032) & -759.619 & -851.365 & 23.803 \\
                        A     & 18.991 & 10 & 644 (845) & -356.137 & -829.539 & 293.317 \\
                        12    & 4.904 & 1 & 186 (185) & -500.482 & -829.231 & 211.165 \\
                        38    & 3.108 & 1 & 69 (69) & -398.545 & -622.414 & 270.931 \\
\hline \hline
                        
                        & $\sigma_{00}$ [$10^{-6}$] & $\sigma_{01}$[$10^{-6}$]  & $\sigma_{02}$ [$10^{-6}$] & $\sigma_{11}$ [$10^{-6}$] & $\sigma_{12}$ [$10^{-6}$] &  $\sigma_{22}$ [$10^{-6}$]  \\
                        \hline
                        & 3780.679 & 5975.700 & 51.481 & 12207.147 & -99.729 & 1117.498 \\
                        & 3960.303 & -4909.646 & -2269.784 & 19788.368 & 2881.083 & 2926.456 \\
                        & 7398.933 & -4951.037 & -2777.662 & 17017.094 & 1157.163 & 3390.401 \\
                        & 1251.076 & -298.331 & -680.632 & 4547.750 & 87.359 & 800.745 \\
                        & 503.237 & 13.463 & -152.413 & 601.277 & 6.550 & 287.669 \\
                        & 284.073 & -362.520 & -202.819 & 1008.998 & 280.335 & 519.652 \\
                        & 12986.602 & 1843.637 & 2415.136 & 20865.956 & 1564.144 & 2472.506 \\
                        & 433.190 & -252.161 & 101.998 & 484.482 & -133.527 & 311.283 \\
                        & 1323.989 & 255.892 & 23.201 & 3916.625 & 210.584 & 369.282 \\
                        & 3344.491 & 202.796 & 1669.993 & 4519.979 & 148.764 & 1285.919 \\
                        & 663.590 & -266.063 & 233.675 & 1217.119 & -29.274 & 247.443 \\
                        & 859.330 & -158.430 & 421.375 & 432.881 & -135.120 & 586.415 \\
                \end{tabular}
\caption{Overview of the characteristics of the extracted substructures. The column `N clusters' lists how many clusters have been joined to form the substructure. The `N members' entry gives the total number of core members in all participating clusters (according to a Mahalanobis distance cut of 2.13 to each of these clusters), and within parentheses the total number of stars in the data set that are assigned to the substructure after a second Mahalanobis distance cut of 2.13 to the structure. 
$\mu$ is the centroid of the structures defined from the original cluster members, where indices $0$ to $2$ reflect $E$, $L_\perp$, and $L_z$ in the scaled IoM space (see paper~I). Each $\sigma_{ij}$ represents the corresponding entries in the covariance matrix. Eqs.~\ref{eq:transf_m} and \ref{eq:transf_s} can be used to transform  the means and the elements of the covariance matrix to physical units, respectively.}
\label{table:substructure_description}
\centering
\end{table*}

The top left panel of Fig.~\ref{fig:Mg_ALL} is for the Helmi streams. It is not possible to come to any conclusions on the chemical properties of this substructure as there is only one star with abundance information. Sequoia and cluster \#64 are shown in the top right panel of Fig.~\ref{fig:Mg_ALL}. It is hard to decipher whether or not \#64 and Sequoia have conclusively different abundance ratios because of the small number of member stars in APOGEE. It is nevertheless interesting  that there is a tendency for Sequoia and \#64 stars to have lower [{Mg}/{Fe}] at $[\mathrm{Fe/H}]\sim -1.5$ than {\it Gaia}-Enceladus, which is consistent with \citet[][]{2019ApJ...874L..35M},  \citet[][]{2020MNRAS.497.1236M}, and \citet{2021arXiv211115423M}. This extremely low [{Mg}/{Fe}] ratio suggests that the progenitor galaxies for the highly retrograde substructures in the nearby halo were less massive and that they have potentially experienced inefficient star formation such that the chemical enrichment by type~Ia supernovae becomes apparent at a lower [Fe/H] than for GE (further supporting our division of Sequoia into three identified clusters). Unfortunately, this scenario is based on information from only about a dozen stars with chemical information. More follow-up chemical abundance studies should be carried out in order to confirm or rule out this hypothesis. Finally, and as already noted on the basis of their MDFs and CaMDs, Sequoia and \#64 have similar stellar populations. 

The two other significantly retrograde substructures (but on lower inclination orbits), Thamnos 1 and Thamnos 2, do not seem to have very low [{Mg}/{Fe}] values (middle left panel of Fig.~\ref{fig:Mg_ALL}). Thamnos~2 has, on average, higher [{Mg}/{Fe}] than {\it Gaia}-Enceladus when compared at a fixed metallicity, while Thamnos~1 seems to be roughly consistent with the {\it Gaia}-Enceladus trend. If Thamnos 1 and Thamnos 2 are debris from small dwarf galaxies, their high [{Mg}/{Fe}] might not be unexpected given their high binding energies. This is because, in order to bring the debris from such low-mass galaxies to the inner Milky Way, the accretion event has to happen quite early on \citep{Amorisco2017}. By that time, there would not have been much time for a significant number of type~Ia supernovae to contribute to the chemical evolution in the system. 

The low-energy cluster \#3 contains stars that overlap both with the hot thick disc and the {\it Gaia}-Enceladus/accreted population in terms of chemical abundance ratios (bottom left panel of Fig.~\ref{fig:Mg_ALL}). We note however that a fraction of its stars fall in between the sequences. 

\subsection{A global view}
\label{new_view}

In this section, we take a step back and present a more global view of the nearby Galactic halo provided by the various substructures we have identified. 

\begin{figure*}
\centering 
\includegraphics[width = 0.95\textwidth]{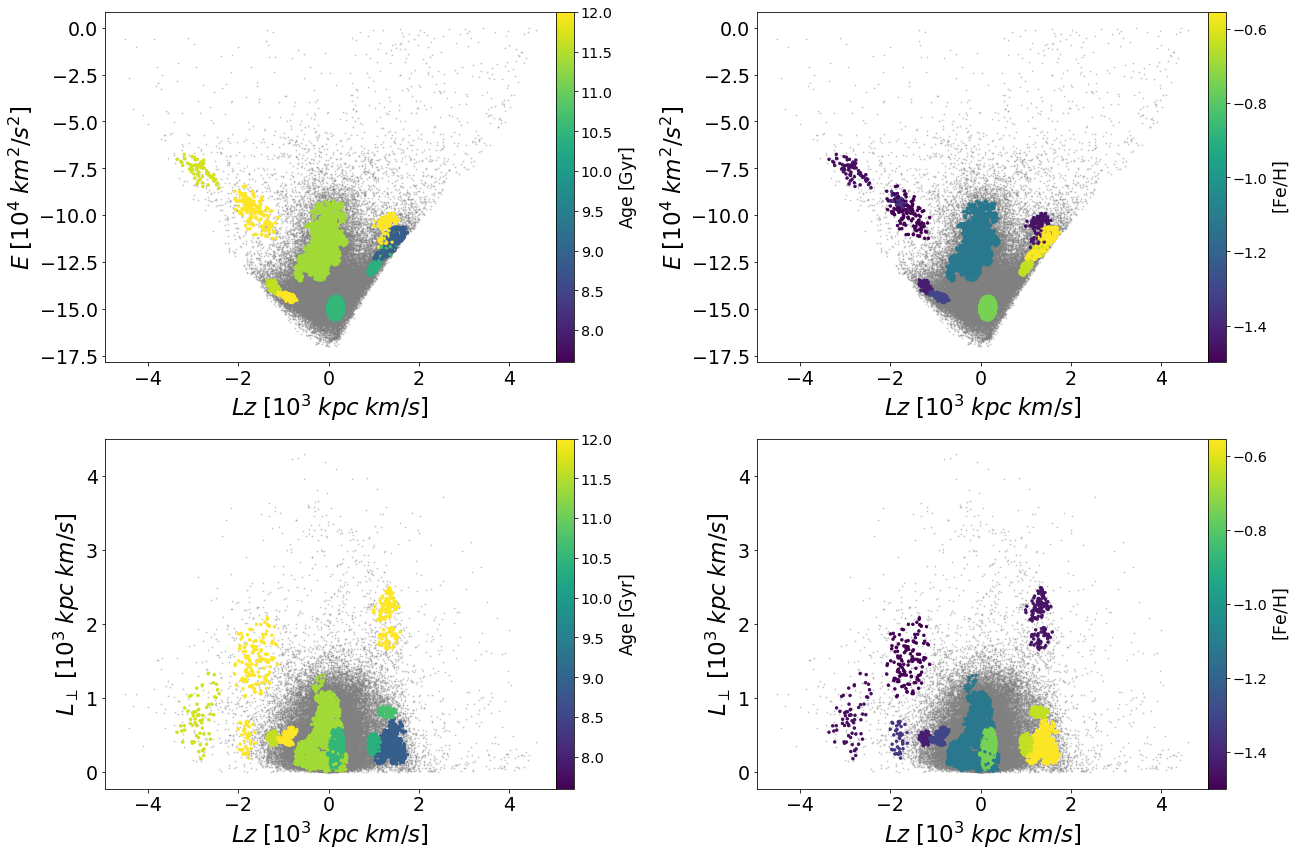} 
\caption{Distribution of the independent substructures detected in this work in the $E-L_z$ (top) and the $L_{\perp}-L_z$ (bottom) planes, colour-coded according to their isochrone fitting age (left) and [Fe/H] (right). This information is displayed in Table~\ref{age_met_tab}. In all panels, the grey background represents the original sample of halo stars analysed in this work.} 
\label{fig:age_met_IoM} 
\end{figure*}

Figure~\ref{fig:age_met_IoM} plots the different substructures in the $E-L_z$ and $L_\perp-L_z$ planes colour-coded according to their age and metallicity as presented in Sect.~\ref{comparing_substructures}. This figure shows an interesting trend. Apart from the already known tendency of metal-poor stars to be on more retrograde orbits \citep[e.g][or more recently, \citealt{2019A&A...631L...9K}]{2007Natur.450.1020C}, we find that the average age of the retrograde substructures tends to be higher than those on more prograde orbits (with the notable exception of the Helmi streams, showing an isochrone average age of $\sim$~12~Gyr). This trend is not the result of an age--metallicity degeneracy (which in fact would go in the opposite sense), because the metallicities from isochrone fitting are fully consistent with those obtained from spectroscopic surveys. 

The trend seen in Fig.~\ref{fig:age_met_IoM} has not yet been reported in the literature, either in observations or in numerical simulations, as far as we are aware. As objects on retrograde orbits show weaker dynamical coupling to the host galaxy, one might expect that such objects would be accreted later (in which case they would have younger populations which is opposite to what is observed). A possible explanation is perhaps that the structures on retrograde orbits were part of a group that fell in together, for example, with GE. In that case, pre-processing in the group would prevent star formation in the smallest objects, and could lead to an outcome such as that observed. 

\section{Conclusions}
\label{conclusions}

In this paper we thoroughly explore the outcome of a data-driven clustering algorithm \citep[see][paper I]{2022arXiv220102404S} applied to a nearby sample of halo stars (up to 2.5 kpc from the Sun) with 6D phase-space information. We performed a detailed analysis of the internal properties (i.e. metallicity distribution functions, CaMDs, and chemical abundances) of the clusters identified by the algorithm in IoM space to robustly assess their independence and to establish and confirm possible links among them. This allowed us to identify 12 large independent substructures encompassing approximately 20\% of the halo stars in our sample. The main results can be summarised as follows:
\begin{itemize}
    \item The large substructures identified include the Helmi streams, Sequoia, {\it Gaia}-Enceladus, Thamnos 1, and Thamnos 2. Given our sample definition and its spatial extent, other known substructures such as Wukong or Aleph are missing. Although we find  clusters (substructure {\bf C2} and cluster \#14; see Sect.~\ref{GE}) with similar energies and more retrograde orbits than GE, we cannot associate these to the substructure known as Rg5 \citep{2020ApJ...891...39Y}, because of their substantially different $L_\perp$. Furthermore, with the information at hand, both substructures are fully compatible with being part of GE. 
    
    \item We find that the most retrograde substructures in our sample (clusters \#62, \#63, and \#64) display kinematic properties that differ from one to another. We hypothesise that stars populating cluster \#64 (the most retrograde one) might stem in part from the metal-poor tail of {\it Gaia}-Enceladus and thus may not be fully related to Sequoia. Proper chemical abundance characterisation of many more stars from these substructures is needed to understand this portion of the halo.   
    
    \item The largest (in number) substructure in our sample is constituted by the
    single cluster \#3. It presents a clear mixture of stars with metallicities characteristic of the (hot) thick disc and a prominent metal-poor component. Our chemical abundance analysis based on APOGEE hints at the presence of an intermediate population between the high-Mg and low-Mg sequences. This suggests the existence of a buried accreted population with high binding energy \citep[but lower than that of Kraken/Heracles,][and hence apparently unrelated] {2019MNRAS.486.3180K, 2021MNRAS.500.1385H}.
    
   \item The large substructure {\bf A} is clearly dominated by the hot thick disc, but also depicts a small contribution of   a metal-poor, halo-like population. This substructure is formed by the union of ten different clusters with compatible stellar populations. The existence of such a large amount of lumpiness within the hot thick disc (also supported by the presence of four additional clusters in a similar region), could be due to the response of a disc to past mergers, as seen in the simulations by \citet{2012MNRAS.419.2163G,2017A&A...604A.106J}. 
   
    \item Most of the independent substructures identified present statistically discernable differences in their MDFs, with some similarities supporting  an origin in systems of comparable size (such as Sequoia and the Helmi streams); although in some cases contamination from neighbouring structures is likely also an issue (e.g. Thamnos 1 and 2).
    
    \item We find that substructures on more retrograde orbits are, on average, older than those in less-retrograde or prograde orbits, in addition to being more metal-poor.
    
    \item We further identify eight small clusters that, based on their stellar populations or their small size, could not be associated to the 12 main independent substructures reported above or identified as an independent substructure on their own. One of these clusters has dynamical and stellar population properties that could support a link with the Sgr streams.
\end{itemize}

In this paper, we demonstrate how the combined analysis of photometry, astrometry, and chemistry in vast samples of stars in the Milky Way can help us to discern the past history of our Galaxy. The advent of new {\it Gaia} data releases, together with upcoming spectroscopic surveys, including chemical abundance follow-ups, will tremendously boost our knowledge of the history of the Milky Way.

\begin{acknowledgements}
We thank the anonymous referee for useful comments improving the original version of this manuscript. We  gratefully  acknowledge  financial  support from a Spinoza prize.  HHK acknowledges financial support from the Martin A. and Helen Chooljian Membership at the Institute for Advanced Study. This work has made use of data from the European Space Agency (ESA) mission {\it Gaia} (\url{https://www.cosmos.esa.int/gaia}), processed by the {\it Gaia} Data Processing and Analysis Consortium (DPAC, \url{https://www.cosmos.esa.int/web/gaia/dpac/consortium}). Funding for the DPAC has been provided by national institutions, in particular the institutions participating in the {\it Gaia} Multilateral Agreement. This research makes use of {\tt python} (\url{http://www.python.org}); {\tt vaex} \citep[][]{2018A&A...618A..13B}, a {\tt Python} library for visualization and exploration of big tabular data sets; {\tt matplotlib} \citep[][]{hunter2007}, a suite of open-source python modules that provide a framework for creating scientific plots; {\tt astropy} \citep[][]{astropy2013, 2018AJ....156..123A}, a community-developed core {\tt Python} package for Astronomy; {\tt numpy} \citep[][]{2011CSE....13b..22V}; and {\tt jupyter notebooks} \citep[][]{Kluyver2016jupyter}.
This work also made use of the Third Data Release of the GALAH Survey \citep[][]{2021MNRAS.506..150B}. The GALAH Survey is based on data acquired through the Australian Astronomical Observatory. We acknowledge the traditional owners of the land on which the AAT stands, the Gamilaraay people, and pay our respects to elders past and present. 
This paper has made as well use of APOGEE DR16 data part of the SDSS IV scheme. Funding for the Sloan Digital Sky Survey IV has been provided by the Alfred P. Sloan Foundation, the U.S. Department of Energy Office of Science, and the Participating Institutions. 
We have made use of RAVE data for this work, see the RAVE web site at \url{https://www.rave-survey.org}. 
Guoshoujing Telescope (the Large Sky Area Multi-Object Fiber Spectroscopic Telescope LAMOST) is a National Major Scientific Project built by the Chinese Academy of Sciences. Funding for the project has been provided by the National Development and Reform Commission. LAMOST is operated and managed by the National Astronomical Observatories, Chinese Academy of Sciences. 
\end{acknowledgements}

\bibliographystyle{aa} 
\bibliography{references}



\onecolumn
\begin{appendix}

\section{Individual clusters properties}
\label{appendix2}

Table~\ref{clusters_prop_tab} shows an overview of the characteristics of the clusters detected by the algorithm described in Paper~I after refining them using a Mahalanobis distance limit of $D_{ij} <2.13$. We encourage the reader to refer to Paper~I for a thorough characterisation of the original clusters. Cluster ID is the cluster number identified by the algorithm; N$_\star$ is the number of stars included in this analysis ($D_{ij} <2.13$);  $\langle{L_{z}}\rangle$ represents the mean value of $L_z$; $\langle{L_{\perp}}\rangle$, the mean value of $L_\perp$; $\langle{E}\rangle$, the mean value of $E$; N$_{\rm [Fe/H]}$ is the number of stars with [Fe/H] determinations from LAMOST LRS, with $\langle{\rm [Fe/H]}\rangle$ its average value and $\sigma$([Fe/H]) its standard deviation.

\begin{table*}[h]
{\centering
\scriptsize
\begin{tabular}{rrrrrrrr}
\hline
Cluster & N$_\star$ & $\langle{L_{z}}\rangle\times$ 10$^{3}$ & $\langle{L_{\perp}}\rangle\times$ 10$^{3}$ & $\langle{E}\rangle\times$ 10$^{4}$ & N$_{\rm [Fe/H]}$ & $\langle{\rm [Fe/H]}\rangle$ & $\sigma$({\rm [Fe/H]}) \\ 
 ID &  & [kpc km/s] & [kpc km/s] & [$\rm km^2/s^2$] &  & dex & dex \\
\hline
1 & 33 & 0.16 & 0.02 & -16.5 & 1 & -0.75 & -- \\ 
2 & 12 & 0.57 & 0.1 & -13.94 & 8 & -0.86 & 0.37 \\ 
3 & 3032 & 0.16 & 0.32 & -14.96 & 2075 & -0.95 & 0.43 \\ 
4 & 14 & 0.74 & 1.18 & -13.67 & 9 & -1.23 & 0.36 \\ 
5 & 65 & -0.13 & 0.64 & -12.93 & 52 & -1.25 & 0.4 \\ 
6 & 122 & -0.16 & 0.42 & -13.14 & 88 & -1.2 & 0.35 \\ 
7 & 91 & -0.22 & 0.19 & -13.03 & 66 & -1.2 & 0.35 \\ 
8 & 70 & -0.86 & 0.52 & -14.45 & 52 & -1.43 & 0.36 \\ 
9 & 17 & -0.2 & 0.73 & -12.64 & 14 & -1.34 & 0.28 \\ 
10 & 58 & -0.33 & 0.28 & -12.8 & 37 & -1.25 & 0.35 \\ 
11 & 50 & -0.96 & 0.41 & -14.25 & 39 & -1.35 & 0.33 \\ 
12 & 186 & 1.01 & 0.37 & -12.78 & 110 & -0.85 & 0.43 \\ 
13 & 110 & -0.4 & 0.11 & -12.65 & 82 & -1.26 & 0.31 \\ 
14 & 77 & -0.67 & 0.21 & -13.12 & 57 & -1.35 & 0.36 \\ 
15 & 194 & 1.18 & 0.3 & -12.1 & 143 & -0.82 & 0.38 \\ 
16 & 21 & 1.36 & 0.24 & -11.99 & 10 & -0.66 & 0.16 \\ 
17 & 81 & 0.07 & 0.42 & -12.1 & 50 & -1.34 & 0.29 \\ 
18 & 66 & 1.36 & 0.54 & -11.7 & 51 & -0.74 & 0.38 \\ 
19 & 81 & -0.55 & 0.29 & -12.41 & 66 & -1.25 & 0.36 \\ 
20 & 14 & -0.27 & 0.06 & -12.14 & 11 & -1.28 & 0.25 \\ 
21 & 38 & -0.21 & 0.97 & -12.61 & 32 & -1.36 & 0.33 \\ 
22 & 43 & -0.06 & 0.46 & -11.51 & 31 & -1.31 & 0.3 \\ 
23 & 52 & 0.16 & 0.82 & -12.29 & 39 & -1.21 & 0.31 \\ 
24 & 345 & -0.08 & 0.72 & -12.22 & 248 & -1.27 & 0.34 \\ 
25 & 84 & -0.14 & 0.2 & -11.81 & 54 & -1.28 & 0.29 \\ 
26 & 62 & 0.11 & 0.35 & -11.3 & 40 & -1.26 & 0.33 \\ 
27 & 108 & 0.21 & 0.2 & -11.26 & 69 & -1.28 & 0.31 \\ 
28 & 30 & 1.39 & 0.23 & -11.61 & 19 & -0.57 & 0.28 \\ 
29 & 53 & 0.03 & 0.1 & -11.6 & 42 & -1.26 & 0.29 \\ 
30 & 15 & 1.42 & 0.66 & -11.81 & 10 & -0.63 & 0.13 \\ 
31 & 39 & 1.38 & 0.36 & -11.67 & 25 & -0.64 & 0.24 \\ 
32 & 230 & 1.53 & 0.31 & -11.1 & 157 & -0.72 & 0.35 \\ 
33 & 14 & 1.44 & 0.14 & -10.98 & 6 & -0.77 & 0.35 \\ 
34 & 9 & 1.57 & 0.18 & -10.81 & 6 & -0.74 & 0.14 \\ 
35 & 59 & -0.25 & 0.5 & -11.9 & 36 & -1.4 & 0.32 \\ 
36 & 123 & 0.04 & 0.94 & -11.89 & 89 & -1.3 & 0.33 \\ 
37 & 78 & -1.23 & 0.45 & -13.79 & 58 & -1.58 & 0.29 \\ 
38 & 69 & 1.29 & 0.81 & -11.93 & 39 & -0.81 & 0.38 \\ 
39 & 47 & 0.18 & 0.64 & -11.3 & 29 & -1.25 & 0.3 \\ 
40 & 10 & 1.68 & 0.11 & -10.82 & 4 & -0.83 & 0.06 \\ 
41 & 31 & -0.09 & 0.06 & -10.92 & 19 & -1.33 & 0.27 \\ 
42 & 34 & 1.46 & 0.59 & -11.1 & 26 & -0.7 & 0.14 \\ 
43 & 49 & -0.4 & 0.39 & -11.94 & 37 & -1.29 & 0.29 \\ 
44 & 29 & -0.08 & 0.59 & -10.52 & 14 & -1.32 & 0.33 \\ 
45 & 22 & -0.4 & 0.17 & -10.63 & 17 & -1.3 & 0.34 \\ 
46 & 22 & 1.51 & 0.54 & -10.79 & 17 & -1.01 & 0.49 \\ 
47 & 58 & -0.14 & 0.56 & -10.89 & 43 & -1.34 & 0.34 \\ 
48 & 46 & -0.0 & 0.15 & -10.34 & 27 & -1.27 & 0.31 \\ 
49 & 72 & -0.23 & 0.15 & -11.16 & 50 & -1.33 & 0.25 \\ 
50 & 50 & -0.05 & 0.75 & -10.94 & 34 & -1.41 & 0.32 \\ 
51 & 24 & -0.1 & 0.67 & -10.35 & 17 & -1.37 & 0.24 \\ 
52 & 43 & 0.25 & 0.06 & -9.8 & 26 & -1.27 & 0.28 \\ 
53 & 88 & 0.06 & 0.81 & -10.37 & 64 & -1.33 & 0.27 \\ 
54 & 28 & 0.15 & 0.83 & -9.66 & 15 & -1.3 & 0.38 \\ 
55 & 43 & -0.06 & 0.37 & -9.91 & 30 & -1.22 & 0.28 \\ 
56 & 76 & -0.04 & 1.01 & -10.79 & 53 & -1.3 & 0.29 \\ 
57 & 26 & -0.08 & 0.66 & -9.78 & 17 & -1.29 & 0.25 \\ 
58 & 20 & -0.25 & 0.73 & -9.73 & 18 & -1.41 & 0.25 \\ 
59 & 23 & -0.22 & 1.19 & -10.47 & 16 & -1.36 & 0.24 \\ 
60 & 95 & 1.3 & 2.26 & -10.3 & 66 & -1.61 & 0.3 \\ 
61 & 48 & 1.34 & 1.81 & -11.29 & 34 & -1.61 & 0.32 \\ 
62 & 42 & -1.82 & 0.47 & -9.28 & 30 & -1.47 & 0.28 \\ 
63 & 135 & -1.62 & 1.51 & -9.91 & 87 & -1.61 & 0.25 \\ 
64 & 67 & -2.86 & 0.72 & -7.61 & 46 & -1.62 & 0.29 \\ 
65 & 12 & 0.9 & 2.71 & -6.99 & 8 & -1.76 & 0.23 \\ 
66 & 13 & 0.59 & 3.06 & -4.11 & 9 & -1.36 & 0.31 \\ 
67 & 38 & 2.02 & 0.13 & -1.72 & 2 & -0.3 & 0.22 \\ 
68 & 29 & 3.95 & 0.16 & -3.55 & 0 & -- & -- \\
\hline \hline
\end{tabular}
\caption{Overview of the characteristics of the clusters analysed in this work.} 
\label{clusters_prop_tab}
}
\end{table*}

\section{Extended substructure summary plots}
\label{appendix1}

In this Appendix we compare the distribution of stars in IoM as well as in velocity planes of the 12 independent substructures identified in this work.

\begin{figure*}[h]
\centering 
\includegraphics[width = 0.95\textwidth]{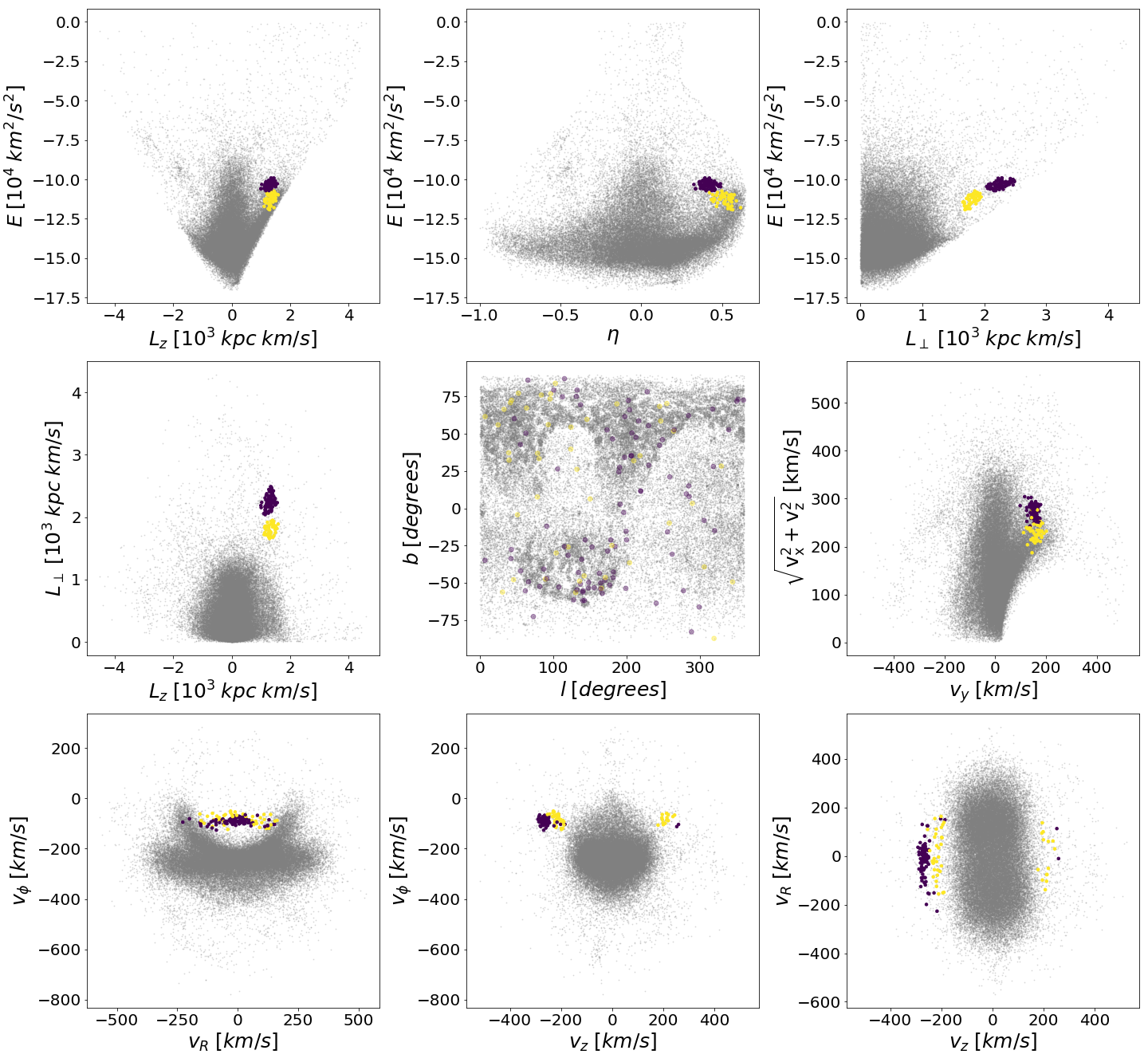} 
\caption{Distribution of stars in several projections in IoM space as well as in velocity space for the Helmi streams (substructure {\bf E}) as identified in this work. In different colours we show the different clusters making up this substructure. For reference, the grey background shows our full halo sample.} 
\label{fig:helmi_summary} 
\end{figure*}

\begin{figure*}
\centering 
\includegraphics[width = 0.95\textwidth]{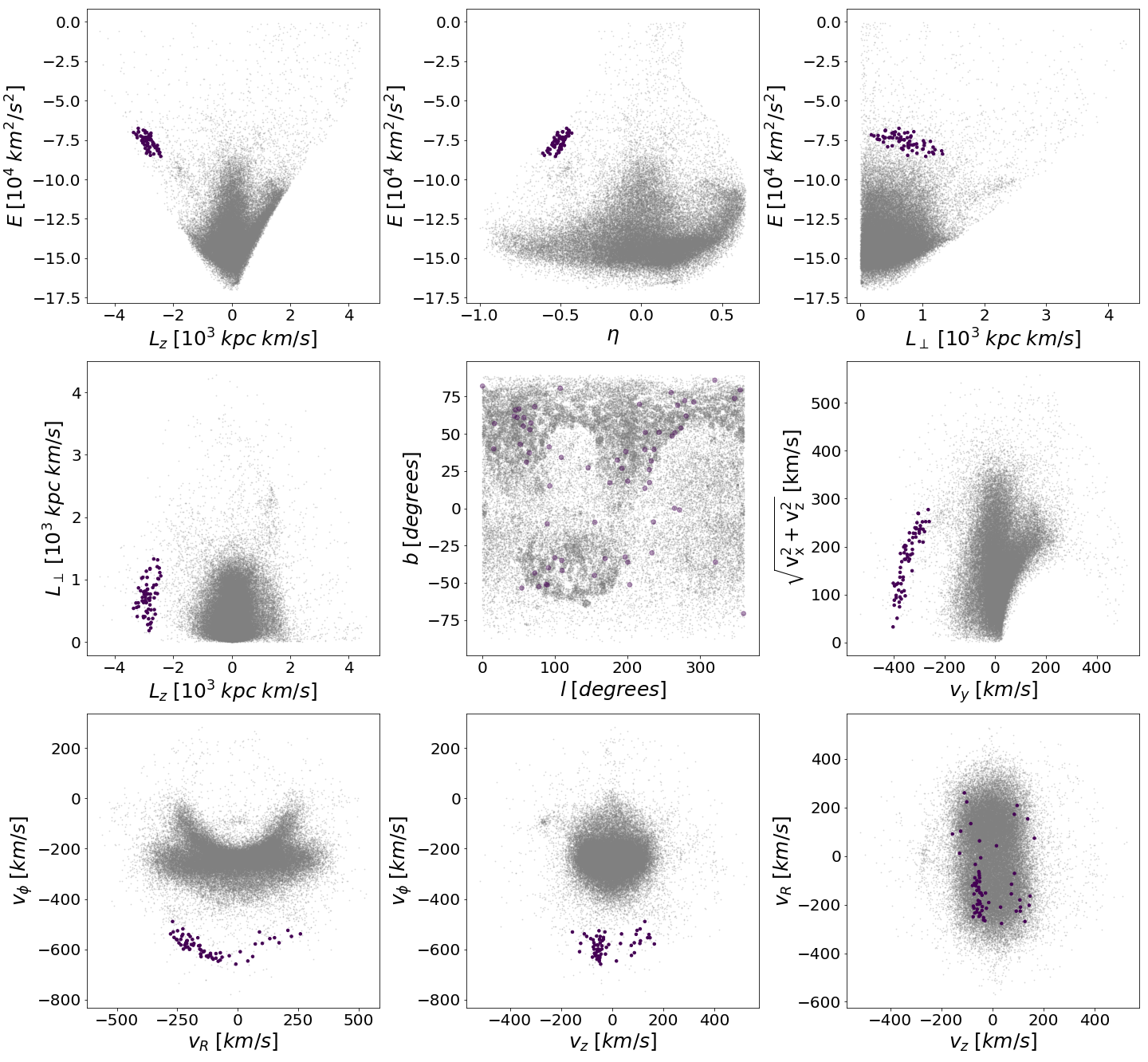} 
\caption{Cluster \#64 in IoM and velocity space. Information as in Fig.~\ref{fig:helmi_summary}.} 
\label{fig:64_summary} 
\end{figure*}

\begin{figure*}
\centering 
\includegraphics[width = 0.95\textwidth]{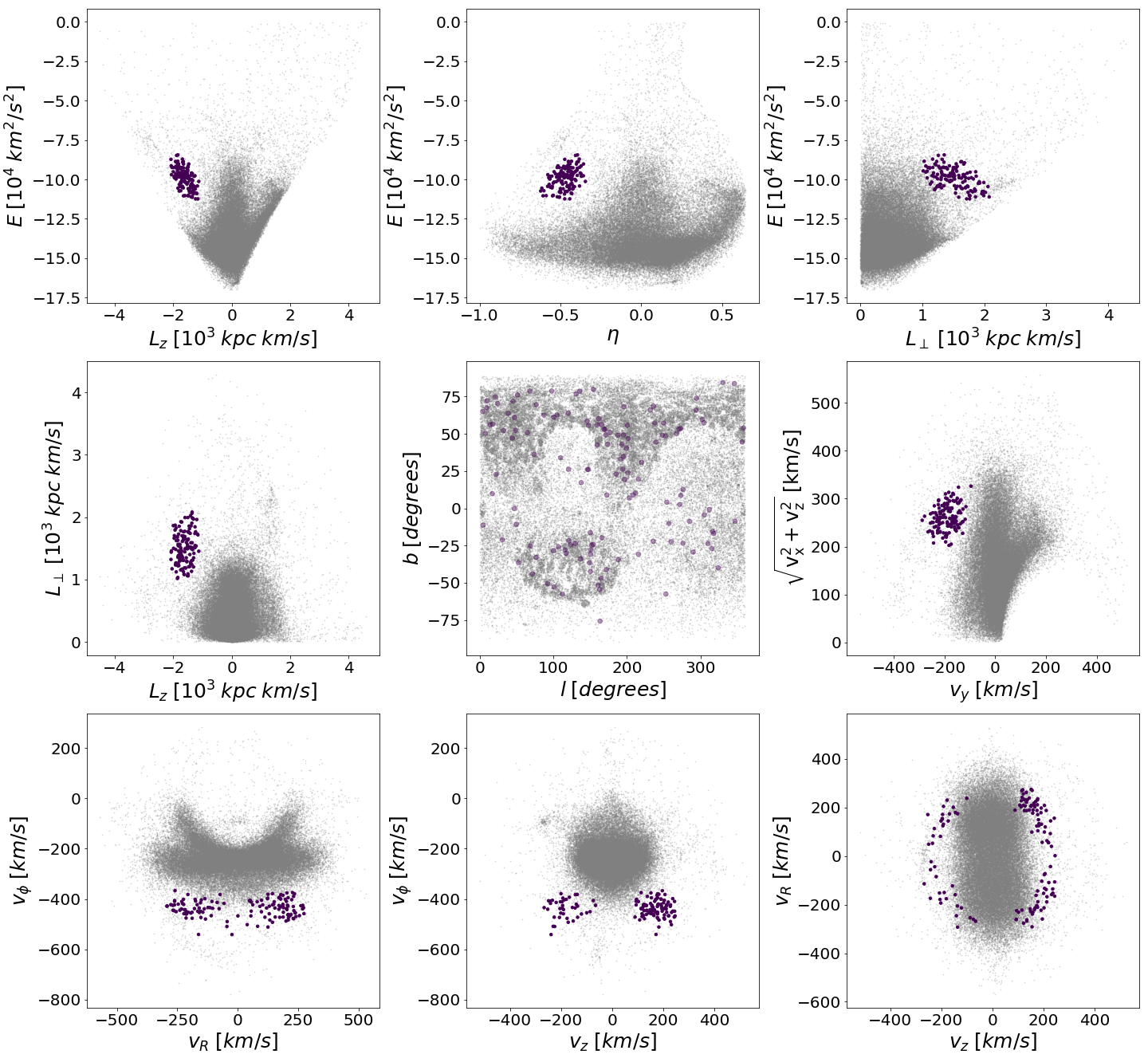} 
\caption{Sequoia (cluster \#63) in IoM and velocity space. Information as in Fig.~\ref{fig:helmi_summary}.} 
\label{fig:sequoia_summary} 
\end{figure*}

\begin{figure*}
\centering 
\includegraphics[width = 0.95\textwidth]{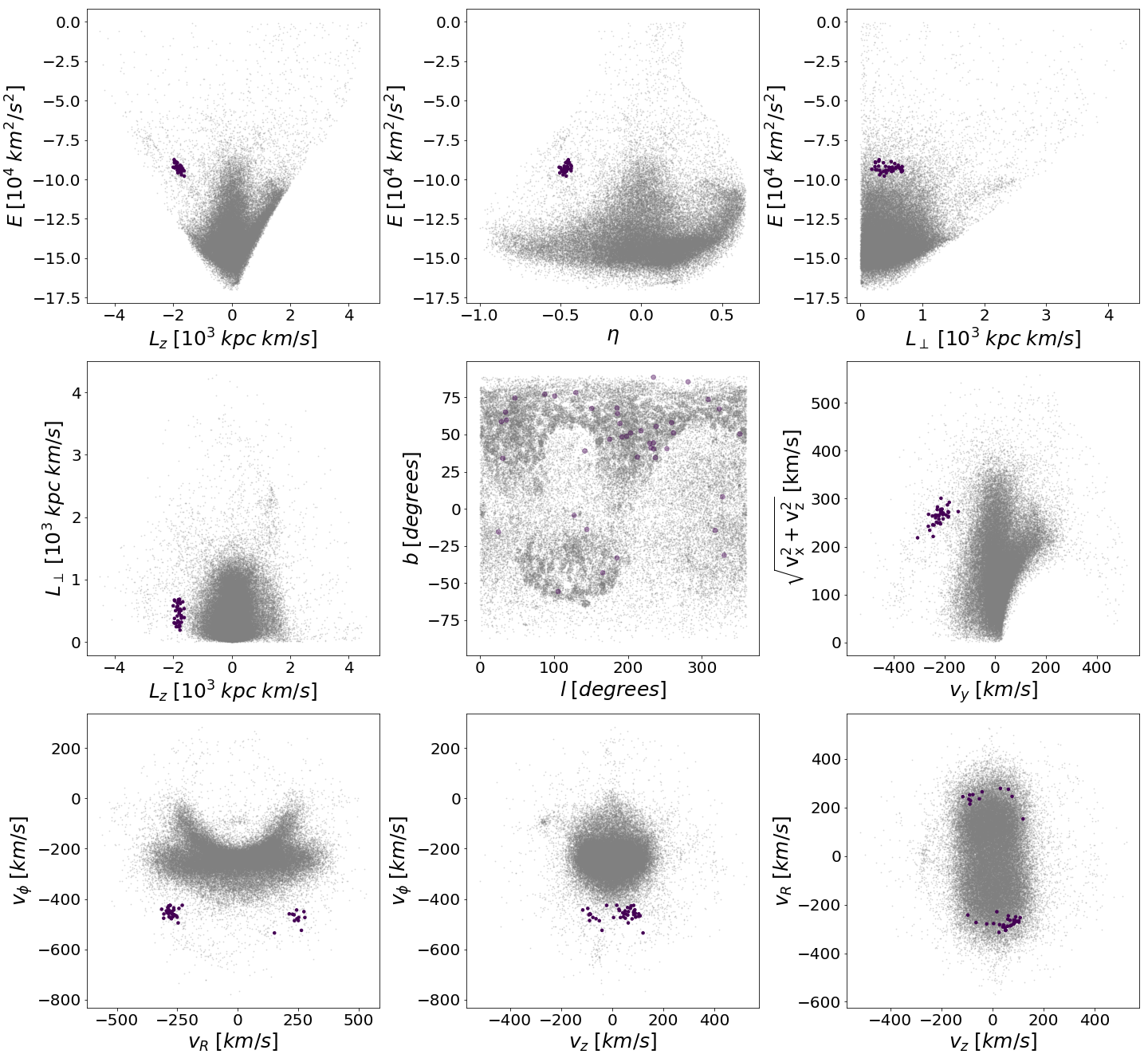} 
\caption{Cluster \#62 in IoM and velocity space. Information as in Fig.~\ref{fig:helmi_summary}.} 
\label{fig:62_summary} 
\end{figure*}

\begin{figure*}
\centering 
\includegraphics[width = 0.95\textwidth]{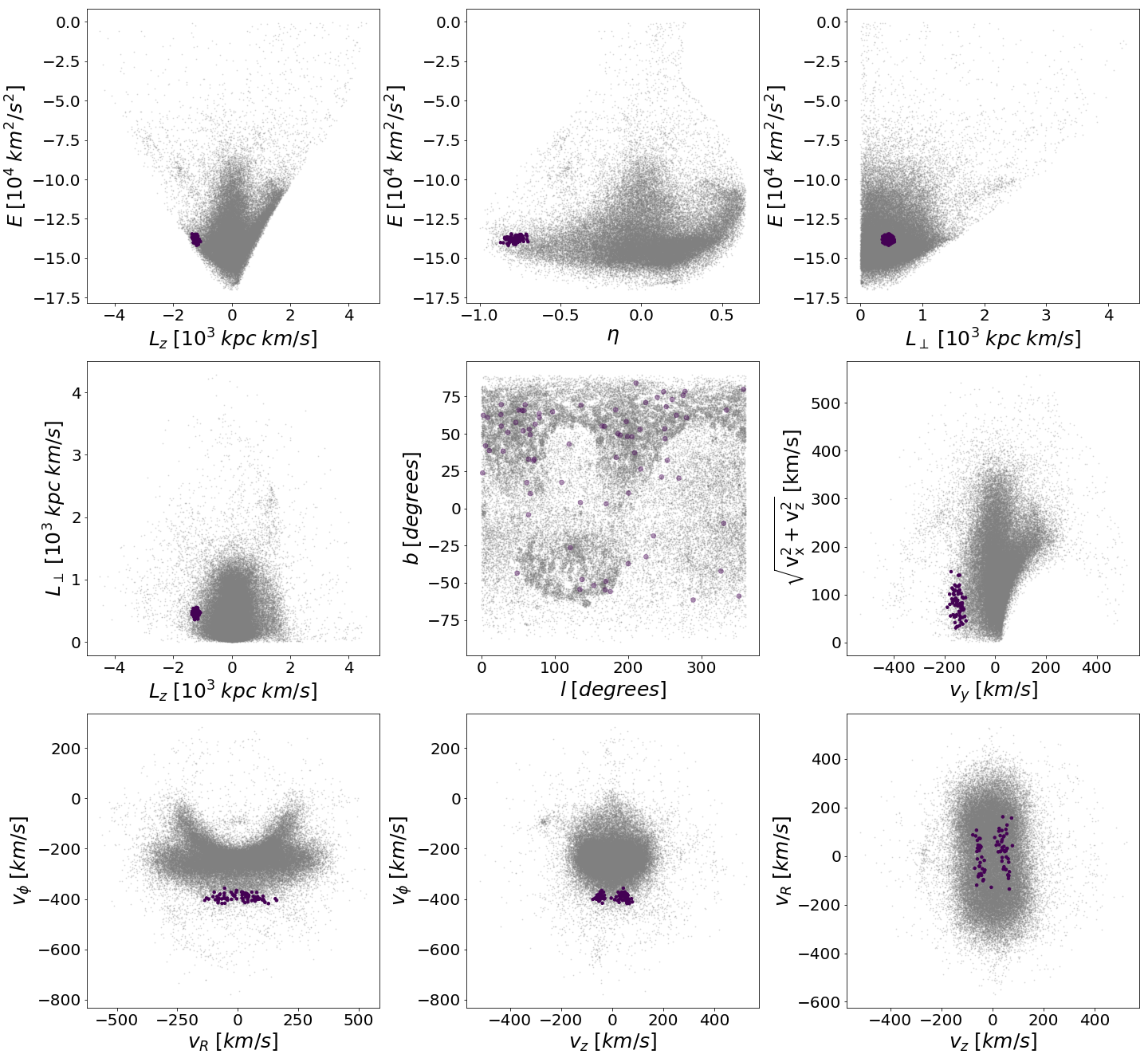} 
\caption{Thamnos 1 (cluster \#37, substructure {\bf B1}) in IoM and velocity space. Information as in Fig.~\ref{fig:helmi_summary}.} 
\label{fig:Thamnos1_summary} 
\end{figure*}

\begin{figure*}
\centering 
\includegraphics[width = 0.95\textwidth]{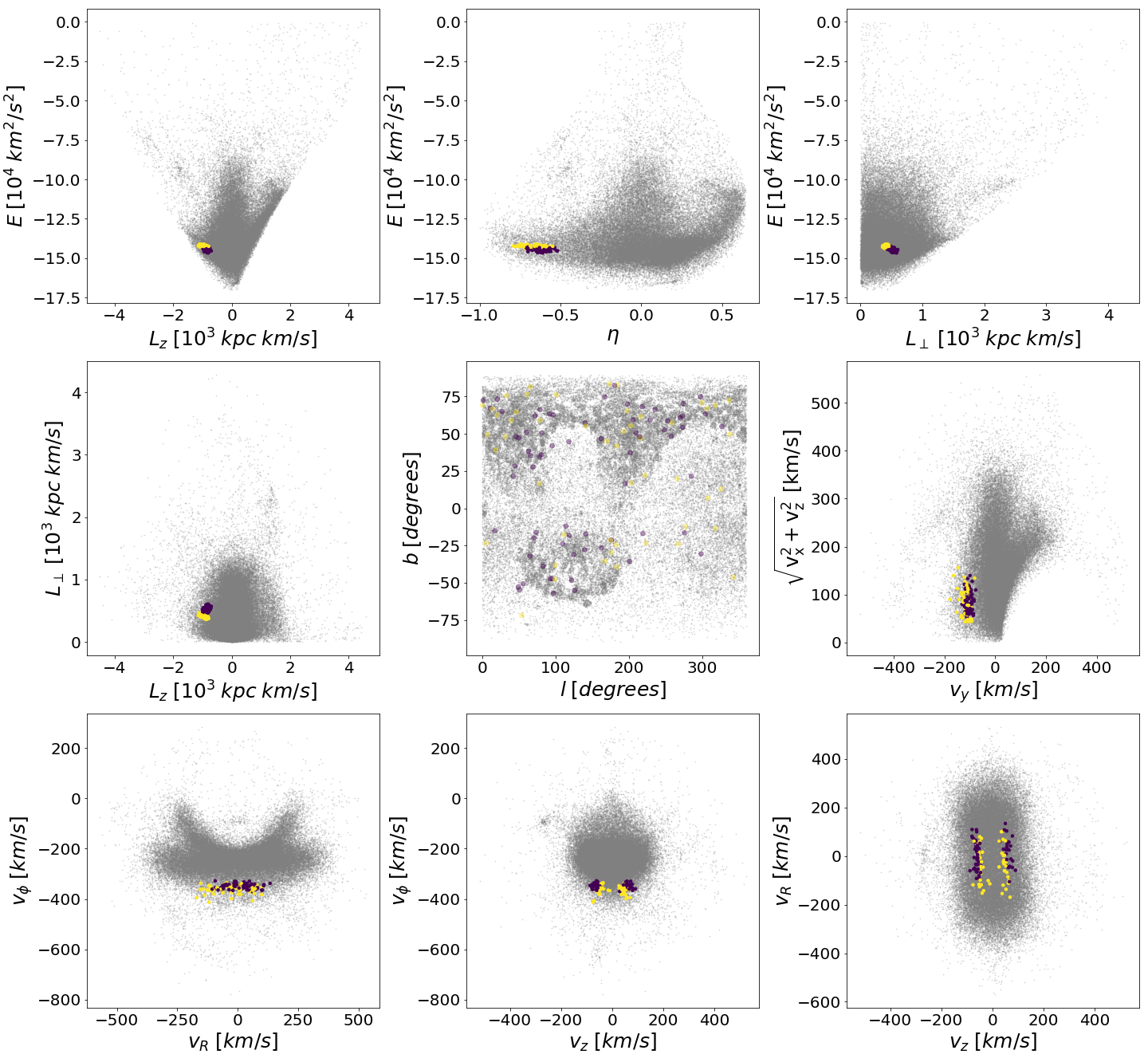} 
\caption{Thamnos 2 (substructure {\bf B2}) stars in IoM and velocity space. Information as in Fig.~\ref{fig:helmi_summary}.} 
\label{fig:Thamnos2_summary} 
\end{figure*}

\begin{figure*}
\centering 
\includegraphics[width = 0.95\textwidth]{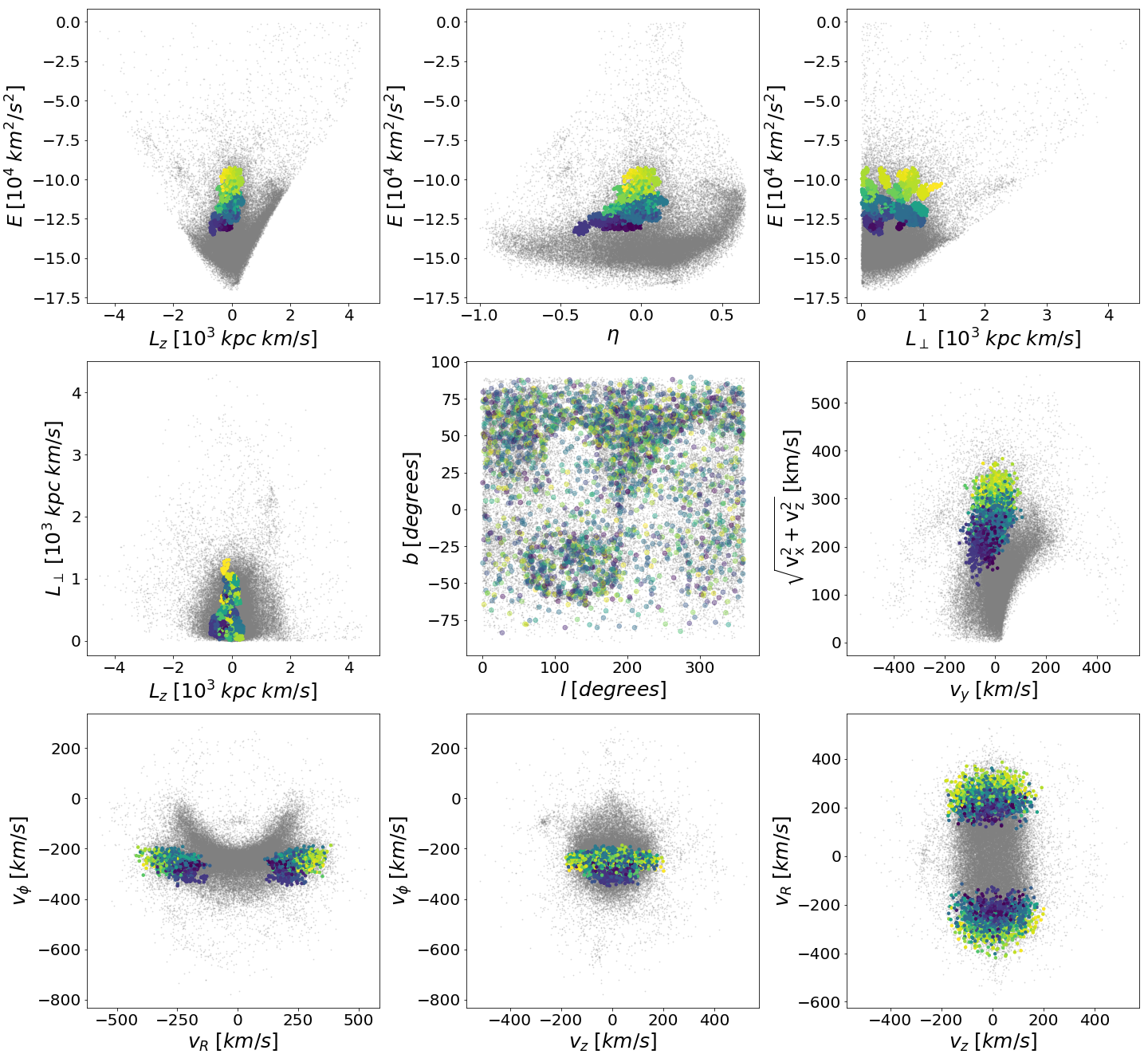} 
\caption{{\it Gaia}-Enceladus (substructure {\bf C}) in IoM and velocity space. Information as in Fig.~\ref{fig:helmi_summary}.} 
\label{fig:GE_summary} 
\end{figure*}

\begin{figure*}
\centering 
\includegraphics[width = 0.95\textwidth]{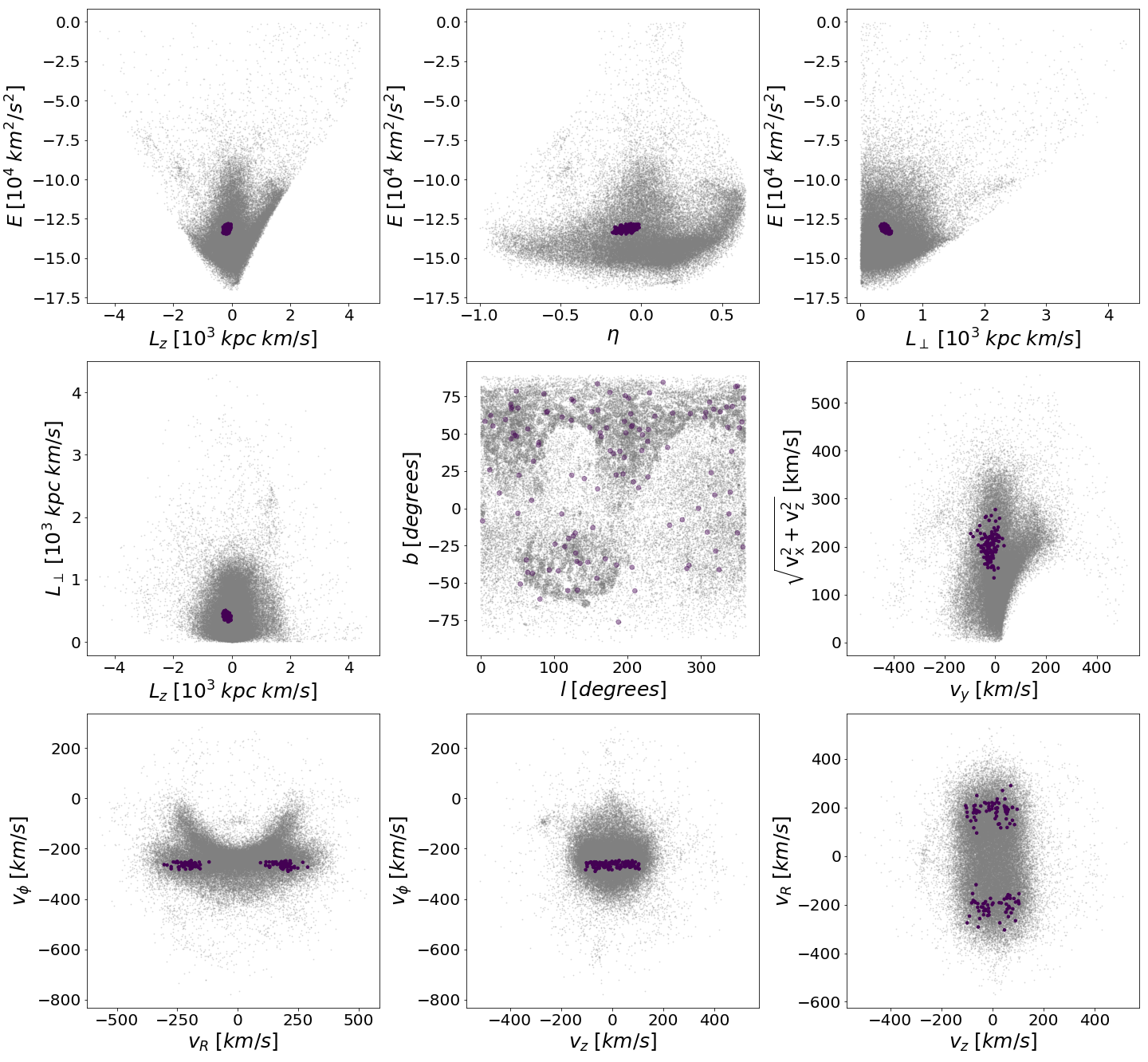} 
\caption{Cluster \#6 in IoM and velocity space. Information as in Fig.~\ref{fig:helmi_summary}.} 
\label{fig:6_summary} 
\end{figure*}

\begin{figure*}
\centering 
\includegraphics[width = 0.95\textwidth]{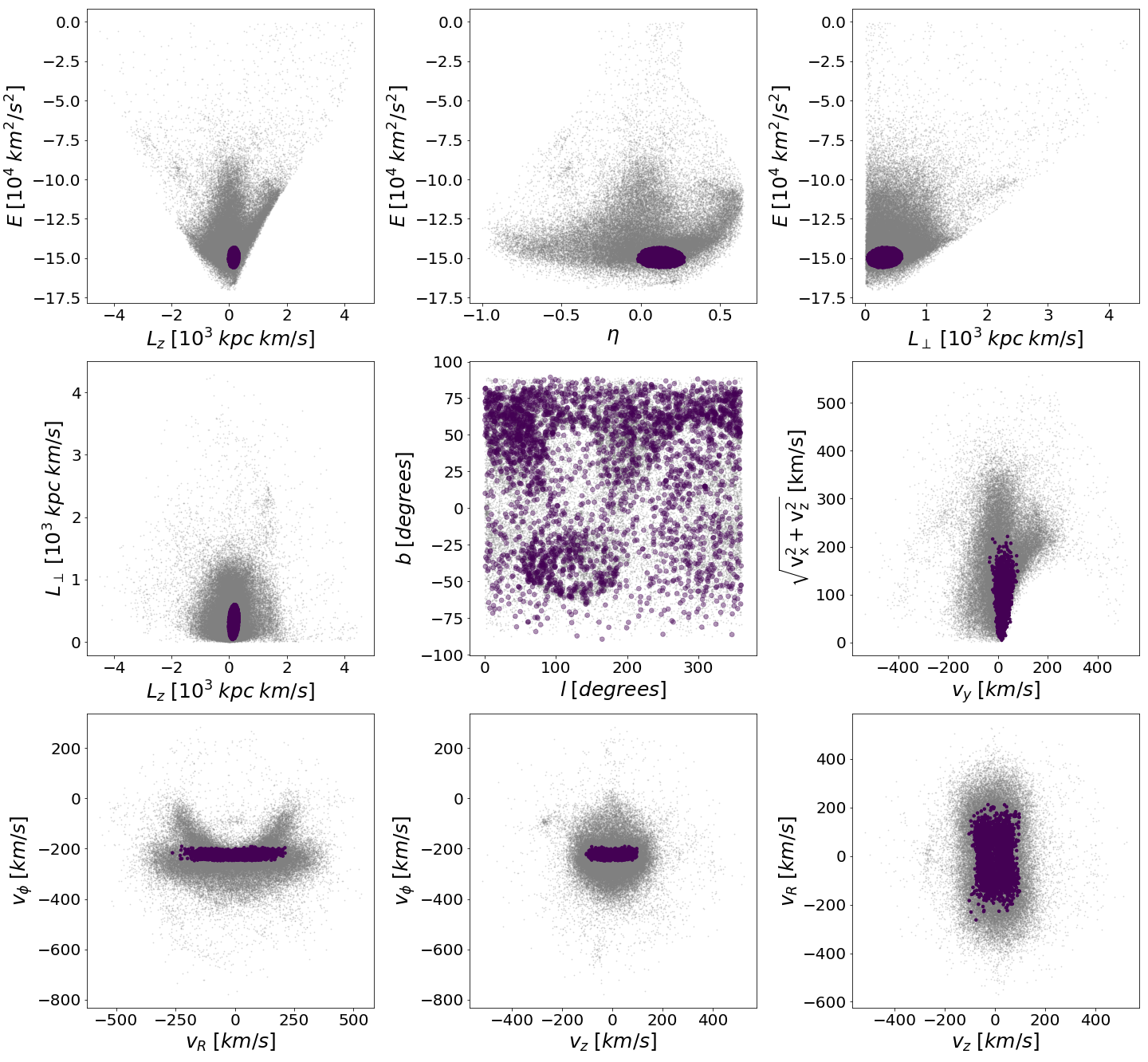} 
\caption{Cluster \#3 in IoM and velocity space. Information as in Fig.~\ref{fig:helmi_summary}.} 
\label{fig:3_summary} 
\end{figure*}

\begin{figure*}
\centering 
\includegraphics[width = 0.95\textwidth]{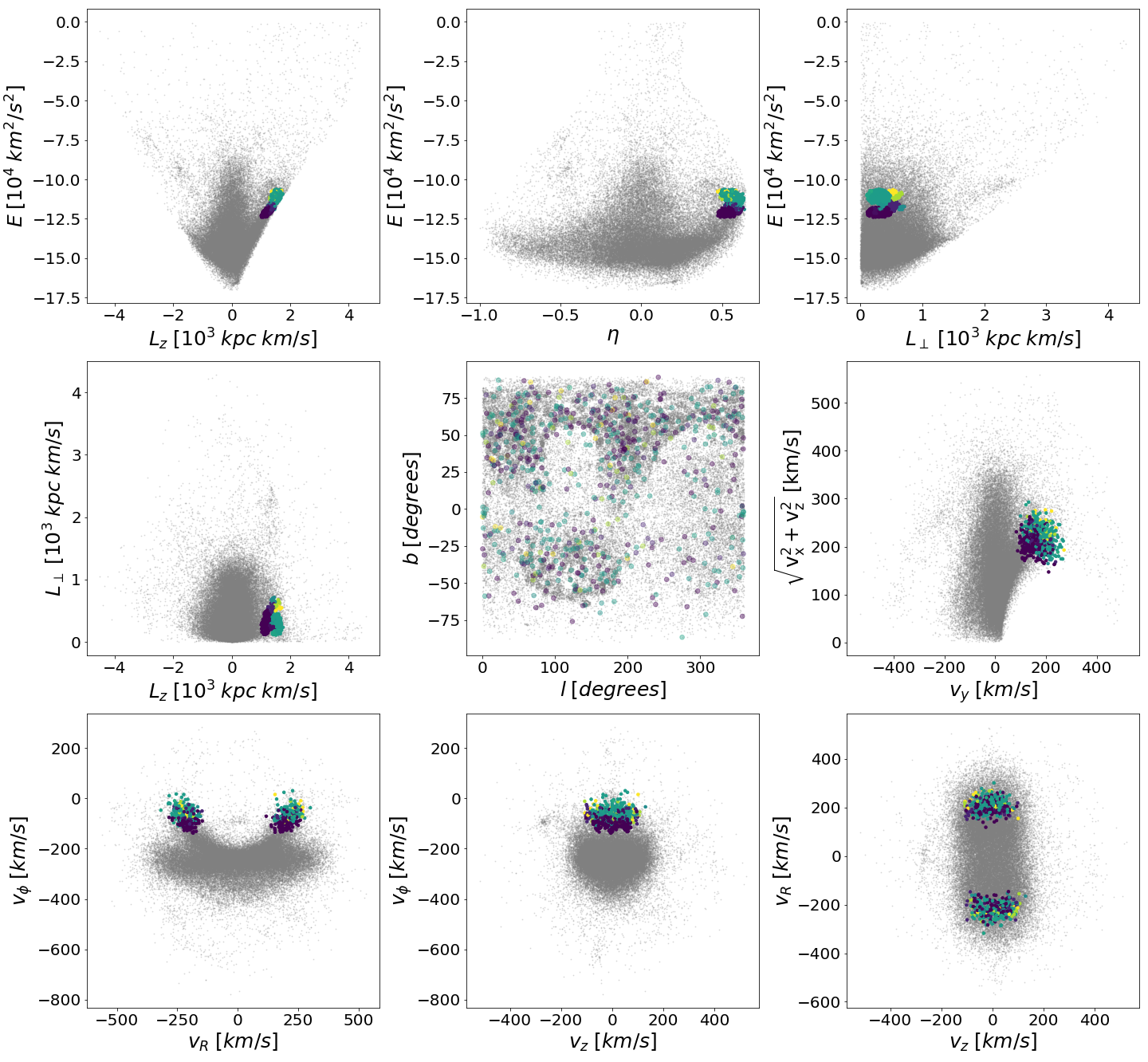} 
\caption{Substructure {\bf A} in IoM and velocity space. Information as in Fig.~\ref{fig:helmi_summary}.} 
\label{fig:A_summary} 
\end{figure*}

\begin{figure*}
\centering 
\includegraphics[width = 0.95\textwidth]{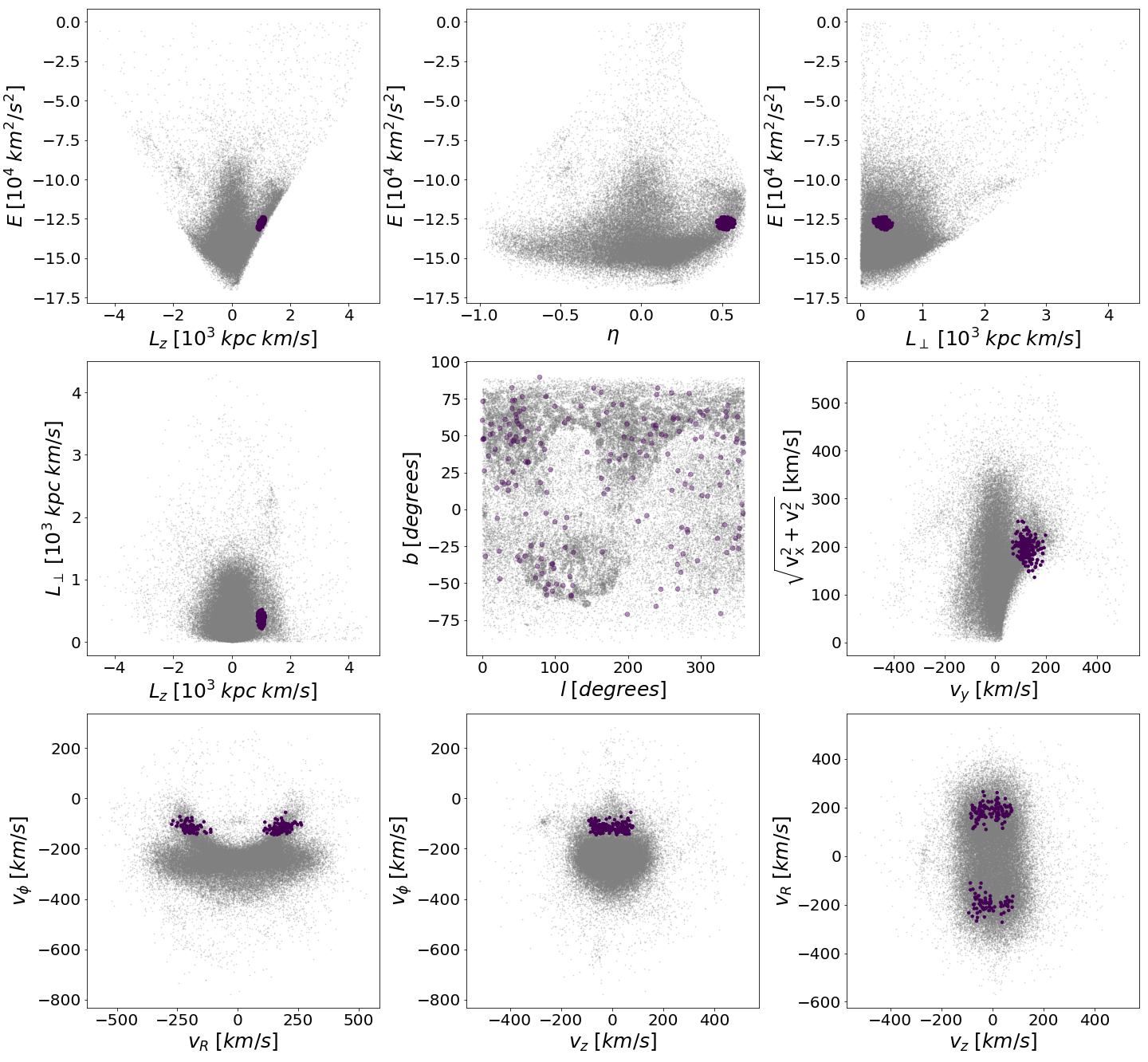} 
\caption{Cluster \#12 in IoM and velocity space. Information as in Fig.~\ref{fig:helmi_summary}.} 
\label{fig:12_summary} 
\end{figure*}

\begin{figure*}
\centering 
\includegraphics[width = 0.95\textwidth]{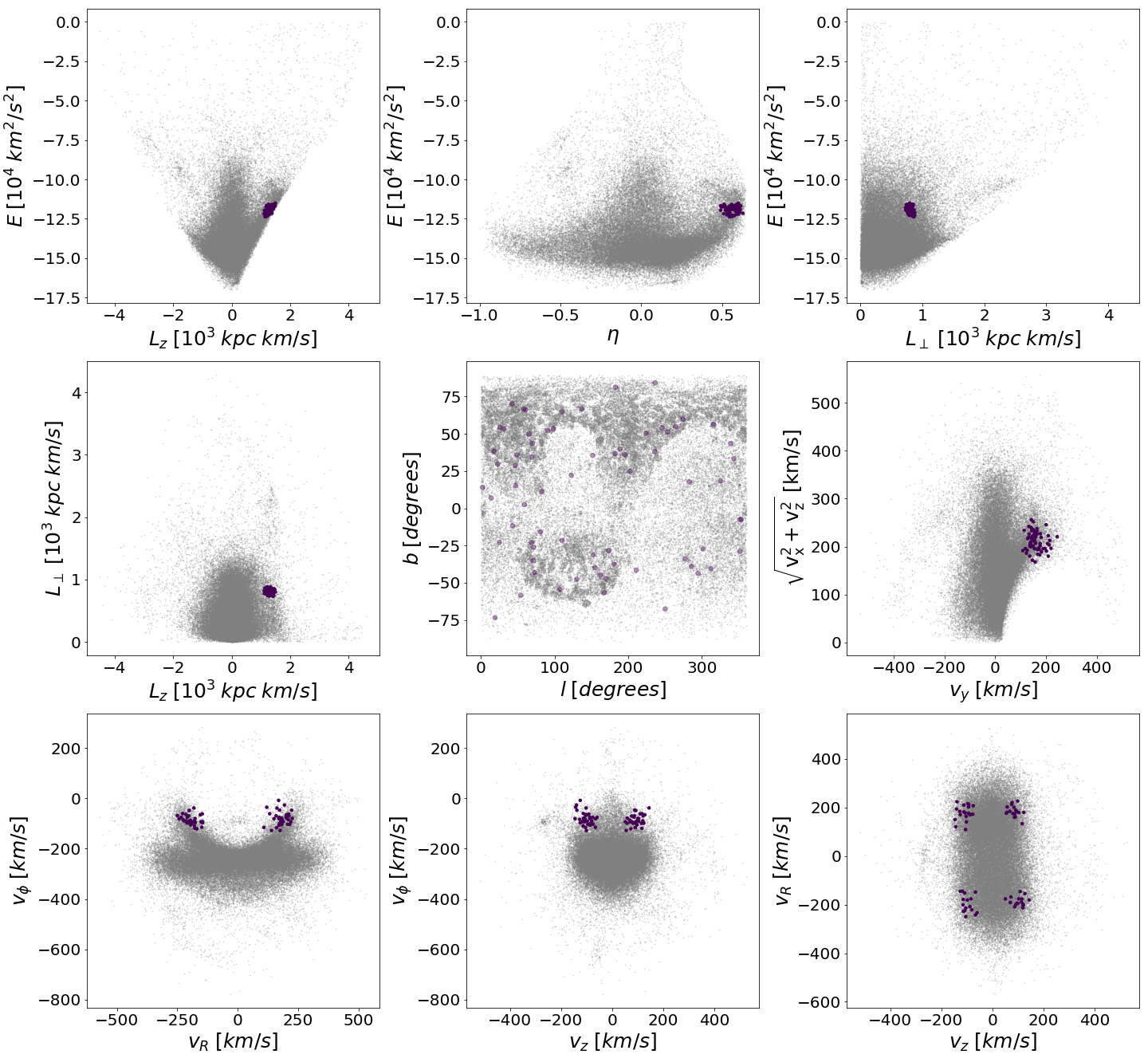} 
\caption{Cluster \#38 in IoM and velocity space. Information as in Fig.~\ref{fig:helmi_summary}.} 
\label{fig:38_summary} 
\end{figure*}

\end{appendix}
\end{document}